\pdfoutput=1
\documentclass[twocolumn,showpacs,preprintnumbers,amsmath,amssymb,superscriptaddress,floatfix]{revtex4}

\usepackage{graphicx}
\graphicspath{{fig/}}
\usepackage{epstopdf}
\usepackage{dcolumn}
\usepackage{bm}
\usepackage{color}
\begin{document}

\preprint{}

\title{Design and Performance of the XENON10 Dark Matter Experiment}

\author{E.~Aprile}
\email[XENON Spokesperson (E. Aprile). Phone: +1-212-854-3258. \\   E-mail address: ]{age@astro.columbia.edu.}
\affiliation{Department of Physics, Columbia University, New York, NY 10027, USA}

\author{J.~Angle}
\affiliation{Department of Physics, University of Florida, Gainesville, FL 32611, USA}
\affiliation{Physik-Institut, Universit\"at Z\"urich, Z\"urich, 8057, Switzerland}

\author{F.~Arneodo}
\affiliation{INFN - Laboratori Nazionali del Gran Sasso, Assergi, 67010, Italy}

\author{L.~Baudis}
\affiliation{Department of Physics, University of Florida, Gainesville, FL 32611, USA}
\affiliation{Physik-Institut, Universit\"at Z\"urich, Z\"urich, 8057, Switzerland}

\author{A.~Bernstein}
\affiliation{Lawrence Livermore National Laboratory, Livermore, CA 94550, USA}

\author{A.~Bolozdynya}
\affiliation{Department of Physics, Case Western Reserve University, Cleveland, OH 44106, USA}

\author{P.~Brusov}
\affiliation{Department of Physics, Case Western Reserve University, Cleveland, OH 44106, USA}

\author{L.C.C.~Coelho}
\affiliation{Department of Physics, University of Coimbra, R. Larga, 3004-516, Coimbra, Portugal}

\author{C.E.~Dahl}
\affiliation{Department of Physics, Case Western Reserve University, Cleveland, OH 44106, USA}
\affiliation{Department of Physics, Princeton University, Princeton, NJ 08540, USA}

\author{L.~DeViveiros}
\affiliation{Department of Physics, Brown University, Providence, RI 02912, USA}

\author{A.D.~Ferella}
\affiliation{Physik-Institut, Universit\"at Z\"urich, Z\"urich, 8057, Switzerland}
\affiliation{INFN - Laboratori Nazionali del Gran Sasso, Assergi, 67010, Italy}

\author{L.M.P.~Fernandes}
\affiliation{Department of Physics, University of Coimbra, R. Larga, 3004-516, Coimbra, Portugal}

\author{S.~Fiorucci}
\affiliation{Department of Physics, Brown University, Providence, RI 02912, USA}

\author{R.J.~Gaitskell}
\affiliation{Department of Physics, Brown University, Providence, RI 02912, USA}

\author{K.L.~Giboni}
\affiliation{Department of Physics, Columbia University, New York, NY 10027, USA}

\author{R.~Gomez}
\affiliation{Department of Physics and Astronomy, Rice University, Houston, TX, 77251, USA}

\author{R.~Hasty}
\affiliation{Department of Physics, Yale University, New Haven, CT 06511, USA}

\author{L.~Kastens}
\affiliation{Department of Physics, Yale University, New Haven, CT 06511, USA}

\author{J.~Kwong}
\affiliation{Department of Physics, Case Western Reserve University, Cleveland, OH 44106, USA}
\affiliation{Department of Physics, Princeton University, Princeton, NJ 08540, USA}

\author{J.A.M.~Lopes}
\affiliation{Department of Physics, University of Coimbra, R. Larga, 3004-516, Coimbra, Portugal}

\author{N.~Madden}
\affiliation{Lawrence Livermore National Laboratory, Livermore, CA 94550, USA}

\author{A.~Manalaysay}
\affiliation{Department of Physics, University of Florida, Gainesville, FL 32611, USA}
\affiliation{Physik-Institut, Universit\"at Z\"urich, Z\"urich, 8057, Switzerland}

\author{A.~Manzur}
\affiliation{Department of Physics, Yale University, New Haven, CT 06511, USA}

\author{D.N.~McKinsey}
\affiliation{Department of Physics, Yale University, New Haven, CT 06511, USA}

\author{M.E.~Monzani}
\affiliation{Department of Physics, Columbia University, New York, NY 10027, USA}

\author{K.~Ni}
\email[Corresponding author (K. Ni). Now at Physics Department, Shanghai Jiao Tong University, China. Phone: +86-21-34202904. \\   E-mail address: ]{nikx@sjtu.edu.cn.}
\affiliation{Department of Physics, Yale University, New Haven, CT 06511, USA}

\author{U.~Oberlack}
\affiliation{Department of Physics and Astronomy, Rice University, Houston, TX, 77251, USA}

\author{J.~Orboeck}
\affiliation{Physik-Institut, Universit\"at Z\"urich, Z\"urich, 8057, Switzerland}

\author{D.~Orlandi}
\affiliation{INFN - Laboratori Nazionali del Gran Sasso, Assergi, 67010, Italy}

\author{G.~Plante}
\affiliation{Department of Physics, Columbia University, New York, NY 10027, USA}

\author{R.~Santorelli}
\affiliation{Department of Physics, Columbia University, New York, NY 10027, USA}

\author{J.M.F.~dos~Santos}
\affiliation{Department of Physics, University of Coimbra, R. Larga, 3004-516, Coimbra, Portugal}

\author{P.~Shagin}
\affiliation{Department of Physics and Astronomy, Rice University, Houston, TX, 77251, USA}

\author{T.~Shutt}
\affiliation{Department of Physics, Case Western Reserve University, Cleveland, OH 44106, USA}

\author{P.~Sorensen}
\affiliation{Department of Physics, Brown University, Providence, RI 02912, USA}

\author{S.~Schulte}
\affiliation{Physik-Institut, Universit\"at Z\"urich, Z\"urich, 8057, Switzerland}

\author{E.~Tatananni}
\affiliation{INFN - Laboratori Nazionali del Gran Sasso, Assergi, 67010, Italy}

\author{C.~Winant}
\affiliation{Lawrence Livermore National Laboratory, Livermore, CA 94550, USA}

\author{M.~Yamashita}
\affiliation{Department of Physics, Columbia University, New York, NY 10027, USA}

\collaboration{XENON Collaboration}
\date{\today}

\begin{abstract}
XENON10 is the first two-phase xenon time projection chamber (TPC) developed within the XENON dark matter
search program. The TPC, with  an active liquid xenon (LXe) mass of about 14 kg, was installed at the Gran Sasso
underground laboratory (LNGS) in Italy, and operated for more than one year, with excellent stability and
performance. Results from a dark matter search with XENON10 have been published elsewhere. In this paper, we
summarize the design and performance of the detector and its subsystems, based on calibration data using
sources of gamma-rays and neutrons as well as background and Monte Carlo simulations data. The results on the
detector's energy threshold, energy and position resolution, and overall efficiency show a performance that
exceeds design specifications, in view of the very low energy threshold achieved ($<$10 keVr) and  the
excellent energy resolution achieved by combining the ionization and scintillation signals, detected
simultaneously. 
\end{abstract}

\pacs{95.35.+d, 29.40.Mc, 95.55.Vj}

\maketitle
\tableofcontents

\section{Introduction}
\label{intro}

The XENON10 detector, a two-phase time projection chamber (TPC) containing 25 kg of pure liquid xenon (LXe),
was built and operated as the first practical prototype of the XENON dark matter search program~\cite{aprile:2005na}.
The R\&D phase which culminated with the XENON10 TPC, involved several smaller size detectors, dedicated to
study many of the properties and  performance characteristics of a two-phase TPC in the energy regime relevant
to the direct detection of dark matter particles scattering off Xe nuclei. The most important  results
obtained with these prototypes are published~\cite{Aprile:2004,Leff,PRL06,Ni:2006th}. 

In addition, several technical developments carried
out during this phase played an important role in establishing the feasibility of the XENON approach for dark
matter detection. Among these, two are particularly important: a) the development by Hamamatsu Photonics of
compact metal channel photomultipliers (PMTs) for the detection of the LXe scintillation light, with
continuous improvement in quantum efficiency and radio-purity~\cite{Aprile:2008ft, Adam:2009ci, Giboni:2007zz}; b)  the development of a pulse tube refrigerator (PTR) optimized for LXe temperature~\cite{Haruyama:2005} to achieve the
required long-term stability of a cryogenics detector filled with a large LXe volume. 

The goal of the XENON10 experiment was to demonstrate the achievable energy threshold, background rejection
power and operation stability of a two-phase TPC at the 10 kg fiducial mass scale, prior to the realization of
a 100~kg scale detector, originally proposed as unit module for a ton scale XENON dark matter search. 

The fast-paced R\&D effort did not emphasize materials radio-purity and XENON10 was built with largely
off-the-shelf components, not screened or selected to minimize backgrounds. The TPC and its associated
cryogenics, purification and data acquisition (DAQ) systems were developed and tested at the Nevis
Laboratories of Columbia University prior to installation at the Italian Gran Sasso Underground  Laboratory
(LNGS)~\cite{LNGS}. The commissioning of the XENON10 detector at LNGS started in April 2006. Following the completion
of the shield system, source calibration and background data were acquired for several months, in stable
conditions. A blind analysis of 58.4 live days of data, acquired between October 6, 2006 and February 14,
2007, and using a fiducial mass of 5.4 kg, excluded previously unexplored parameter space, setting improved upper
limits on both spin-independent~\cite{Angle:2007uj} and spin-dependent~\cite{Angle:2008we} cross-section for dark matter
scattering off nucleons.

The XENON10 results,  and the unusually fast time scale in which they were achieved, validated the scientific reach of a position
sensitive, homogeneous and self-triggered LXeTPC  for dark matter direct detection. 
New and improved LXe dark matter detectors, with a target mass at the 100 kg scale, are currently in operation ~\cite{Aprile:2009zzc}
or under construction~\cite{XMASS,LUX}. They promise to advance the field with significant improvement in sensitivity over the
next few years.  In this paper we wish to summarize the development and performance of the XENON10 detector
and associated systems. The plan of the paper is the following. In Section 2 we describe the design of the
detector, cryogenics, purification and slow-control systems, followed by a description of the shield and data
acquisition systems. In Section 3 we discuss the data processing, the algorithms and cuts used to infer the
parameters relevant for a dark matter search. In Section 4, we present results from gamma and neutron
calibration of the XENON10 TPC. Finally in Section 5, details on the analysis of XENON10 data used for the published dark
matter search results are discussed. The appendix to the paper deals with a summary of the sources of backgrounds in
the XENON10 experiment, in comparison with Monte Carlo predictions.

\section{The XENON10 Detector}

\subsection{Principle of Operation}

\begin{figure}[htb]
    \includegraphics[width=0.5\textwidth]{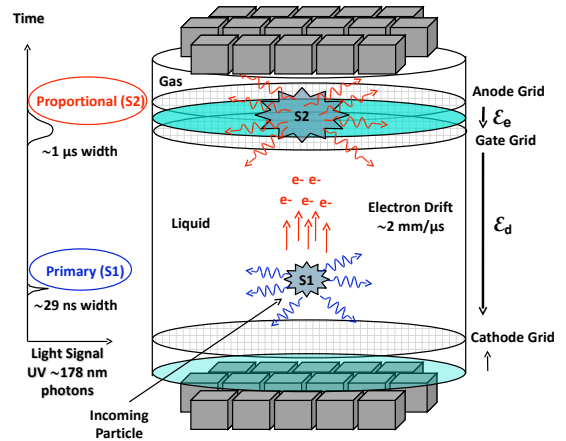}
    \caption{(Color online) Principle of operation of the XENON10 two-phase TPC. A particle interaction in the liquid produces primary scintillation light ($S1$) and ionization electrons. The electrons drift under an electric field $\varepsilon_d$   (about 1 kV/cm) until they are extracted from the liquid into the gas where they are accelerated by an electric field $\varepsilon_e$ (about 10 kV/cm) producing  proportional scintillation light ($S2$).  Two arrays of photomultipliers, one in the liquid and one in the gas,  detect simultaneously the $S1$ and $S2$  light signals.}
    
    \label{fig:tpccartoon}
\end{figure}

The principle of operation of the XENON10 two-phase (liquid-gas) time projection chamber (TPC) is shown
schematically in Fig.~\ref{fig:tpccartoon}. The energy deposited by a particle interaction within the active
LXe volume is detected by the simultaneous measurement of  ionization electrons and of primary scintillation photons ($S1$), with a wavelength of 178 nm~\cite{Jortner65}, produced by the de-excitation to
the ground state of excited diatomic Xe molecules (Xe$^{\ast}_2$)~\cite{SKubota:79:free:elec}. Both direct excitation of atoms and electron-ion recombination lead to the formation of Xe$^{\ast}_2$. 

Under the application of an electric field, $\varepsilon_d$, on the order of 1 kV/cm, the electrons which escape recombination with the parent positive ions, drift towards the liquid-gas interface with a velocity of about 2~mm/$\mu$s. Once they reach the liquid surface, they are extracted into the gas,
where a much stronger electric field, $\varepsilon_e$, accelerate them  leading to secondary scintillation photons emission ($S2$).  In the XENON10 TPC, photomultiplier tubes (PMTs), coupled directly to the sensitive liquid and gas regions, are used to detect both $S1$ and $S2$ signals. As shown in \cite{Aprile:2004}, the extraction yield is 100\% for a field
greater than 10 kV/cm. The number of secondary scintillation photons emitted in the
gas is proportional to the number of  electrons, and thus to the energy deposited in the liquid by the incoming particle. We refer to this signal as proportional scintillation signal.  The following equation expresses the dependence of this signal on the operation conditions:
\begin{equation} \label{eq:prop}
	N_{\gamma}=\alpha N_e(\varepsilon_e/p-\beta)p x \text{.}
\end{equation}
Here $p$ is the gas pressure and $x$ is the distance traveled by the
electrons in the gas which largely determines the width of the $S2$ pulse. The units for $p$, $\varepsilon_e$ and $x$ are bar, kV/cm and cm, respectively. \(N_e\) is the number of electrons extracted from the liquid to the gas phase, $N_{\gamma}$
is the number of proportional scintillation photons produced. $\alpha$ (the amplification factor) and $\beta$ (the threshold of the reduced field for proportional light production) are experimentally determined values. For the case of Xe, the reported values for $\alpha$ vary from 70 to 140 and $\beta$ values are between 0.8 and 1.0 kV/cm/bar~\cite{Bolozdynya:1999,Aprile:2004,Monteiro:2007}. The number of secondary photons generated by one drifting electron is sufficiently large to be detected by PMTs, thus a two-phase detector is single-electron sensitive. In XENON10 operating conditions, the values of $p$, $\varepsilon_e$ and $x$ were 2.1~bar, 12 kV/cm and 0.25~cm, respectively,  yielding $\sim$100 photons per extracted electron. 

The different amplitude of the $S1$ and $S2$ signals associated with nuclear recoils, such as produced by WIMP or neutron
interactions, and with electron recoils, such as produced by background beta and gamma-rays, provides the basis for
background discrimination in a two phase TPC such as XENON10. Since electron diffusion in LXe is small~\cite{aprile:2009}, the
proportional scintillation pulse is produced in a small volume with the same $XY$ coordinates as the
interaction site, allowing 2D localization. The $Z$ coordinate is inferred from the drift time measurement, and the known electron drift velocity at the operating field. The TPC approach and the strong self-shielding property of LXe ($\rho\approx 3$~g/cc, Z = 54), make the XENON concept 
very powerful for background identification and rejection based on event topology: multiple scatter 
events from the overwhelming majority of gamma rays and from neutrons can be identified and rejected. 
The sensitive target can be ``fiducialized" to keep only the inner core free of background. 


The capability to accurately identify the 3D position on an
event-by-event basis, also allows to correct the position dependence of both the direct and proportional
scintillation signals, which in turn results in improved energy resolution. For a recent review of the
properties of LXe and its response to radiation we refer to~\cite{aprile:2009}.

  One of the challenges of a two-phase detector is to efficiently detect the small number of primary scintillation
photons associated with the low energy events of interest in a dark matter search.  Because of the large refractive index of LXe \cite{Solovov:04} and consequent total internal reflection at the liquid/gas interface, the presence of PMTs at the bottom of the active drift volume is essential for efficient $S1$ light collection.    

When the XENON concept was proposed, the development of PMTs  capable to operate directly in LXe had just began. The typical quantum efficiency of these early PMTs at 178 nm  was in the 10\% range. To achieve high light collection of the $S1$ signal and thus a low energy threshold required for a dark matter search, the baseline XENON detector design used a CsI photocathode in place of a common cathode, and  photomultipliers or other type of photodetectors in the gas region~\cite{NSF:01}. During the XENON R\&D phase, different photodetectors were tested\cite {SiPM, APD} and also charge-sensitive readouts based on the Gas Electron Multiplier concept \cite{FSauli:97:GEM}. The operation of CsI photocathodes to detect the direct light in two phase prototypes 
confirmed \cite{CsI_IEEE} the high QE first measured by the Columbia group~\cite{CsI_old}. However, the
complexity of operating  a TPC under the very strong electric field required by the CsI photocathode, while suppressing positive photon feedback, and the fast improvement of the performance of Hamamatsu PMTs for LXe operation, led us to abandon the CsI photocathode approach. Prior to adopting the Hamamatsu one-inch square R8520 PMT  for XENON10, a two-inch diameter R9288 PMT was extensively tested~\cite{Aprile:2004,CsI_IEEE}. 

\subsection{Electrodes Assembly and Electric Fields}

\begin{figure}[htb] \centering
    \includegraphics[width=0.40\textwidth]{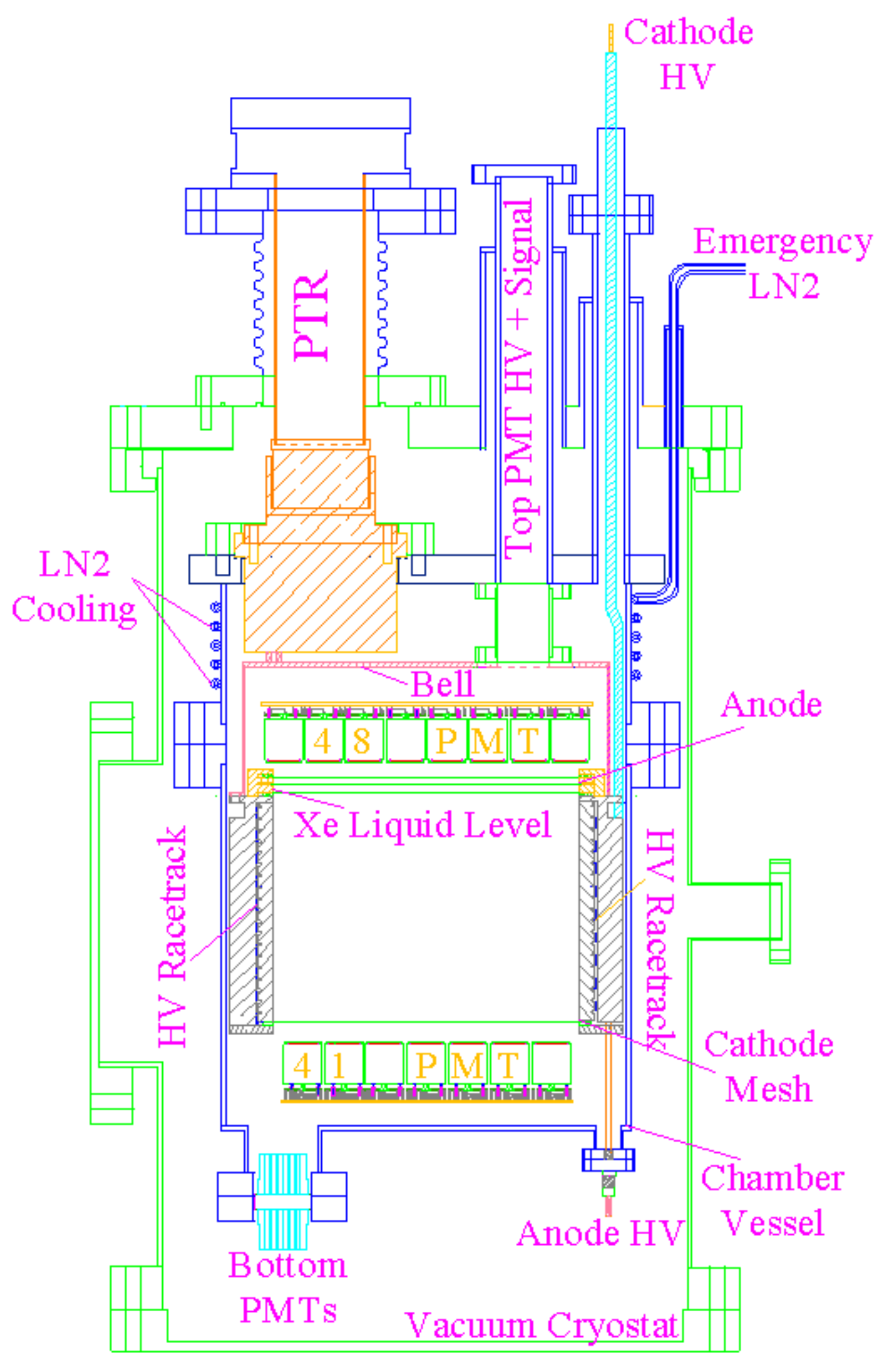}
	\caption{(Color online) Schematic drawing of the XENON10 detector.} 
    \label{fig:xenon10draw}
\end{figure}

\begin{figure}[htb] \centering  
    \includegraphics[width=0.45\textwidth]{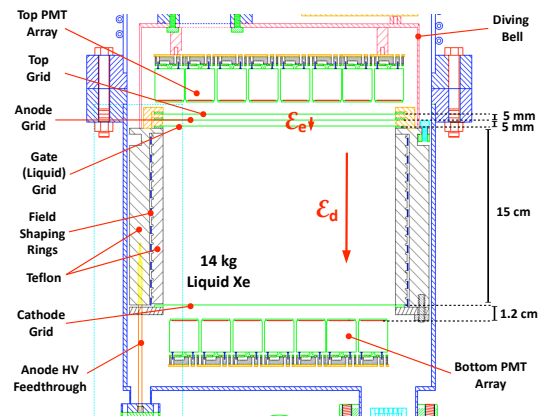}    
	\caption{(Color online) A close-up view of the XENON10 TPC structure.}
    \label{fig:xenon10-TPC-only}
\end{figure}

\begin{figure}[htb]
   \includegraphics[width=0.45\textwidth]{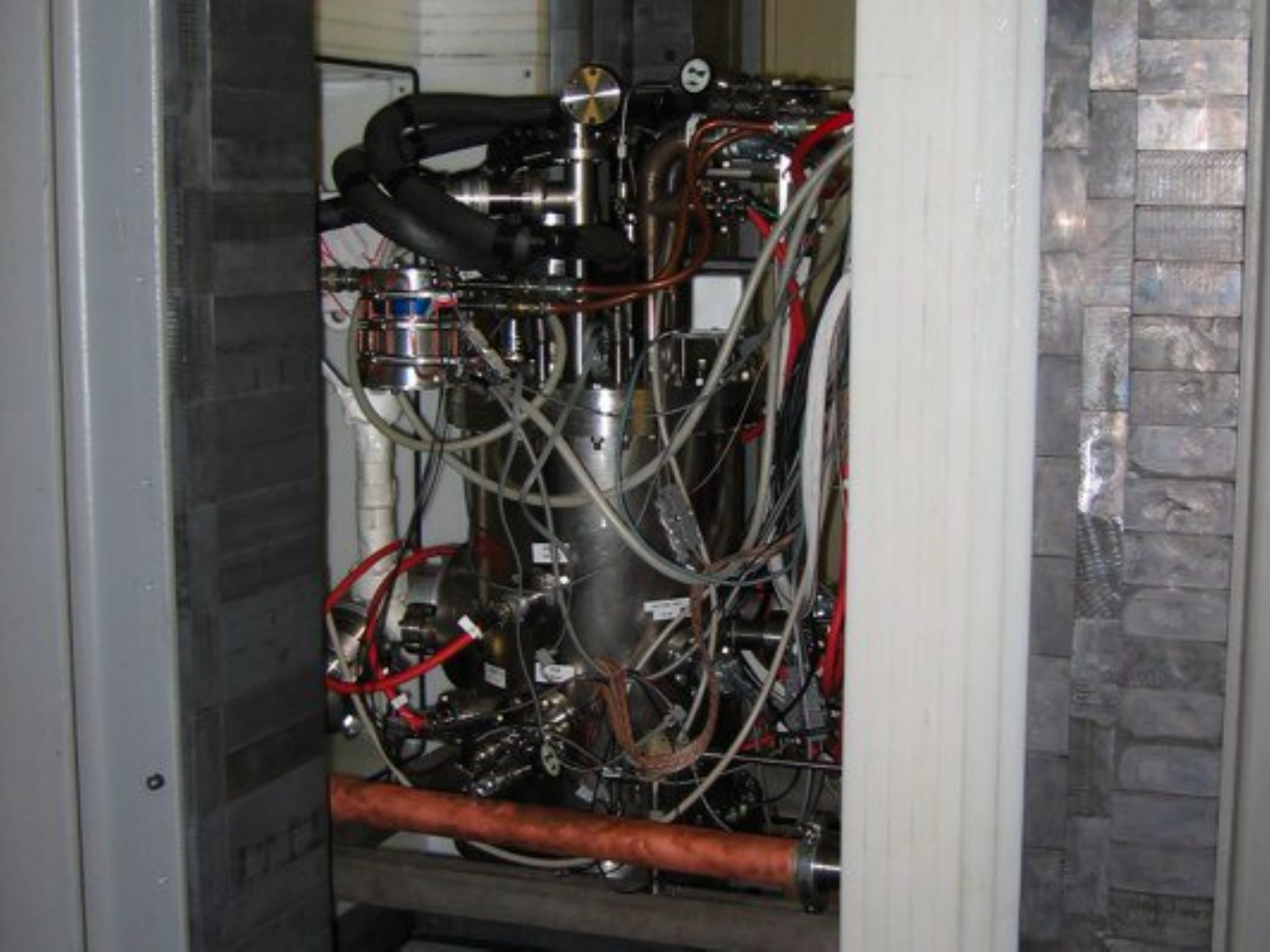} 
   \caption{(Color online) Photo of the XENON10 detector and shield, with the shield door open.}
   \label{fig:shield}
\end{figure}

Fig.~\ref{fig:xenon10draw} shows a 2D  mechanical drawing of the XENON10 detector assembly, and Fig.~\ref{fig:shield} shows a picture of the detector mounted on the movable door of the passive shield (see section \ref{sec:shield}). A closer view of the inner TPC structure is shown in Fig.~\ref{fig:xenon10-TPC-only}.  The TPC active volume is defined by a polytetrafluoroethylene (PTFE) cylinder with an inner diameter of 20~cm and a height of 15~cm for a total active mass of about 14~kg of LXe. PTFE is used as UV light reflector \cite{Yamashita:04} and as electrically insulating support for the TPC structure.
  
To produce the electric fields in the liquid and gas regions, four wire meshes, also referred to as grids, are used; two in the liquid (cathode and gate grid) and
two in the gas (anode and top grid).  The 0.203 mm thick electro-formed meshes, are made of  electropolished 304 stainless
steel,with a bar width of 0.182~mm and 2.0~mm $\times$
2.0~mm square holes, insuring good optical transmission.  
The cathode, at the bottom of  the PTFE cylinder, and the gate grid define the TPC
15~cm drift distance. Field shaping rings, made of 0.5 mm thick copper and spaced by 0.76 cm, are mounted outside the PTFE cylinder to insure a uniform electric field across the drift volume.  

A 5~mm gap separates the gate grid from the anode and the anode from the top grid. The liquid level is between anode and gate grid and determines the extraction field. The liquid level, 2.5~mm above the gate grid, is  kept constant by the use of a pressurized cylinder closed on the top, similar to a diving bell. The pressure is provided by the return of the gas circulating through the purifier (see Section~\ref{sec:instr:recsys}). Custom-made capacitors are used to monitor the liquid level: one cylindrical capacitor is used to measure the
liquid level and four parallel-plate capacitors are used to measure the inclination of the detector,
to ensure that the liquid level is parallel to the grids and thus that the extraction field is uniform. One of the parallel-plate capacitors is
filled with PTFE and is used as a reference capacitor: the capacitances of other parallel-plate capacitors are
compared with the reference value to control the inclination of the detector. The XENON10 cryostat was equipped with leveling feet which could be adjusted from outside the shield structure to achieve the required degree of leveling.

\begin{figure}[htb]
   \includegraphics[width=0.45\textwidth]{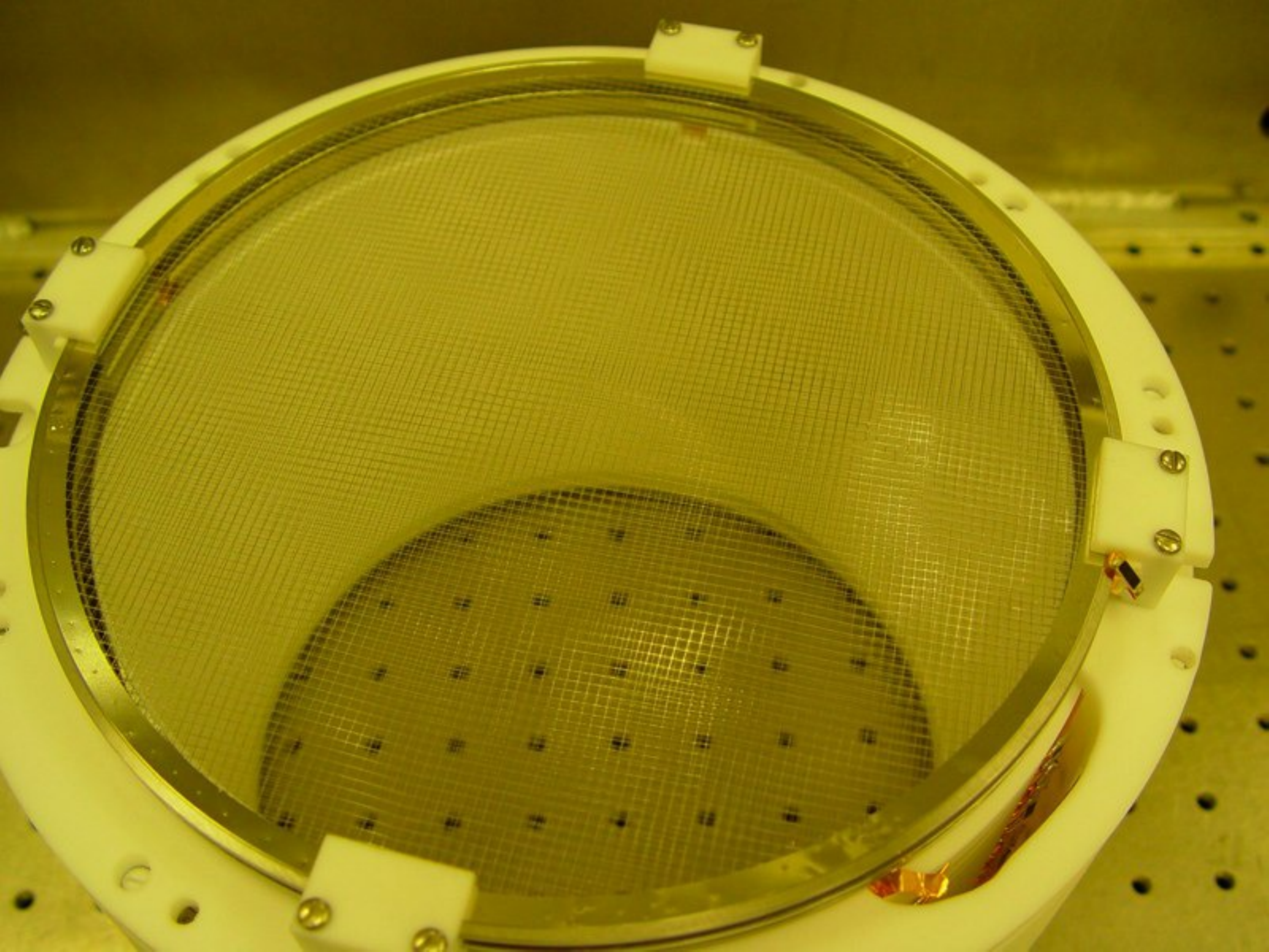} 
   \caption{(Color online) Photo of the assembled XENON10 TPC structure.}
   \label{fig:meshes}
\end{figure}

To minimize passive LXe outside the TPC, the PTFE cylinder is surrounded by an outer PTFE
cylinder, with cut-outs for the resistors of the voltage divider network, directly mounted on the field shaping
wires. A photo of the assembled TPC structure, with the top three meshes clearly visible, is shown in Figure
\ref{fig:meshes}. For most of the data acquired with XENON10, the top and gate grids were biased to -1.15~kV, with the
anode and cathode at  $+3.5$~kV and -13 kV, respectively. With these voltages, the drift field in the liquid
was 0.73 kV/cm while the field in the gas was 12 kV/cm. The high voltage for the gate, anode, and top meshes
was provided by CAEN 1733 and 1833 power supplies ~\cite{caen}, while the higher voltage for the cathode was provided
by a Heinzinger power supply~\cite{Heinzinger}. Commercial HV feedthroughs and cables carried these
voltages to the cryostat. To carry the high voltages from the vacuum of the cryostat to the cathode mesh, a
custom-made PTFE insulated feedthrough was used. Inside the TPC structure and below the LXe level,
bare wires were used to carry the HV to the upper meshes, taking advantage of the excellent dielectric properties of LXe and of PTFE.

\begin{figure}[htb] 
    \includegraphics[width=0.35\textwidth]{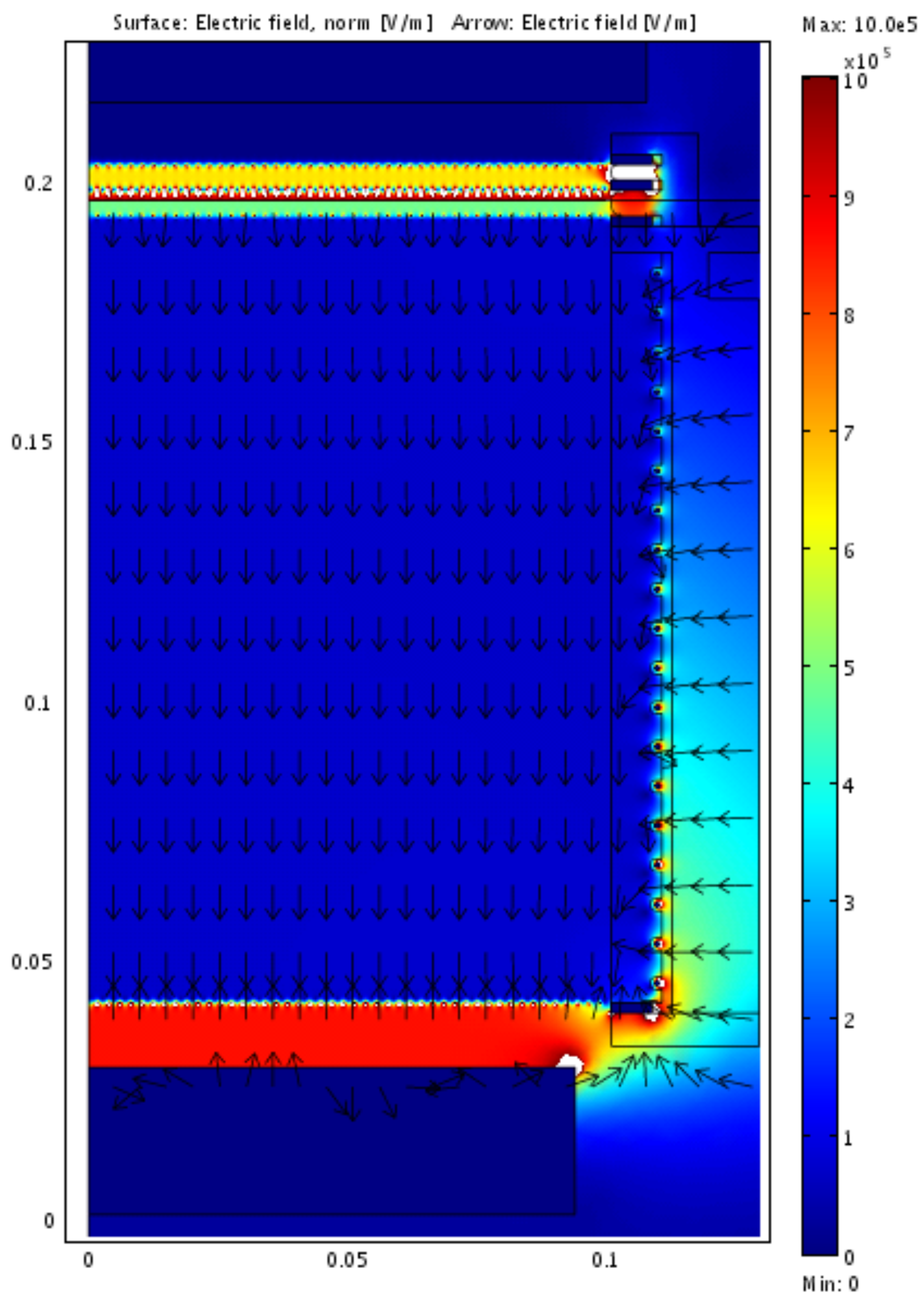} 
    \caption{(Color online) Simulated electrostatic field in XENON10. Arrows and colors indicate the field direction and strength, respectively. The field cage was optimized
    for field uniformity in the sensitive region. The region below the cathode shows a strong reversed field.}
    \label{fig:fieldsimuation}
\end{figure}


Results from electric fields simulations for the XENON10 TPC are shown in Fig.~\ref{fig:fieldsimuation}, for a
drift field of 1~kV/cm. The field is uniform up to  3~mm from the PTFE wall.  For the dark matter
search results,  only events with at least 10~mm distance from the edge of the active volume were accepted.
Gamma-ray calibration data confirmed that no charge was lost for events in this region. On the bottom of the
detector, a region of reversed field direction exists between the cathode  and the bottom PMT array, separated by a distance of 1.2~cm. The
field reversal extends  slightly into the sensitive volume, affecting a region less than 0.9~mm above the
cathode.

The same set of field simulations was used to establish that the displacement of the true $XY$ event position
and the mean $XY$ position from the proportional scintillation was negligible ($<0.6$~mm). The impact of the
field distribution in the gas on the energy resolution of the detector was also studied. For the inner region
of the meshes, it was  found that the relative path length variation of electrons traversing the gap from the
liquid-gas surface to the anode mesh is $\leq 20\%$ FWHM. 

\subsection{PMTs and Calibration System}

The active LXe volume is viewed by 89 PMTs: the bottom array of 41 PMTs is located 1.2~cm below the cathode, fully immersed in LXe. The 48 PMTs of the top array are in the gas. The XENON10 PMTs are 1" $\times$ 1" square Hamamatsu R8520-06-AL
~\cite{hamamatsu} designed to work in LXe and optimized for low radioactivity. They have a
bialkali photocathode and a quartz window, with a typical quantum efficiency $>20\%$ at 178 nm. They
are compact (only 3.5 cm tall), metal channel PMTs,  with 10 multiplication stages and a total amplification
of a few $10^6$. They are certified for operation at -100$^\circ$C and up to 5 bar. The radioactivity of these PMTs was  measured with a ultra sensitive HPGe detector~\cite{Laubenstein}
at the Gran Sasso Low Background facility~\cite{Arpesella}. Their contamination in
$^{238}$U/$^{232}$Th/$^{40}$K/$^{60}$Co was measured as low as (0.25 $\pm$ 0.04)/(0.21 $\pm$ 0/05)/(9.3 $\pm$
1.1)/(0.59 $\pm$ 0.05) mBq/PMT.

\begin{figure}[htb]
    \includegraphics[width=0.45\textwidth]{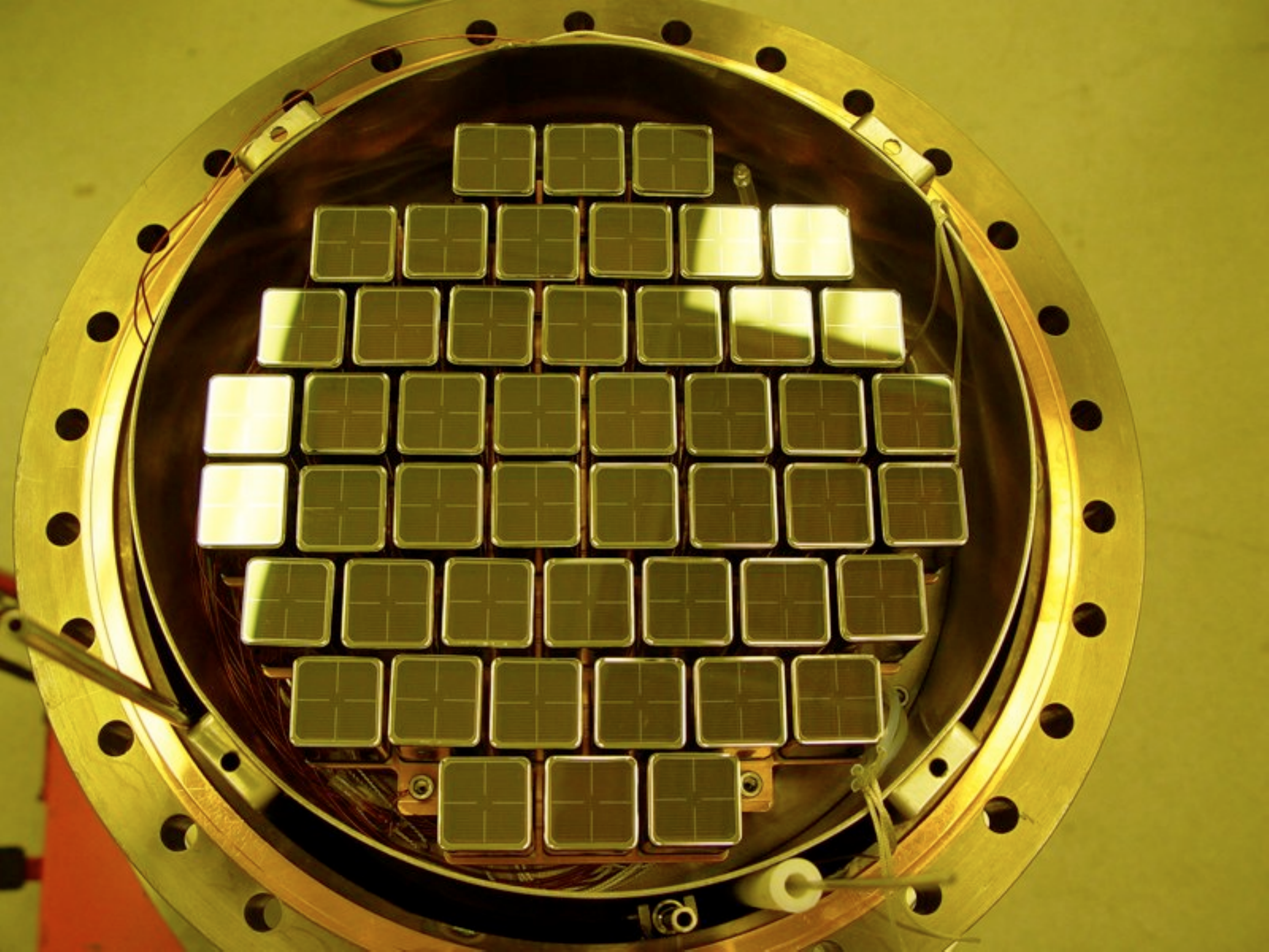} 
    \caption{(Color online) XENON10 top PMT array,  with the cathode HV feedthrough and liquid level meter also visible.}
    \label{fig:pmtarray}
\end{figure}

A photo of the top array, mounted inside the ``diving bell" structure which defines the liquid level, is  shown in Fig.~\ref{fig:pmtarray}. The custom-made cathode HV feedthrough and the level meter
are also visible.
The high voltage for the PMTs is provided by CAEN 1833 power supply boards~\cite{caen}; this system
featuring single channel control, allowed to easily equalize the gain of the PMTs, by adjusting individual voltages.

The HV divider for the PMTs was mounted on a Cirlex substrate, with surface mount
components. The Cirlex boards for each of the two arrays are mounted on copper plates with PTFE spacers for
insulation. Despite the small mass, the ceramic of the resistors and capacitors contribute the largest
fraction of the radioactive background of the HV divider boards. Heating of the LXe by the resistive dividers is kept to a minimum by choosing high
resistance values (10 M$\Omega$), allowed because of the low event rate and low light level during dark matter search data-taking. The typical power consumption of the XENON10 PMTs was about 5 mW per PMT.


Negative HV was applied to the cathode of each PMT, via unshielded Kapton wire. The average HV was about 750~V, with a maximum of 900 V.   The signal was carried out with
50 $\Omega$ teflon cables (RG178U), stripped of their shielding, to avoid trapped air which would impact the liquid purity. PMTs signal and HV
cables were carried outside by 48 pin Burndy feedthroughs~\cite{Kyocera}. The radioactivity of these
feedthroughs was reduced by replacing the ceramic plugs holding the pins with plugs made out of polyether
ether ketone (PEEK).

The 89 PMTs were routinely calibrated with the goal of equalizing the gain and to correct for differences in
the quantum efficiency. The gain equalization was performed using light emitting diodes (LEDs). Two LEDs were
used: one was mounted close to the top array, to equalize the bottom PMTs; the other was mounted close to the
bottom array, to equalize the top PMTs. Both LEDs were covered with a PTFE cap acting as a diffuser  to ensure
a uniform illumination of all the PMTs in each array. The LEDs were driven by a pulse generator, which
synchronously activated also the main DAQ trigger: the pulse duration was 6 $\mu$s, but the charge on each
channel was integrated only in a 1~$\mu$s window, 4~$\mu$s after the activation of the LED, to minimize
electronics noise generated when the LED was switched on. 

\begin{figure}[htb]
    \includegraphics[width=0.45\textwidth]{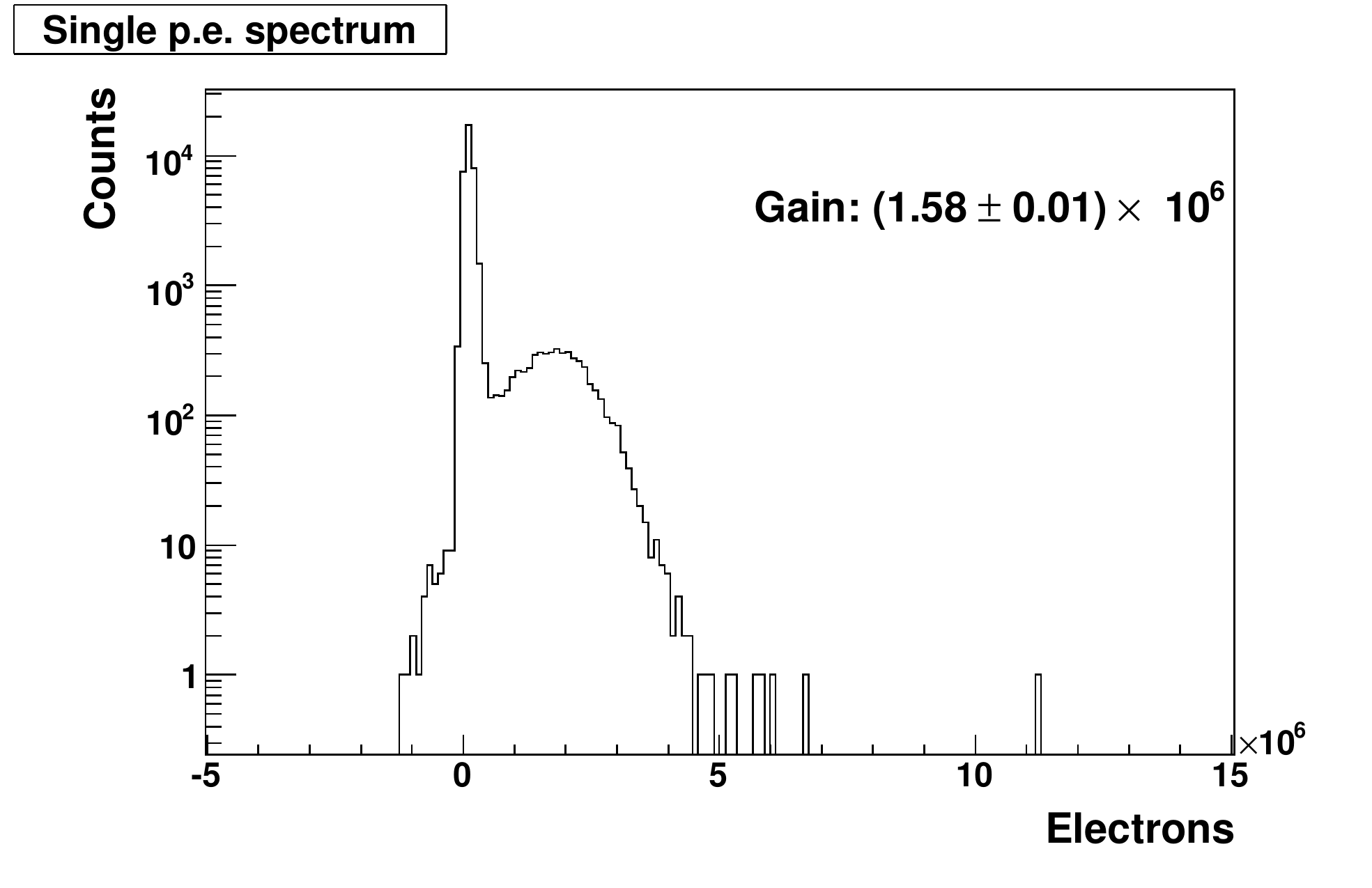}
    \caption{Single photoelectron spectrum of  a XENON10 top PMT, acquired during a routine LED calibration. Clearly visible is also the noise peak.}
    \label{fig:single_pe}
\end{figure}

The gain measurement was performed in single photoelectron (p.e.) regime. Under the assumption that the p.e.
spectrum is the sum of a Gaussian noise peak and a Gaussian single p.e. peak, each with a Poisson-like
distribution, and with a single photoelectron to noise ratio of 1/5,  less than 2\% of the events
had 2 or more photoelectrons (a typical single p.e. spectrum is shown in figure~\ref{fig:single_pe}). To
ensure that the single p.e. condition was met for all the PMTs, the calibration data were acquired under
different illumination conditions. The spectra for each PMT were fitted with a function which included the noise
Gaussian, the single p.e. Gaussian and a third Gaussian accounting for the multiple p.e. contribution.

\begin{figure}[htb]
    \includegraphics[width=0.45\textwidth]{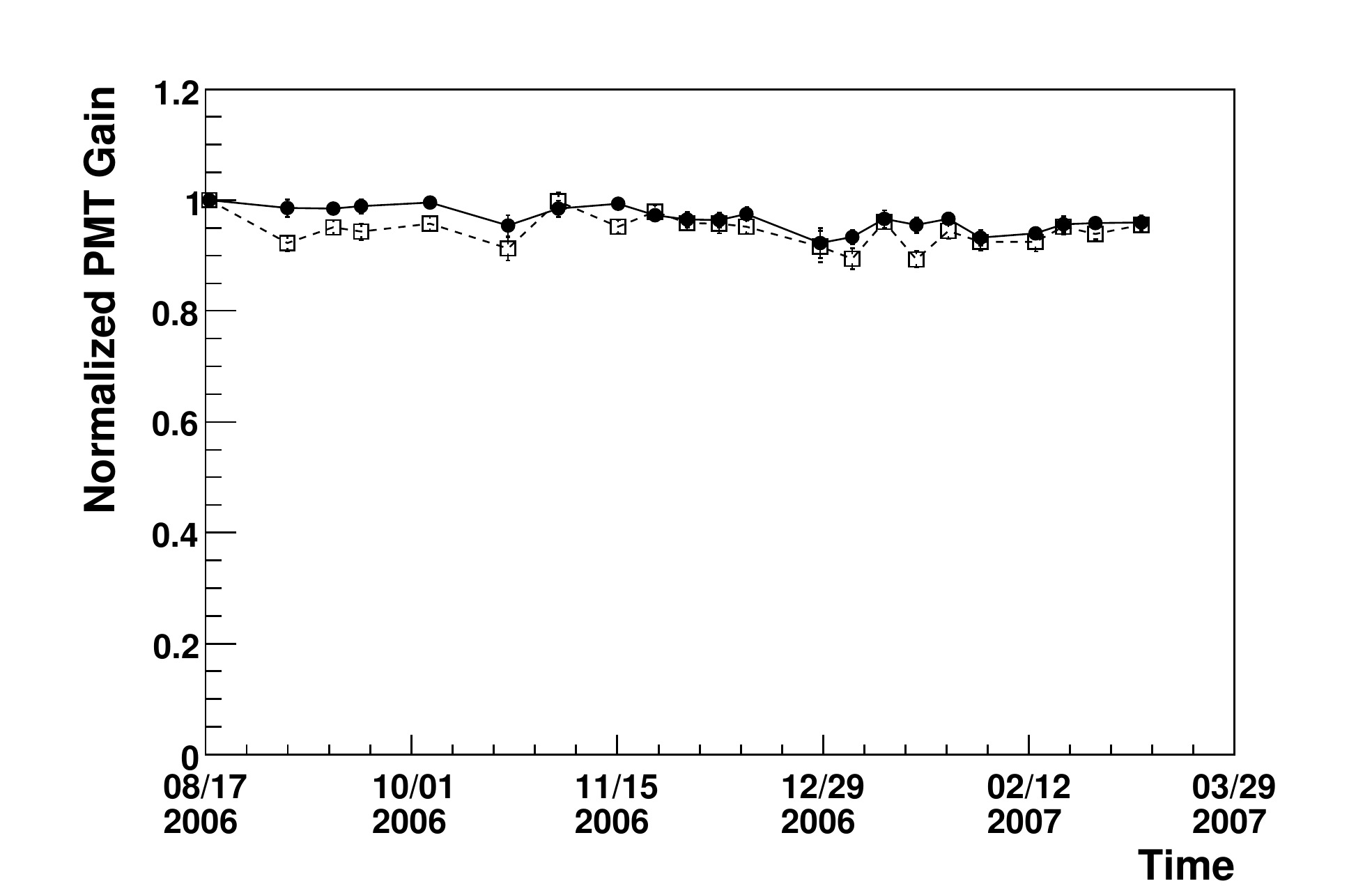}
    \caption{Stability of the gain of two typical XENON10 PMTs during the WIMP search data taking period. }
    \label{fig:gain_stability}
\end{figure}

After measuring the gain for each channel, the amplification factor  was equalized for all PMTs, by adjusting
the single PMT operating voltage. After the equalization, the gains were measured again and the resulting
amplification factors were then fed  to the XENON10 data analysis programs (see Sec.~\ref{sec:data_processing})
and used as a conversion factor between the measured charge and the actual p.e. number. During normal
operation, such a gain measurement was repeated at least once a week, to ensure the stability of the PMTs
gain. The time evolution of the gain, measured during the WIMP search data taking, is shown in
Fig.~\ref{fig:gain_stability} for two typical PMTs. The variation in PMT gain was $< 2\%$.


\subsection{Cryogenic System}

\begin{figure}[htb]
    \includegraphics[width=0.45\textwidth]{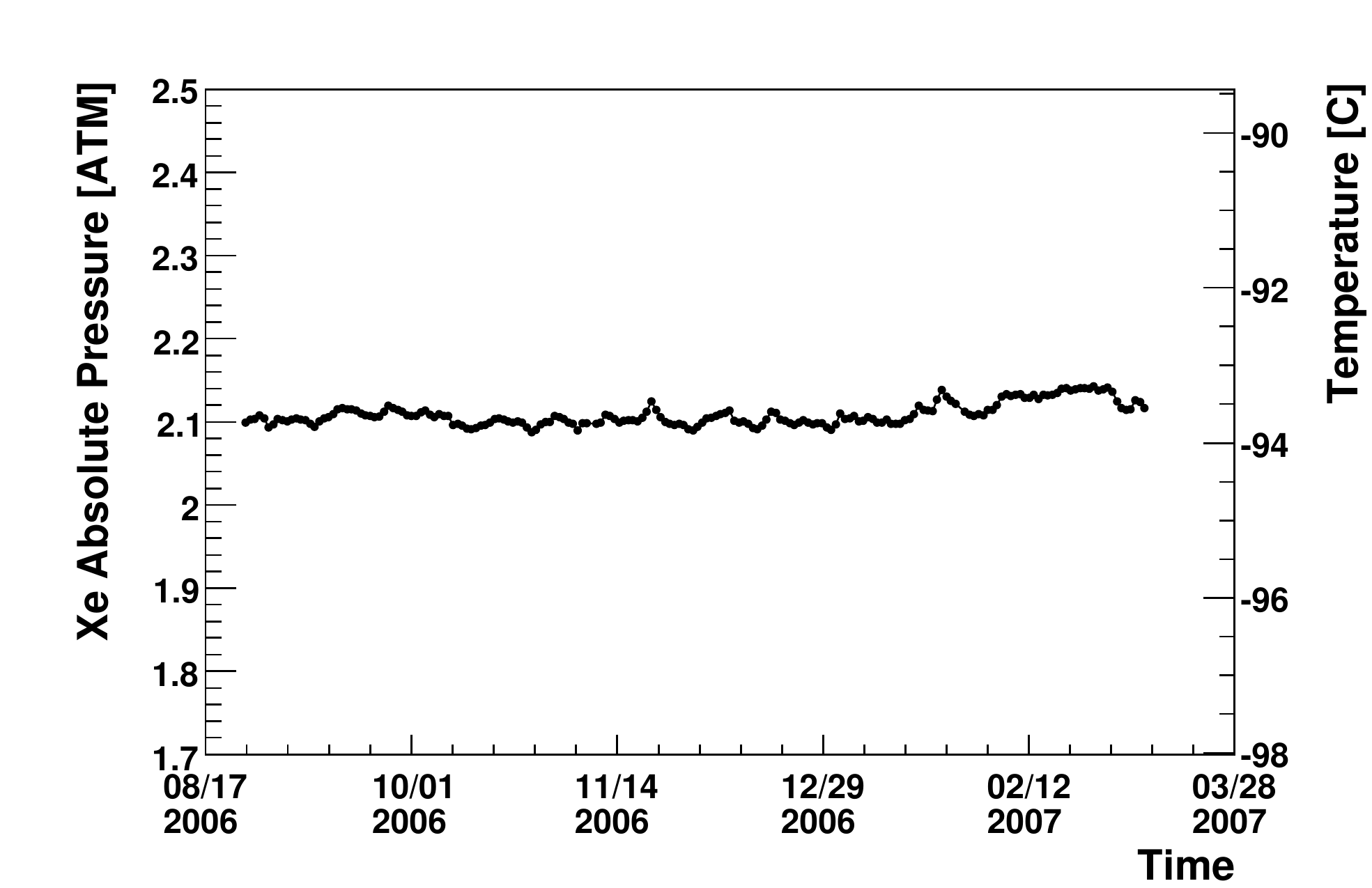}
        \caption{ Stability of the liquid Xe temperature and corresponding vapor pressure in the XENON10 detector during the calibration and WIMP search data-taking period.}
        \label{fig:press_stab}
\end{figure}

A reliable and stable cryogenics system was an essential requirement for the XENON10 experiment since both PMTs
gain and the proportional light yield vary with temperature. The XENON10 TPC vessel containing the 25 kg of LXe, made of stainless steel (SS), was wrapped with aluminized mylar and surrounded by a vacuum cryostat, also made of SS, with typical vacuum in the mTorr range. The cryogenics system was based on a Pulse Tube Refrigerator (PTR) with 100 W of cooling power at $165 \, \mathrm{K}$, with a 3.5 KW
compressor. This type of cryocooler was developed  specifically for Xe liquefaction and re-condensation in the
LXe scintillating calorimeter of the MEG experiment~\cite{Haruyama:2005}. To meet the ultra-high purity requirement for the XENON10 TPC, the cryocooler was placed  outside of the detector vessel, mounted on the vacuum cryostat top flange and attached to a
copper plate which worked as re-condenser for the evaporated Xe gas. The cooling power was sufficient  to
cool-down the detector, liquefy the gas and re-condense it.  The desired temperature of the LXe was adjusted
by a Proportional-Integral-Derivative (PID) controlled heating element on the cold head of the PTR.  Due to
the narrow temperature margin of less than 4 degrees between the liquid and solid phase of Xe, temperature
control during liquefaction is especially important.  The typical operating temperature for XENON10 was
-94$^\circ$C with a stability better than 0.05$^\circ$C. At this temperature the Xe vapor pressure is
$\sim$2.1 atm. With this cryogenic system we have been able to achieve the level of long term stability
required for a dark matter experiment. Fig. \ref{fig:press_stab} shows the stability of the LXe temperature and corresponding Xe vapor pressure during several months of continuous operation of XENON10.

The XENON10 cryogenics system included an independent LN$_2$ system for emergency cooling. A pressure rise in the detector, above a set value, would automatically trigger a flow of  LN$_2$  through a cooling coil next to the PTR. The TPC pressure sensor and the solenoid valve control were powered by a dedicated uninterruptible power supply. The emergency cooling system would keep detector in a safe sate even in the extreme case of power failure and no access to the laboratory for more than 24 hours. We note that the large Xe mass and high thermal capacity stabilizes the temperature and the pressure against abrupt changes. A rupture disk would finally burst if the absolute pressure in the detector would rise above 3.5 atm, to avoid damage to the PMTs.

\subsection{Xe Purification and Recirculation System}
\label{sec:instr:recsys}
The 15 cm electron drift gap in the XENON10 detector imposes ultra-high purity requirements on the LXe, in
order to minimize charge loss by electron attachment to impurities molecules.  The overall concentration of
electronegative impurities (Oxygen equivalent) must be well below 1 part per billion (ppb). Moreover, stable
operation of the detector demands that this high purity be maintained over long periods of time. For detector
and gas system ultra high vacuum materials and procedures were used, and for cleaning the Xe gas  a single purifier,
a high temperature SAES getter (PS4-MT3-R) \cite{SAES} was used. 

\begin{figure}[htb] 
   \includegraphics[width=0.45\textwidth]{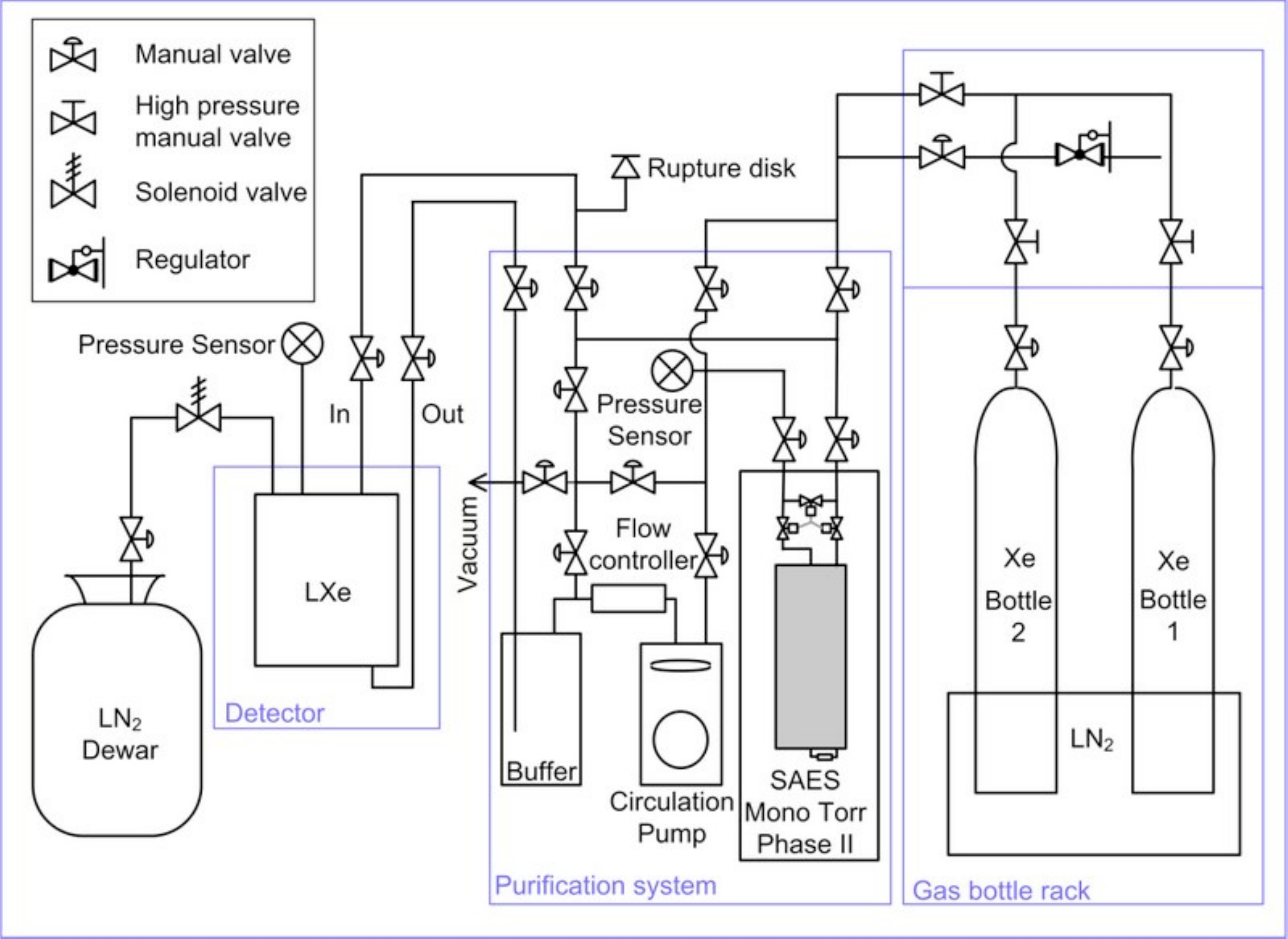} 
   \caption{Schematic of the Xe purification system with closed-loop re-circulation.}   
   \label{recir_sys} 
\end{figure}

However, the PMTs limited the bake-out temperature of the detector to 70$^\circ$C.  In addition, LXe is a good
solvent due to Van der Waals interactions with impurity molecules. Thus, once pure Xe is liquefied in the
detector, the purity level can easily deteriorate. We therefore implemented a closed circulation system to
continuously clean the gas through the getter, after initial liquefaction. Such a system was successfully developed and tested at Columbia on several smaller scale prototypes, prior to adapting it to XENON10. For gas circulation a double-diaphragm pump (KNF-N143.12E) \cite{KNF} was used. With a gas flow rate of 2.6 slpm, limited  largely by the PTR cooling power, it would take about one month to reach the  purity level required for negligible electron attenuation.  A schematic of the gas purification and recirculation system used on XENON10 at LNGS  is shown in Fig.~\ref{recir_sys}.  


\subsection{Slow Control System}

A Slow Control System (SCS) was developed to monitor all essential run-time parameters of the XENON10
experiment, controlling the status of various hardware components, triggering alarms  and transmitting
important parameters. 

The SCS monitors over 290 parameters, which include gas pressure, cryogenic temperatures, flow rate, liquid
level, grids and PMTs high voltages, DAQ trigger and acquisition rate, room temperatures and detector
inclination. Figures \ref{fig:press_stab} is an example of the detector parameters monitored by the SCS during the dark matter
run.  

The SCS consists of 4 parts: (i) server, (ii) monitor client, (iii) alarm client, and (iv) history plotter. The server establishes communication with all the different instruments monitored, makes all the parameters available for the clients, and stores the information to disk.
The monitor client allows each user access to all the parameters monitored over the last 24 hours. The alarm client triggers the alarm system in case one of the preassigned parameters falls outside the allowed range. Alarms are sent via email and as text messages to the cell phones of the appropriate personnel. In addition to this automated system, the LNGS personnel continuously monitors the gas pressure. The history plotter allows the user to access the information stored for each parameter for any particular
time.

The SCS is a platform independent software developed exclusively for the XENON10 experiment using the Java programming language. The communication between the different SCS components is done with Java Remote Method Invocation (RMI).

\section{The XENON10 Shield}
\label{sec:shield}

\begin{figure}[htb]
    \includegraphics[width=0.45\textwidth]{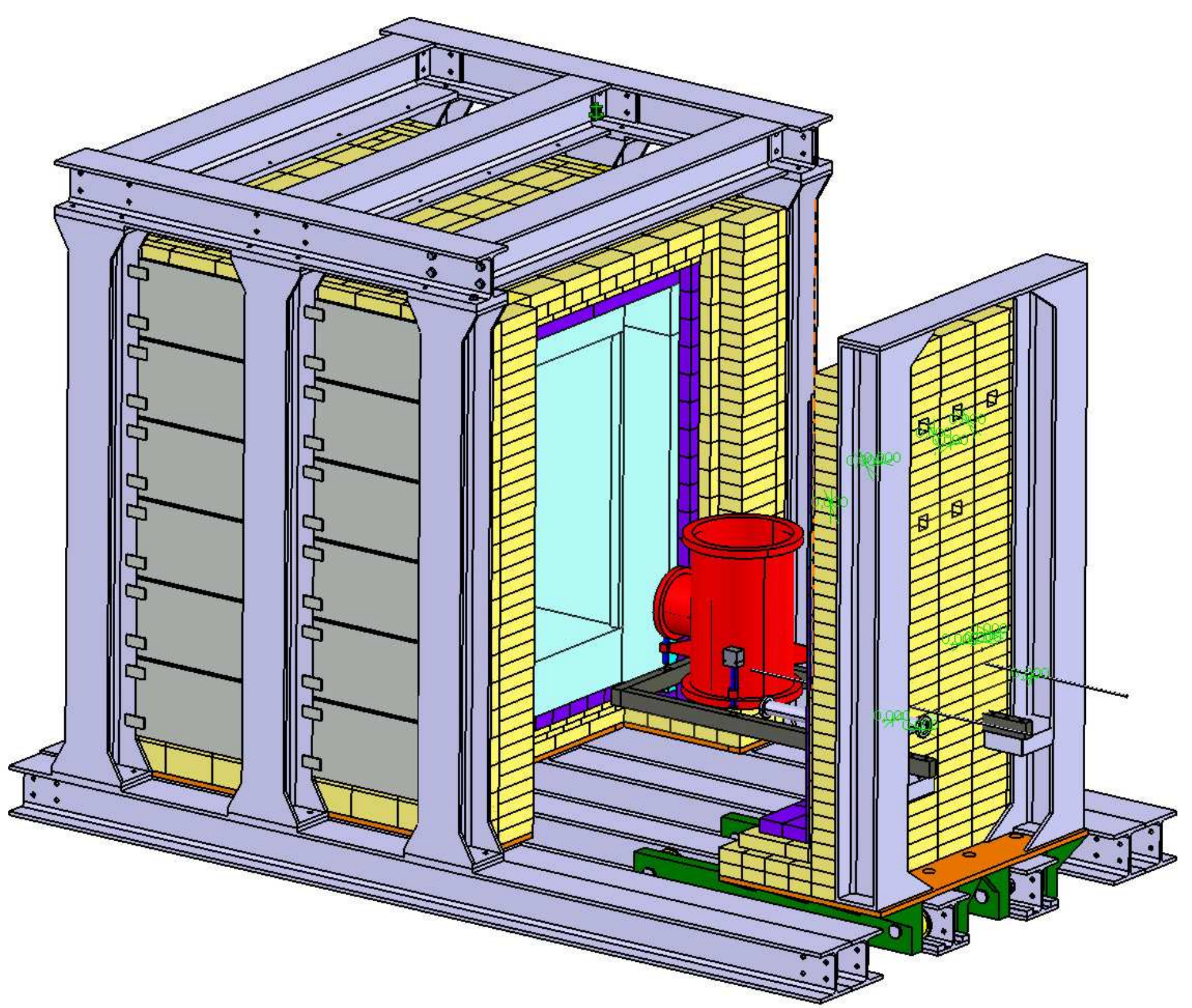}
    \caption{(Color online) The XENON10 shield structure. Common and low-radioactivity Pb bricks are shown in yellow and
        purple respectively, while polyethylene is light blue. Not shown are the 20cm of polyethylene on
        the "door" and the 15cm below the shield structure. }
    \label{fig:shield_dwg}
\end{figure}

The XENON10 detector is protected from external background by a cubic steel-framed structure, consisting of
$20$~cm high-density polyethylene (HDPE) inside of $20$~cm Pb; a schematic is shown in
Fig.~\ref{fig:shield_dwg}. The Pb was supplied in $5\times10\times20$~cm bricks and was stacked so as to avoid
any line-of-sight penetration along the cracks. The outer $15$~cm of Pb have an activity in $^{210}$Pb of
about $560\pm90$~Bq/kg, while the inner $5$~cm are a low-background Pb obtained from \emph{Fonderies de
Gentilly}~\cite{Fonderies}. It was measured at a germanium counting facility \cite{Laubenstein,Arpesella} to
have an activity in $^{210}$Pb of ($17\pm5$)~Bq/kg. Specific activities for various shield components are
shown in Table \ref{tab:radio01}.

\begin{figure}[htb]
    \includegraphics[width=0.45\textwidth]{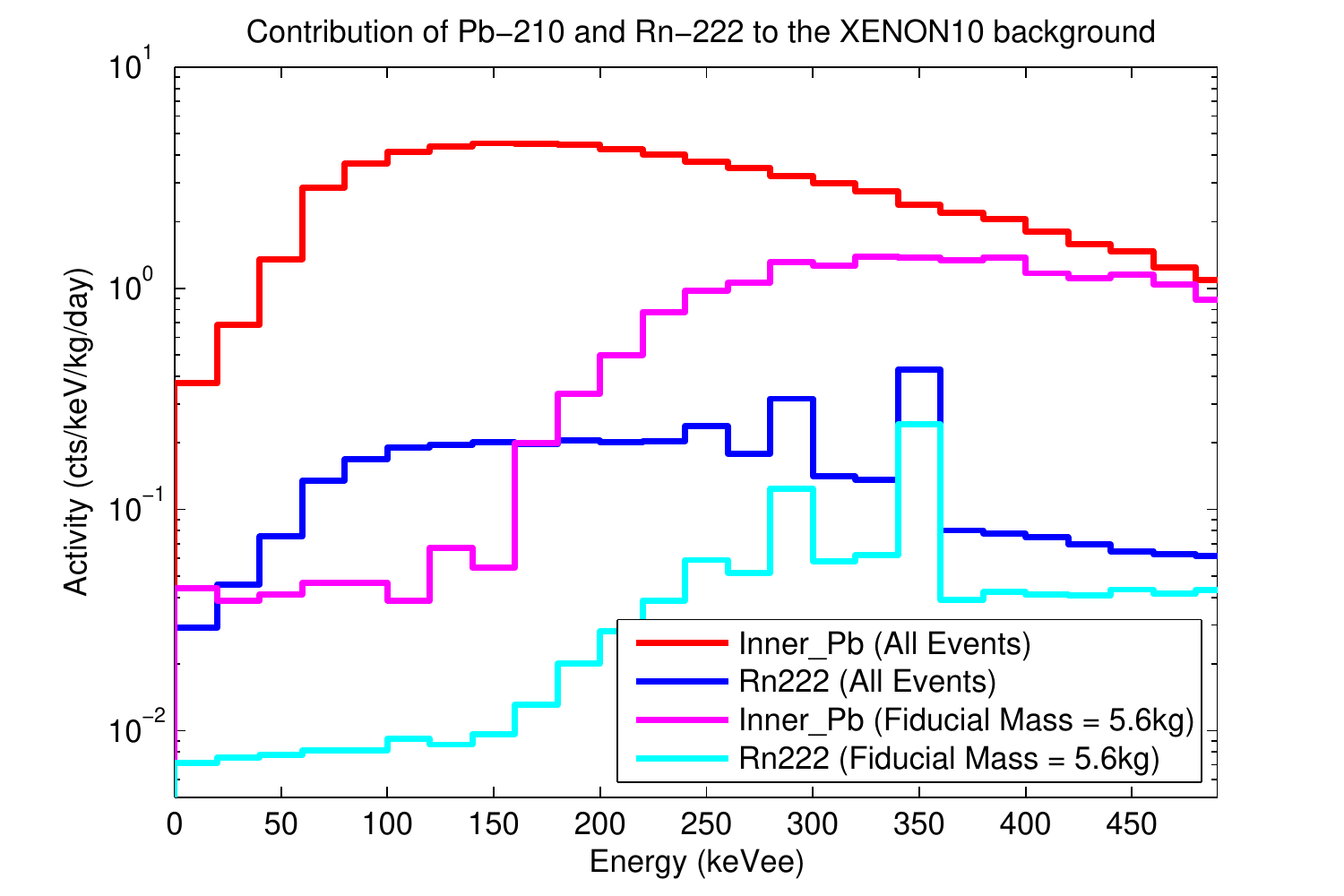}
	\caption{(Color online) Simulated contributions of $^{210}$Pb (17 Bq/kg) and $^{222}$Rn (5 Bq/m$^3$) to the XENON10
		background, for the entire 14~kg active volume (red, blue) and also for a 5.6~kg fiducial volume
		(cyan, magenta); the fiducial volume corresponds to $2$~cm Xe self-shielding (radial) and $3$~cm each
		(Top/Bottom).}
    \label{fig:Pb210}
\end{figure}

A Monte Carlo simulation shows that $20$~cm Pb results in an attenuation of the external gamma flux larger than $10^5$
while its internal activity leads to a contribution of less than 0.32~cts/keVee/kg/day (dru) ($E < 25$~keVee) to
the raw event rate inside the shield cavity. With the additional self-shielding of the outer $2$~cm of Xe (as
used in the WIMP search data analysis \cite{Angle:2007uj}), the contribution of the Pb activity to the electron recoil background
drops below $0.05$ dru ($E < 25$~keVee), becoming sub-dominant (see Fig.~\ref{fig:Pb210}). 

The $20$~cm of HDPE provides reduction of external fast-neutron flux by a factor of 90.  Placing the Pb
outside the HDPE has the benefit that the neutron scattering on the Pb reduces the energy of the
neutrons, thus increasing the efficacy of the HDPE.  The dominant source of neutrons is expected from
$(\alpha,n)$ and fission reactions in the surrounding rock, as well as from cosmogenic production in the rock and Pb.
The ambient neutron flux from natural radioactivity in the rock has been measured to be about
$4\times10^{-6}$~n/cm$^2$/s \cite{Belli:1989,Wulandari04}, and from this the integrated rate of neutron interactions in
XENON10 is calculated to be $<0.015$~neutron/month in the dark matter region of interest.  The muon flux in the
cavern was measured to be $1.06\pm0.03$~muon~/m$^2$/hour \cite{Sorensen:2008th};  from this, and the muon-induced neutron yield in rock and in Pb, the
expected rate for cosmogenic neutrons is expected to be about 13 times higher \cite{LDV:2009th} than from natural radioactivity.

\begin{table*}
\caption{\label{tab:radio01} Results from radioactive measurements for relevant shield components (values in
    mBq/kg).}
\begin{ruledtabular}
\begin{tabular}{lcccccccc}
 & $^{238}$U & $^{232}$Th & $^{210}$Pb & $^{235}$U & $^{40}$K & $^{137}$Cs & $^{60}$Co & $^{54}$Mn\\
\hline 
Pb (inner) & $<3.9$ & $<6.8$ & ($17 \pm 5$)~$\times 10^3$ & $<20$ & $<28$ & $<0.85$ & $<0.19$ & \\
Pb (outer) & $<1.6$ & $<5.7$ & $(560 \pm 90)$~$\times 10^3$ & $<51$ & $14 \pm 6$ & $<2.1$ & $<1.1$ & \\
Polyethylene & $<5.2$ & $<6.6$ & & $<3.2$ & $<64$ & $<2.6$ & & \\
Steel (I-beams) & $7.8 \pm 3.2$ & $<4.1$ & & $<3.1$ & $<15$ & $<1.4$ & $170 \pm 3$ & $3.3 \pm 0.6$ \\
Steel (side panels) & $<3.5$ & $<4.7$ & & $<1.8$ & $<26$ & $<2.2$ & $2.4 \pm 0.7$ \\
Steel (ceiling plate) & $<8.3$ & $<8.7$ & & $<5.7$ & $<42$ & $<2.8$ & $2.9 \pm 1.2$ & $2.0 \pm 0.8$ \\
\end{tabular}
\end{ruledtabular}
\end{table*}

The internal shield cavity is $107.5\times$90$\times$90~cm, of which the detector occupies about $30\%$ of the
volume.  Care was taken not to leave any trapped air pockets between the lead and polyethylene, or anywhere
else in the structure.  An additional $15$~cm HDPE is also present below the shield
structure.  All seams were coated with a low-radioactivity silicon-gel, and a $10$~mm thick rubber seal runs in a
$5$~mm deep trench along the perimeter of the polyethylene on the door;  thus the cavity is able to sustain a
slight over-pressure, and boil-off N$_2$ from a low-pressure dewar is passed continuously into the top of the
cavity at a rate of $1.5 \pm 0.1$~standard liter/min, and allowed to exit at the bottom.  The concentration of
$^{222}$Rn in the shield cavity, typically about $130$~Bq/m$^3$, was reduced by the N$_2$ purge to a value of  $<5.5$~Bq/m$^3$ (consistent with the sensitivity floor of the RAD-7 radon measuring device~\cite{RAD7}).  The purge process occurred in about $30$ hours.  The expected contribution to the electron
recoil background ($E < 25$~keVee) from this level of Rn activity is $7\times10^{-3}$~dru in the inner
fiducial volume (Fig.~\ref{fig:Pb210}).

The shield structure is secured by steel panels along the outer walls, with bolts passing
 through the HDPE sheets. The XENON10 detector is supported by stainless steel I-beams on a single wall,
which also serves as a door and can be moved laterally on rails.  All cryogenic and electronic feed-throughs
pass through this wall, using a Z-shape where possible, and as far off-center as possible, to minimize
line-of-sight.  The overall diameter of each opening was also minimized, and gaps in the Pb (after placing the
cables) were filled with low-radioactivity Pb shavings. Silicon-gel was used to fill holes in the HDPE, also ensuring
air-tightness in these locations.  During the construction all components were cleaned and prepped with
Ethanol before mounting them on the shield structure. no cables or pipes have to be disconnected when opening the door to access the cryostat.

\section{The XENON10 Data Acquisition System} 
\label{sec:daq}

To maximize the information available from time structure and amplitude of the primary (S1) and secondary (S2)
pulses, the signal from 89 PMTs are individually digitized by an ADC array at 105~MHz (about 10~ns/sample).
The S1 pulses from the direct scintillation light of a typical background event in the energy region of
interest have 4-27~p.e.  distributed over a width of $<$200~ns (with a characteristic decay time of
$\sim$27~ns \cite{Doke:1999}). The S2 pulses from the secondary proportional scintillation have $\times200$
the number of photoelectrons distributed over about $1~\mu$s. The distance between the S1 pulse and the S2
pulse vary from 0 to 80~$\mu$s, the maximum drift time for electrons extracted in the liquid. Events with
multiple scatters in the active Xe volume will have more than one S2 pulse. To guarantee that the S1 pulse and
all S2 pulses associated with an event are digitized, a full waveform of 160~$\mu$s is recorded for each
event.

\begin{figure}[htb]
    \includegraphics[width=0.45\textwidth]{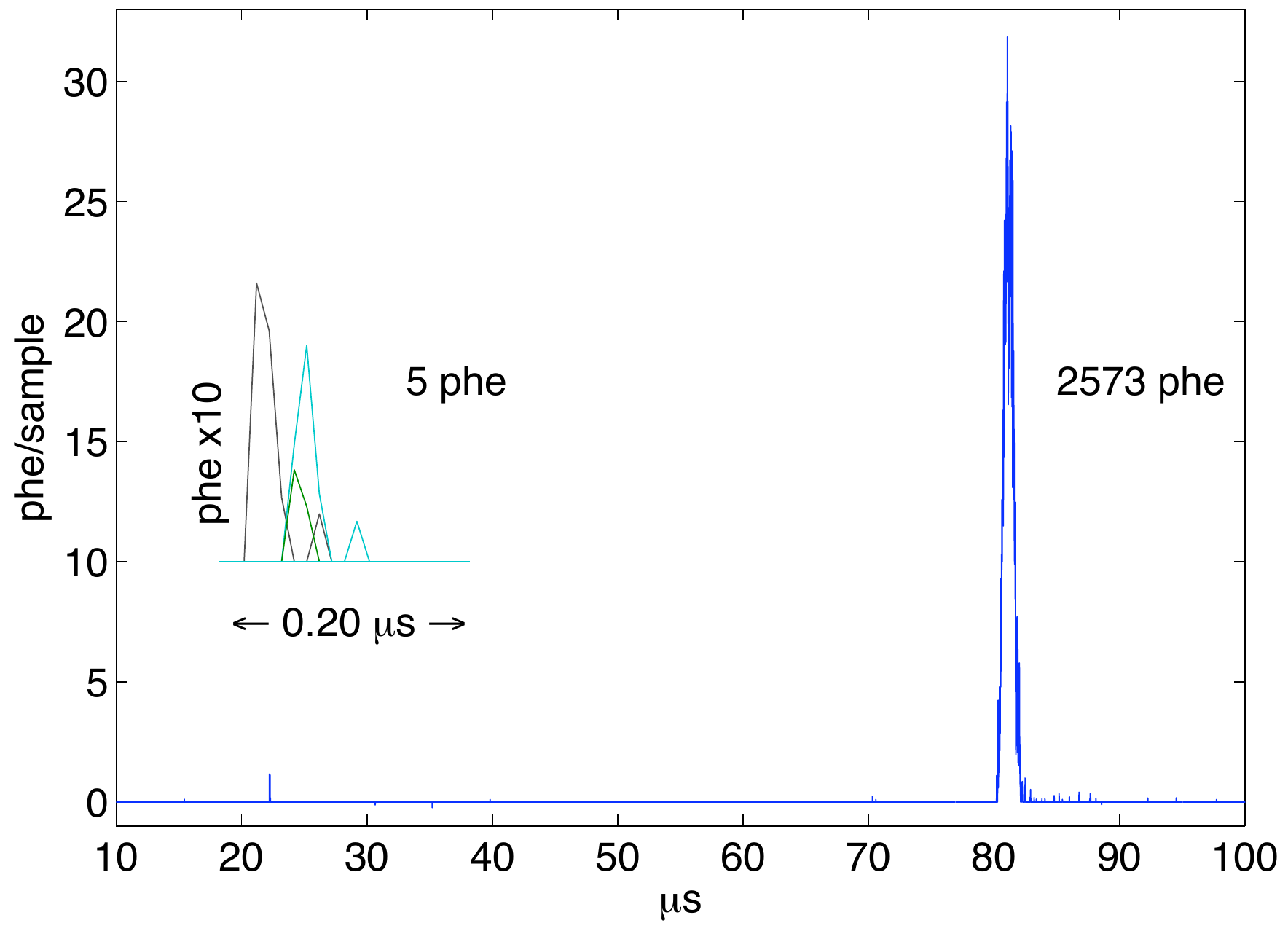}
    \caption{(Color online) S1 and S2 pulses for 2.3~keVee gamma event in XENON10 background data.  Signals from each PMT hit
        are plotted separately. An enlarged view of the S1 pulse (5~p.e.) is also shown. See \cite{Sorensen:2008th} for more details.}
    \label{fig:sample_waveform}
\end{figure}

The digitizers are \emph{Struck 3301} fast ADC (VME bus) modules~\cite{struck}, with sampling frequency
105~MHz and 14-bit ADC resolution (+0.1~V $\rightarrow$ -1.9~V).  The signal from each PMT is amplified using
Phillips PS776 $\times$10 amplifiers, and a single-pole RC filter (30~MHz) at the input to the ADC prevents
aliasing. The filter also has a slight shaping effect; because a single photo-electron pulse width is similar to the inverse sample rate of the digitizer, this improves the single photoelectron
sensitivity (see Fig. \ref{fig:sample_waveform}). An on-line software baseline-flattening algorithm sets small
fluctuations in the baseline ($<\pm$8 digitizer bins) to zero, allowing for very efficient lossless data
compression by a factor of $\sim$20 using open source software (gzip).  The data is not further modified or
processed before being recorded to disk.

\begin{figure}[htb]
    \includegraphics[width=0.40\textwidth]{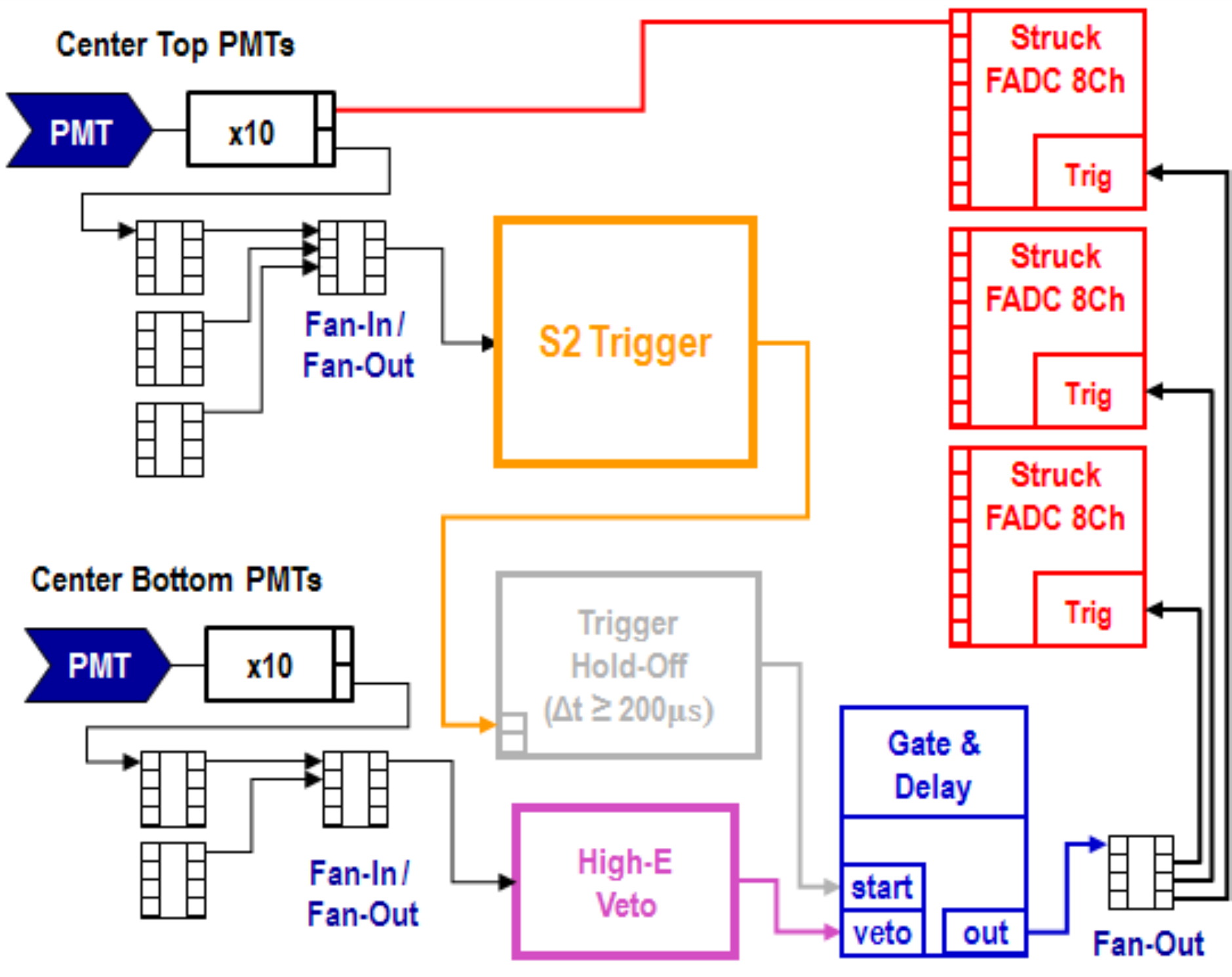}
        \caption{(Color online) Schematics for the XENON10 Data Acquisition System electronics.}
        \label{fig:daq_overall_diagram}
\end{figure}

The sum of the 30 central PMTs of the top array provides the global S2-sensitive event trigger (see Fig.
\ref{fig:daq_overall_diagram}).  The summed signal is amplified $\times10$ by a CAEN N968 with integration time constant
1~$\mu$s; this value was selected based on the width of the S2 signal.  The shaped signal is then passed to a CAEN voltage discriminator.  The S2 trigger was verified to have 100\% efficiency for a square pulse voltage input of width 1 $\mu$s and a step size 16 mV.  Since a typical photoelectron from a PMT with $2\times10^6$ gain has an integrated area of 16~mV$\cdot$ns, this is equivalent to 100 photoelectrons.  Considering that about 55\% of S2 scintillation is recorded by the top PMT array, it is inferred that the trigger efficiency is 100\% for S2 events of 182~phe or greater.  Because a single ionized electron extracted from LXe typically yields $24\pm5$~photoelectrons (after taking into account light collection and PMTs quantum efficiency), 182~photoelectrons in the S2 signal corresponds to $7-8$ ionized electrons.  This is the expected charge from an event with about 1~keV nuclear
recoil equivalent energy~\cite{PRL06}.  The trigger signal is distributed from the discriminator to all ADC
modules simultaneously using a Fan-Out chain.  The S2 trigger provides a lower threshold than is possible with
an S1-based trigger (as is evident from Fig. \ref{fig:sample_waveform}); moreover, an S2-trigger allows for
elimination of S1-only events due to scatters in the charge-insensitive regions of the detector, thus reducing
the trigger rate. A trigger hold-off is used to prevent the system from triggering on after-pulsing following
a large S2 pulse. The hold-off requires that no S2 trigger candidates occur for the previous 200~$\mu$s before
allowing a trigger to start acquisition.  The hold-off time is included in computing the acquistion dead time.
For gamma calibration data runs, a high energy S1 veto is added to reduce the trigger rate.  The veto is based
on the sum of the 9 center bottom array PMTs and eliminates events with an S1$>$150~keVee.  Using the S2-trigger and
no high energy veto, the acquisition rate during WIMP search mode is $\sim$2.5~Hz, leading to a dead time of
7\%.


\section{The XENON10 Data Processing and Analysis}
\label{sec:data_processing}

\subsection{S1 \& S2 Pulse Identification}

The raw data for an event consist of 88 waveforms of 16350 samples and is reduced to physical parameters
using two fully separate and parallel analysis chains.  One is written in ROOT \cite{rootwebsite}, and the other uses the commercial MATLAB software;  both were specifically developed for the XENON10 data analysis.  The high-level structure of the data reduction is similar in both cases, and proceeds in three stages: i) preprocessing the waveforms, ii) searching for pulses, and
finally iii) computing the reduced quantities associated with each scintillation pulse.

In the preprocessing stage, the samples of each PMT waveform are zeroed if they are below a threshold of $\sim$1
mV, which is much lower than the single photoelectron level and which depends on the noise characteristics of the
channel. The waveforms of all channels are then added into a total waveform that is used to search for S1
and S2 pulses.

\begin{figure}[htbp]
\includegraphics[width =0.45\textwidth]{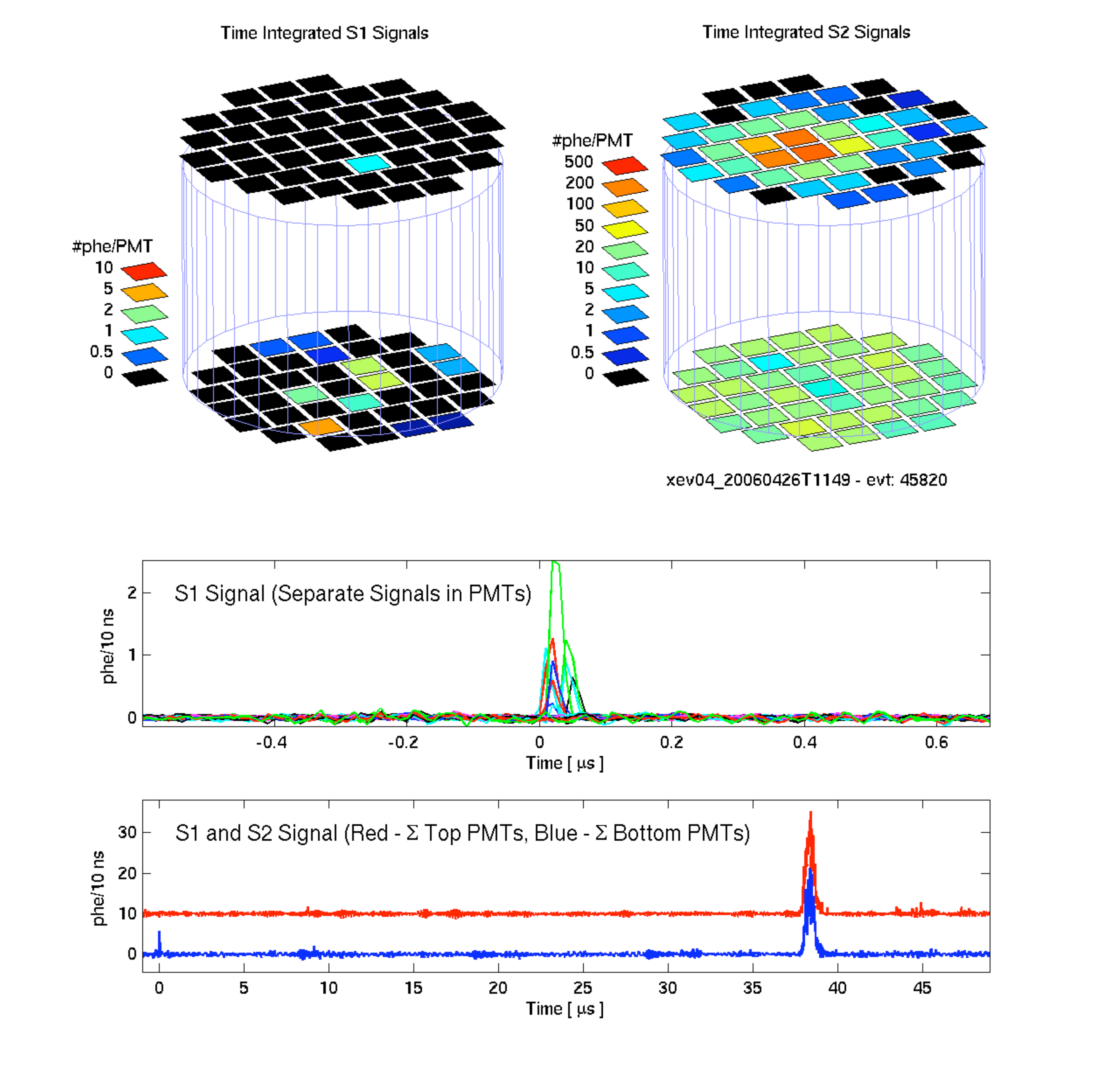}
\caption{\label{fig:typevent}(Color online) A typical event showing $S1$ and $S2$ and their hit patterns on the two arrays of PMTs (from~\cite{LDV:2009th})}
\end{figure}

The pulse searching stage operates in two steps: it first looks for S2-like pulses in the entire waveform
and then looks for S1-like pulses that precede the first S2-like pulse. Fig.~\ref{fig:typevent} shows a typical events with $S1$ and $S2$ and their hit patterns on the PMTs.

The S2 pulse finding algorithm starts by applying a digital filter to the entire waveform to smooth out the
high frequency components in order to facilitate the detection of the extent of pulses. It then searches the
filtered waveform for regions where the signal exceeds a threshold of 10 mV for at least $0.6 \, \mu s$, a
time interval large enough to contain at least one S2 pulse, and for which the preceding and following $0.2 \,
\mu s$ have an average signal less than 5\% of the maximum within the interval. Because of the long
afterpulsing tails that follow large S2 pulses the interval above threshold will often contain multiple
pulses. The algorithm then recursively searches for S2 like pulses within that interval. This is done by
computing the extent of any potential pulse by starting from its maximum sample in the interval and going
backward in the trace until either the signal drops below 0.1\% of its maximum or the slope of the signal
changes sign. This defines the left boundary of this pulse. The same procedure is repeated going forward in
the trace to find the right boundary. If the pulse found has a FWHM larger than $0.4 \, \mu s$ it will be
considered as an S2 pulse and its location and boundaries will be saved. The recursive search for S2 pulses
then continues within the interval (excluding the regions where any pulse might already have been found). The
parameters of the four largest S2 pulses are kept.  A key difference in the MATLAB-based analysis was the development of a custom double box filter \cite{Dahl:2009th}, with the width of the two boxes set to 10 and 100 digitizer samples, respectively.  These choices correspond to the approximate width of S1 and S2 signals.  This filter proved extremely adept at correctly identifying the smallest S2 pulses.

The S1 pulse finding algorithm searches the total waveform for signal excursions of at least 12 mV above the
baseline. The boundaries of the potential pulse are defined as the points where the signal drops
below 0.1\% of its maximum. If the $0.5 \, \mu s$ preceding the pulse and the 50 ns following the pulse
respectively have an average signal less than 0.5\% and 4\% of the maximum, and if the decay time of the pulse
is faster than 200 ns, then its location and boundaries will be saved. The parameters of the two largest S1
pulses identified are kept.

Once all the pulses have been identified, the reduction enters its last stage, computing the quantities of
interest. Each PMT waveform is integrated over the boundaries of the S1 and S2 pulses identified and
the values are converted into photoelectrons using the PMT gains. For each pulse a number of quantities are
computed: FWHM, height, mean arrival time, number of PMT coincidences, number of digitizer channels
saturating, etc. The $XY$ position of each S2 pulse is computed using two different position reconstruction
techniques (see section \ref{sec:posrec}). A few parameters of the entire waveform are also computed: its total
area, the average and RMS of the baseline before any pulses, and finally the total S1 and S2 signals.

\subsection{Event Position Reconstruction}
\label{sec:posrec}

An important feature of the XENON10 detector is its ability to localize events in 3D, enabling background reduction with fiducial volume cuts and multiple scatter events rejection. 
Due to the small spread of the drifting electrons by diffusion in LXe \cite{El_diff}, the proportional scintillation signal is localized to a region with the same $XY$ coordinates of the interaction site.  Near the edge of the TPC, field non-uniformities lead to systematic radial displacements towards the center of about 1~mm. Charge focussing effects in the extraction region between gate mesh and anode lead to charge displacements of less than half the mesh pitch, i.e., $< 1$~mm. 
The event $z$-coordinate, along the electric field direction, is inferred from the time difference between the primary and secondary signals, with a resolution which is $<1$~mm given the precision of the time measurement. Events with energy depositions at different $z$ positions will produce more than one S2 signal and can thus be readily identified and rejected.


To obtain the $XY$ event localization, with a few mm resolution, much smaller than the dimension of the PMTs, the light pattern measured by the top PMT array must be de-convolved. Two light pattern recognition algorithms were developed and tested on XENON10 data: a
minimum $\chi^2$ algorithm and a neural network ($NN$) based algorithm.

The minimum $\chi^2$ algorithm compares the measured light pattern with simulated S2 signals, calculating the values
\begin{equation}
	\chi^{2}(x,y) = \sum_{i=1}^{M}\frac{[S_{i}-s_{i}(x,y)]^{2}}{\sigma^{2}_{i}}
\end{equation}
for all simulated $XY$ positions. In this formula $S_{i}$ and $s_{i}$ are the measured and
simulated S2 signals (in numbers of photoelectrons) on the $i$-th PMT, $M$ is the total number of
PMTs (48 in the top array), and $\sigma_{i}$ takes into account the uncertainties of both the measured
and simulated signals. The minimum value of the
$\chi^{2}(x,y)$ function is used as an estimator for the true position in the $XY$ plane. If the statistical uncertainty
in the simulated S2 pattern is small, the main contribution to $\sigma_{i}$ is the fluctuation in the measured
signal, which includes the statistical fluctuation on the photoelectric emission from the PMT photocathode
$\sigma_{pe,i}$ and the fluctuation $\sigma_{g_{i}}$ on the gain $g_{i}$. If the number of detected
photoelectrons is sufficiently large, $\sigma_{pe,i}$ can be approximated as $\sigma_{pe,i} \approxeq
\sqrt{S_{i}}$ while $\sigma_{g_{i}}$ is measured for each PMT through its single photoelectron spectrum
\begin{equation}
	\sigma^{2}_{i} = \sigma^{2}_{pe,i} \left[1 + \left(
	\frac{\sigma_{gi}}{g_{i}} \right)^{2} \right] = S_{i} \left[1 + \left(
	\frac{\sigma_{g_{i}}}{g_{i}} \right)^{2} \right]
\end{equation}

The second algorithm is based on a fast artificial neural network. Its adaptive properties 
enable estimation of the vertex position even with a degraded input pattern, e.g., due to non-functioning PMT channels.
The speed of the algorithm is nearly independent of the dimension of the detector.

Due to the cylindrical geometry of the active volume, we chose an algorithm based on two different 
sub-networks, working simultaneously to estimate the polar coordinates, radius $r$ and polar angle $\theta$. The NNs are feed-forward  multilayer perceptrons with two hidden layers. We applied an hyperbolic tangent as activation function for the neurons in the hidden
layers, with linear output for the output neuron. The network was trained, using the backpropagation rule, on a set of $4\times 10^{4}$ simulated events, with the same simulation code used for
the $\chi^2$ algorithm.  The events were randomly generated in the $XY$ plane with a mean value of  $2.5\times
10^{3}$ pe/events. This number is sufficiently large  to minimize statistical fluctuations in the
total number of detected photoelectrons, and is comparable to the typical S2 signals in the energy
range of interest. 

\begin{figure}[!t]
\begin{center}
\includegraphics[width=0.4\textwidth]{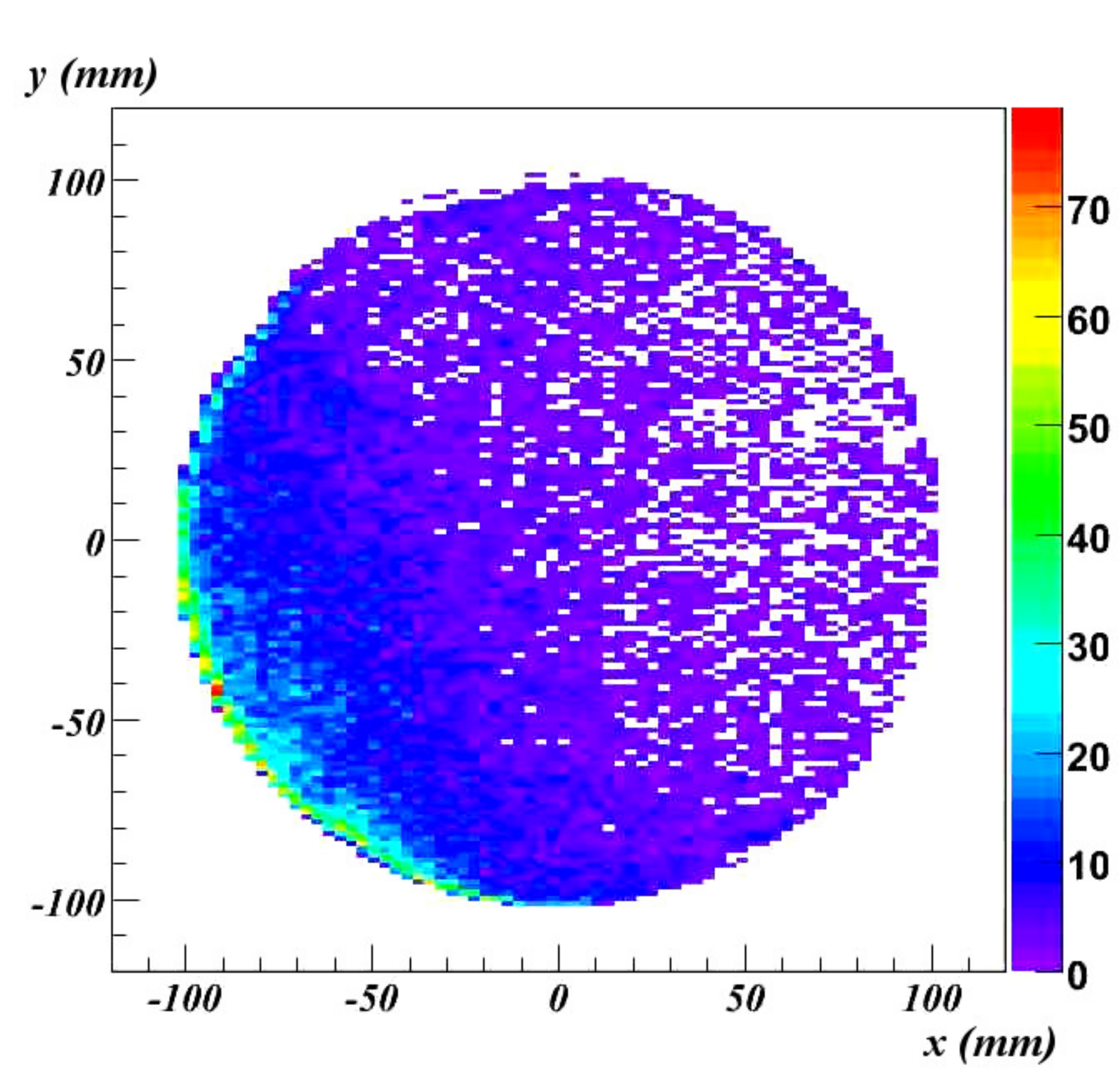}
\caption{\small{(Color online) $XY$ map of the reconstructed vertices with an external $^{137}$Cs
source: the position of the source can be clearly inferred. }}
\label{fig:NNCs}
\end{center}
\end{figure}

\begin{figure}[!t]
\begin{center}
\includegraphics[width=0.45\textwidth]{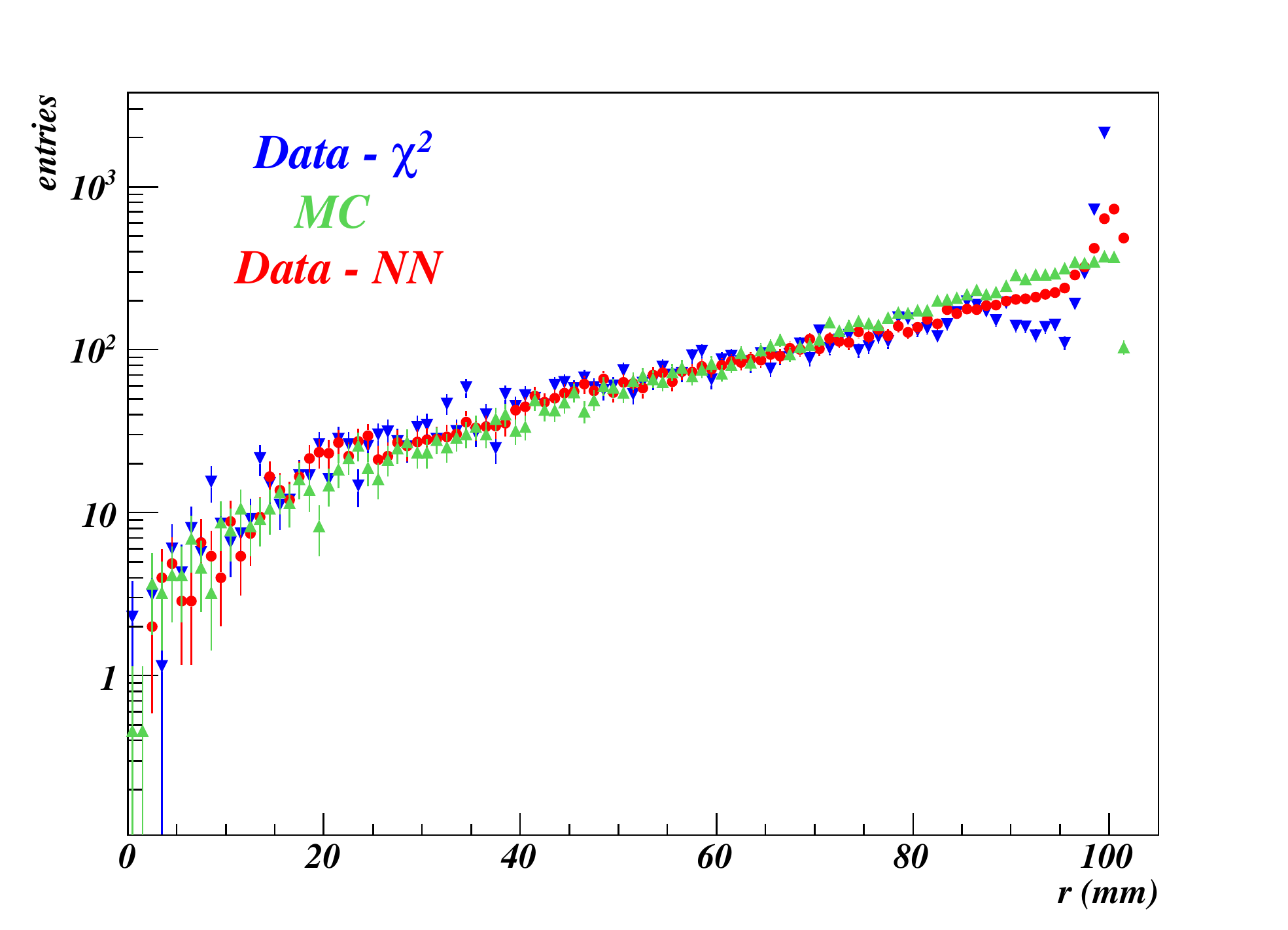}
\caption{\small{(Color online) Data-MC comparison between the radial distribution of the
vertices in the case of a $^{137}$Cs source irradiation of XENON10: the MC distribution (greeen) is plotted
with the real data distributions obtained through NN(red) and
$\chi^{2}(blue).$  }}
\label{fig:Csconf}
\end{center}
\end{figure}

The input for both NNs is a vector with the top PMT signals in number of
photoelectrons, normalized by the sum of their signals. In this way the output of the NNs is
almost independent of signal size.  The MC-estimated error on the 
radial position $r$ is $\sim 0.7$~mm for $r<80$~mm, and rises to $1.2$~mm near
the edge of the active volume.  Fig.~\ref{fig:NNCs} shows a map of the reconstructed interaction vertices 
produced by a Cs source placed near the cryostat. The source location is apparent. In fig.~\ref{fig:Csconf} the radial distributions of the
Cs-interaction vertices obtained though NNs and $\chi^{2}$ reconstruction algorithms are compared with the corresponding MC 
simulation. Clearly the NN distribution is in better agreement with the MC
simulation near the edge of the detector, where reflection of scintillation light on the Teflon renders  
the vertex reconstruction more
difficult. Both NN and $\chi^{2}$ distributions show very good agreement
with the simulated data for $r<80$~mm. 

\subsection{Basic Quality Cuts}
\label{sec:qc}

A number of basic quality cuts were defined to remove events that do not correspond to real energy depositions
in the detector, events with unphysical parameters, events with misidentified features, etc.. Many of the basic
quality cuts are based on the known physical properties of the direct scintillation light of LXe and on the
characteristics of proportional scintillation of gaseous Xe.

\begin{figure}[htbp]
    \includegraphics[width =0.45\textwidth]{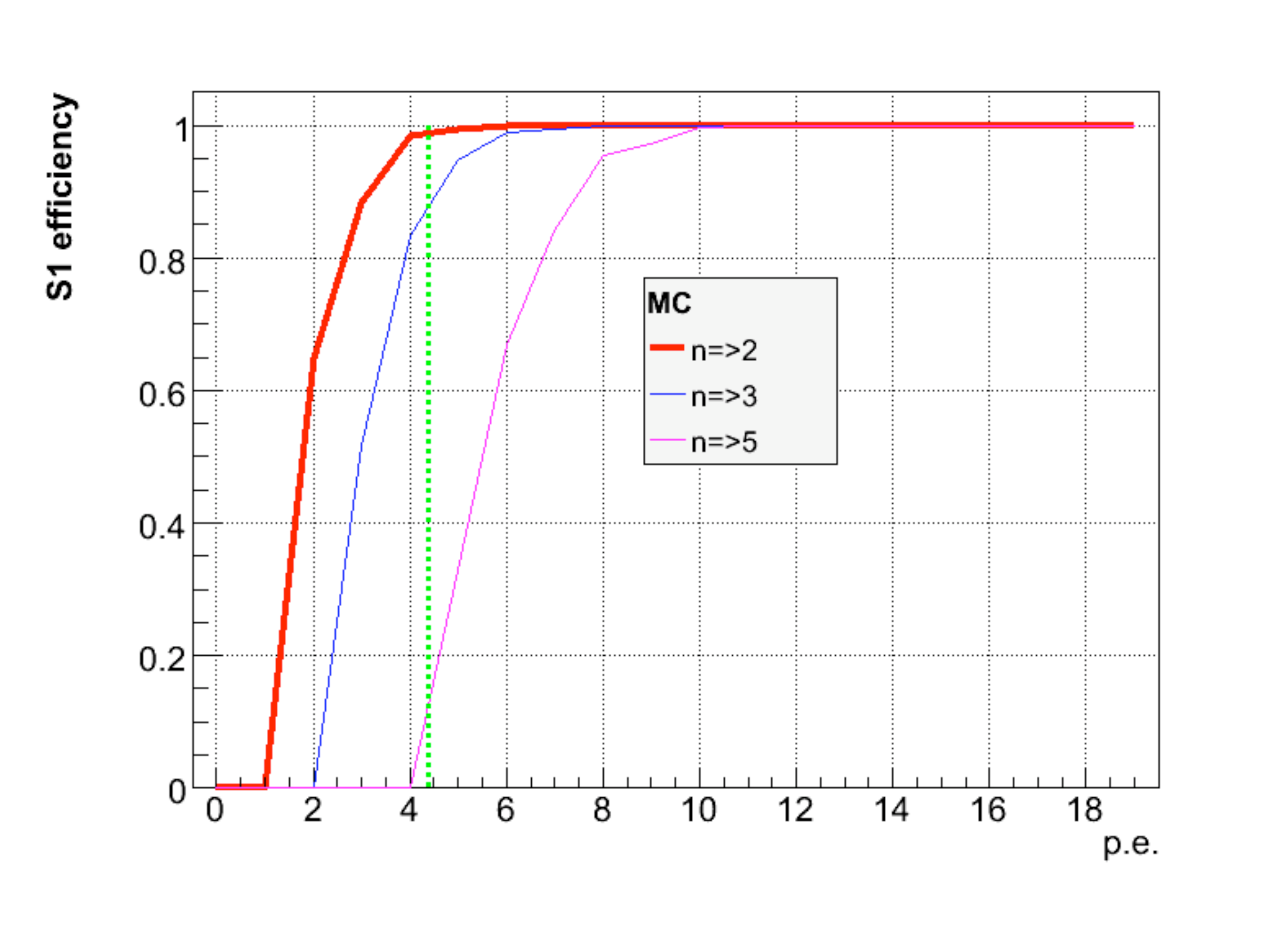}
	\caption{\label{fig:trg_eff} (Color online) Probability of S1 identification as a function of the number of required
	coincidences.}
\end{figure}

The basic quality cuts applied to the S1 signal are on the number of PMT coincidences, on the width and on the
mean arrival time. In order to clearly distinguish low energy S1 signals from random single photoelectrons,
the minimum coincidence requirement is set to 2 PMTs having a signal above 0.35 pe. The efficiency of this cut
is shown in figure \ref{fig:trg_eff}. In addition, the width at 50\% height of the S1 pulse is required to be
larger than 20~ns and smaller than 150~ns (note that this is the width in the digitized trace, after the
anti-aliasing filter of the FADC as described in section \ref{sec:daq}) and its mean arrival time is required
to be shorter than 60 ns. These cuts ensure that the S1 signal is not confused with noise or with small
single electron S2 signals, for example.

Multiple basic quality cuts are also applied to the S2 signal. They consist of a cut to reject events with
more than one S2 pulse, a cut on the width, a cut to remove events with ADC saturation and a cut on the
proportion of the S2 signal seen by the top and bottom PMT arrays. As WIMPs are not expected to scatter more
than once in the detector, events with more than one S2 pulse are usually discarded. This is also necessary
for other analyses, when looking at the spatial dependence of the direct scintillation for example (section
\ref{sec:actxe}). In terms of pulse shape, the FWHM of the S2 signal is required to be larger than 0.5 $\mu
\mathrm{s}$ but less than 1 $\mu \mathrm{s}$, consistent with what is expected from the proportional
scintillation gap. Events for which the S2 signal peaks at a higher voltage than the input range of the FADC
are also removed, although this effect is absent at energies lower than 50 keVee. Monte Carlo simulations
have shown that the expected fraction of S2 light measured by the top (bottom) PMT array should be about 55\%
(45\%) and is consistent with what is observed in calibration data. Consequently, events where the ratio of S2
light seen on the top and bottom arrays falls outside the expected range are also rejected. For example, very
large S1 signals from alpha particles misidentified as S2 pulses would be cut since the bottom array would
measure a much larger proportion of the signal.

Another cut, intented to remove events in which the $XY$ position has been incorrectly inferred, is also
applied. The cut requires the $\chi^2$ between the S2 signal distribution from data and the Monte Carlo
simulated distribution at the position inferred from the Neural Network algorithm to be less than a threshold
value. 

Finally, an additional important parameter, related to the quality check of the pulse finding algorithm, is defined as
\begin{equation}
	\log \left({\frac{S1+S2}{A-S1-S2}}\right)
\end{equation}
where $A$ is the total area of the trace and $S1$ and $S2$ are the total signals identified. This parameter is
analogous to a signal to noise ratio since it will be very large when all the features of the waveform have
been identified and are included in the total $S1$ and $S2$ signals whereas it will be very low if some
features have been not been identified. 

\section{The XENON10 Calibration Data Results}

\subsection{Gamma Calibration}

To achieve a reliable and accurate detector calibration, with minimum impact on WIMP search exposure time, the
XENON10 shield was designed to allow introduction of external calibration sources without exposing the
detector cavity to outside air.  The gamma calibration sources used were $^{57}$Co, $^{137}$Cs, $^{60}$Co and $^{228}$Th.  The gamma
source calibration data are used not only to determine the energy scale based on S1 (or S2) signals, but also
to define the detector's response to electron recoils from background events.

\subsubsection{S1 light yield and energy scale}

\begin{figure}[htbp]
	\includegraphics[width=0.45\textwidth]{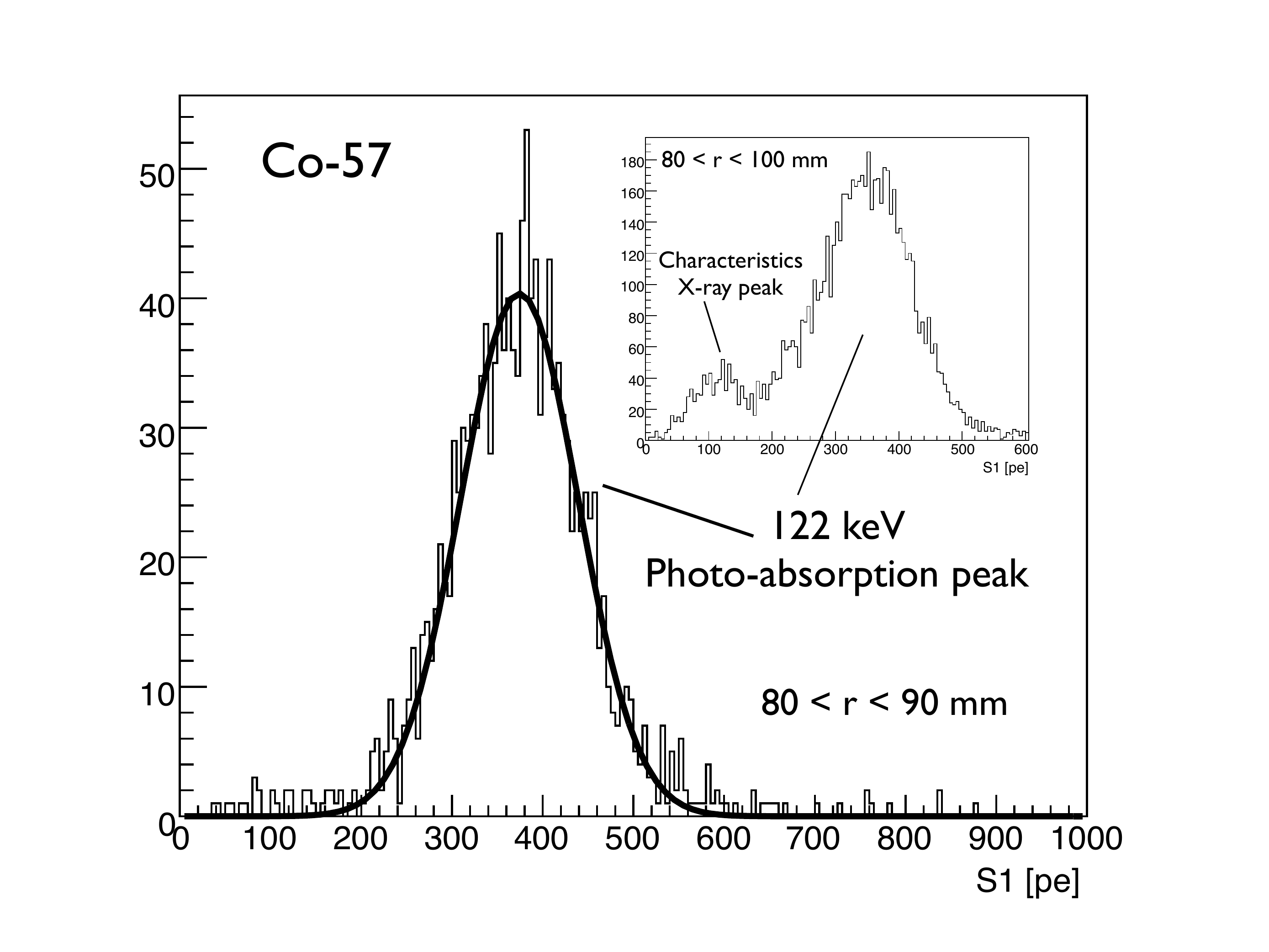}
	\caption{S1 scintillation spectrum from $^{57}$Co gamma calibration.}
	\label{fig:Co57}
\end{figure}

The direct scintillation (S1) light yield in XENON10 depends on the event energy and position as well as on the electric field strength in liquid xenon.  The operating field in XENON10 was 0.73 kV/cm. Measurements of
the S1 yield at this field were carried out with gamma sources.  In the bulk region ($r <$ 5 cm), the 662 keV
photo-absorption peak of $^{137}$Cs gives an average light yield of 2.2$\pm$0.1 p.e./keV. For 122 keV gammas from $^{57}$Co source, the light
yield is about 3.1 p.e./keV, for events with radial positions between 8 and 9 cm, as shown in
Fig.~\ref{fig:Co57}. The different light yield for different gamma ray energies is due to the different
contribution of the electron-ion recombination~\cite{Yamashita:04}. We were also able to observe the 30~keV characteristic X-ray peak for events near the edge of the sensitive volume. 


\subsubsection{S2 yield and liquid xenon purity}

\begin{figure}[htb] 
   \includegraphics[width=0.49\textwidth]{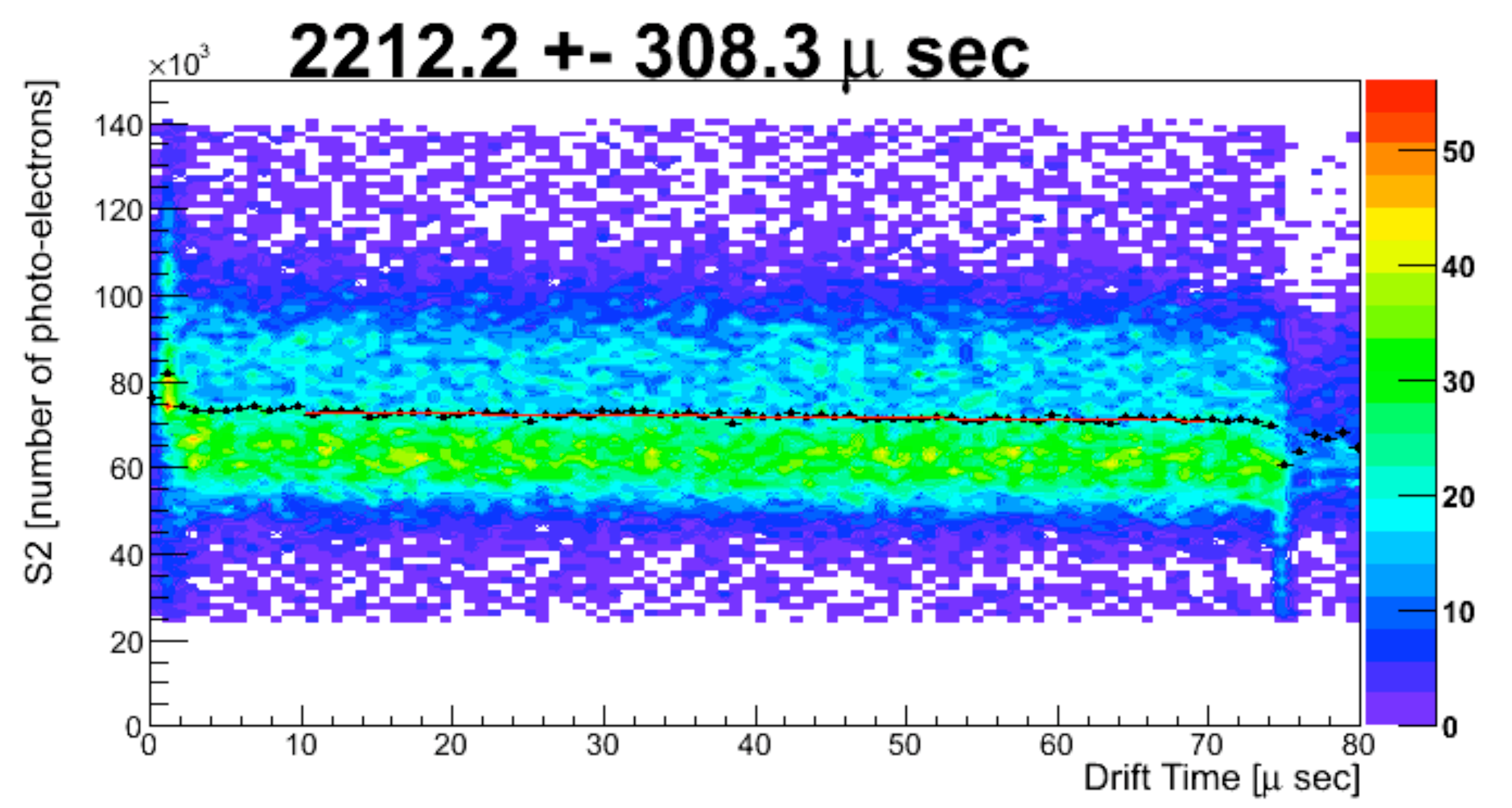} 
   \caption{(Color online) Determination of the electron lifetime from calibration data. The band is the S2 signal measured by the bottom PMT array as a function of drift time for fully-absorbed 164~keV gamma events from the activated Xe calibration of XENON10.}   
   \label{ws004_etime}
\end{figure}

The proportional scintillation light S2 is proportional to the number of electrons liberated in the liquid
 by an ionizing event. The number of electrons which reach the liquid-gas interface depends strongly on the liquid xenon purity, which can be inferred from a measurement of the electron lifetime $\tau$~\cite{Bakale:1976}. The presence of impurities reduces the number of electrons produced in the liquid at time t=0 ($N_e(0)$) according to the relation: $N_e(t) = N_e(0) e^{-t/\tau}$. 
 To determine the electron lifetime, we used the data from $^{137}$Cs calibration, measuring the attenuation with drift time of the S2 signal associated with the full energy peak of 662~keV gamma rays. Throughout the dark matter search period an electron lifetime longer than 2~ms was measured, corresponding to $\ll1 \, \text{ppb}$ O$_2$ equivalent impurity concentration in the LXe.  A similar purity level was inferred during a subsequent calibration of XENON10 with gamma rays from neutron-activated Xe gas, carried out shortly after the first dark matter search run was completed.  Fig.~\ref{ws004_etime} shows the electron lifetime measurement using S2 from the 164~keV gamma events uniformly distributed in the XENON10 active volume. The electron lifetime inferred from this calibration is $2.2 \pm 0.3 \,\mathrm{ms}$, confirming the excellent purity level achieved in XENON10 during the dark matter search. 

\begin{figure}[htbp]
\includegraphics[width =0.45\textwidth]{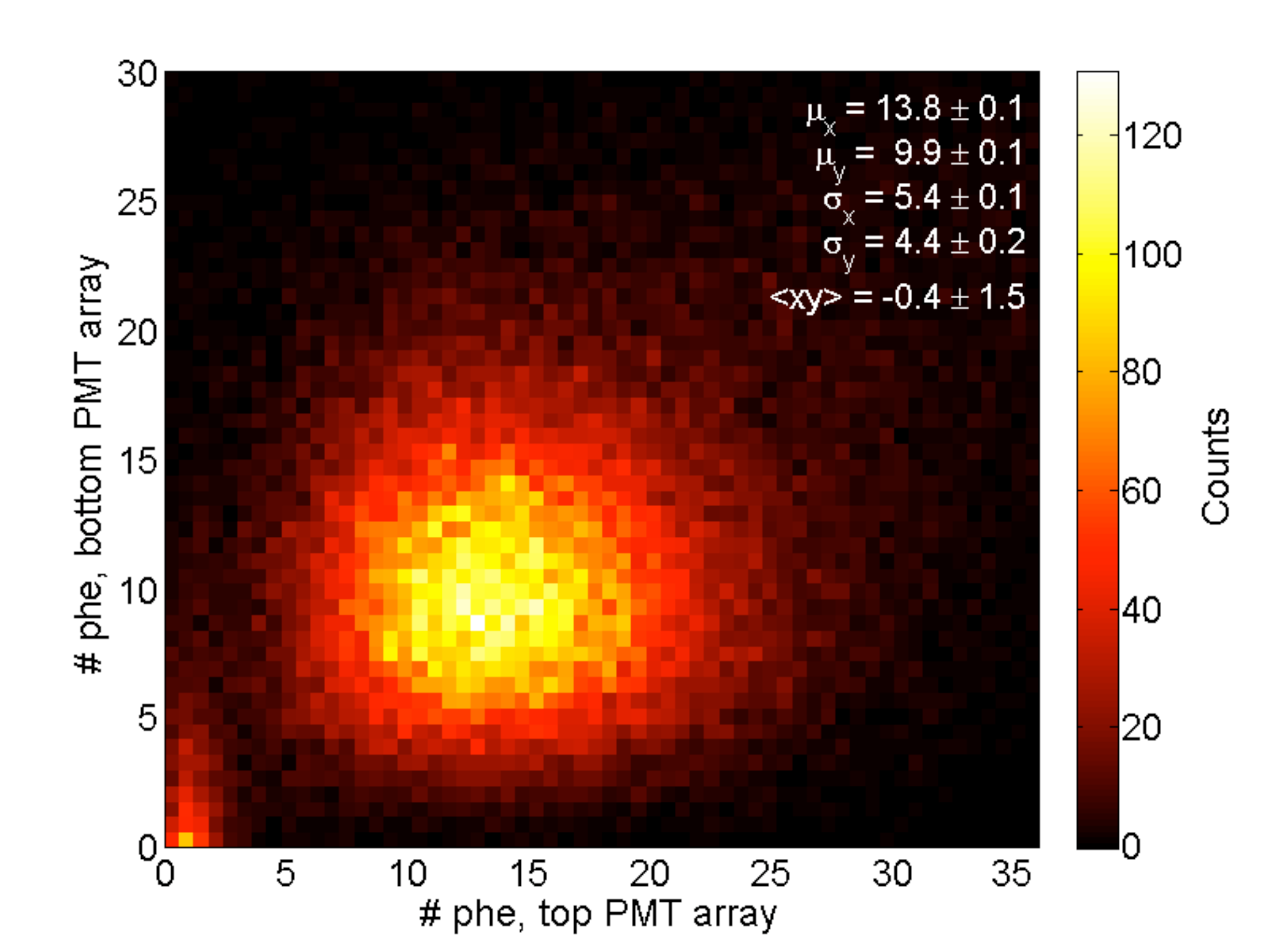}
\caption{\label{F-singleelectron}(Color online) S2 from single electrons seen in the neutron calibration data, with the signal detected by the top PMT array on the x-axis, and bottom PMT array on the y-axis.  These events were selected based only on pulse width (full width 0.2~$\mu$s - 3.0~$\mu$s) plus a requirement that they would be at least 10~$\mu$s from the trigger region (either before or after).}
\end{figure}

As previously discussed, the amplitude of the S2 signal, for a given number of ionization electrons extracted from the liquid, depends on several factors: the gas pressure $p$, the gas gap $x$, the electric
field $E$ across the gap, the light collection efficiency of the PMTs, and the quantum
efficiency $Q_E$ of the PMTs. The observed number of photoelectrons per electron drifting in the gas can be found from single electron pulses within the data. During the operation of the XENON10 detector, a class of pulses due to single electron emission from the liquid to
the gas phase was observed~(Fig.~\ref{F-singleelectron}). From this observation, we estimate that, on average,
each electron produces 13.8 p.e. on the top PMT array and 9.9 p.e. on the bottom PMT array respectively.

\subsubsection{Position-dependence of $S1$ and $S2$ Signals}
\label{sec:actxe}

The S1 light collection efficiency (LCE) has a strong dependence on the event position, due to the effects of
total internal reflection at the liquid-gas interface, the solid angle, the optical transmission of the grids
and the teflon reflectivity. In order to
obtain an accurate energy calibration, data from both Monte Carlo simulation and internal sources were used to study the
position dependence of the signals. 

\begin{figure}[htb]
	\includegraphics[width=0.35\textwidth]{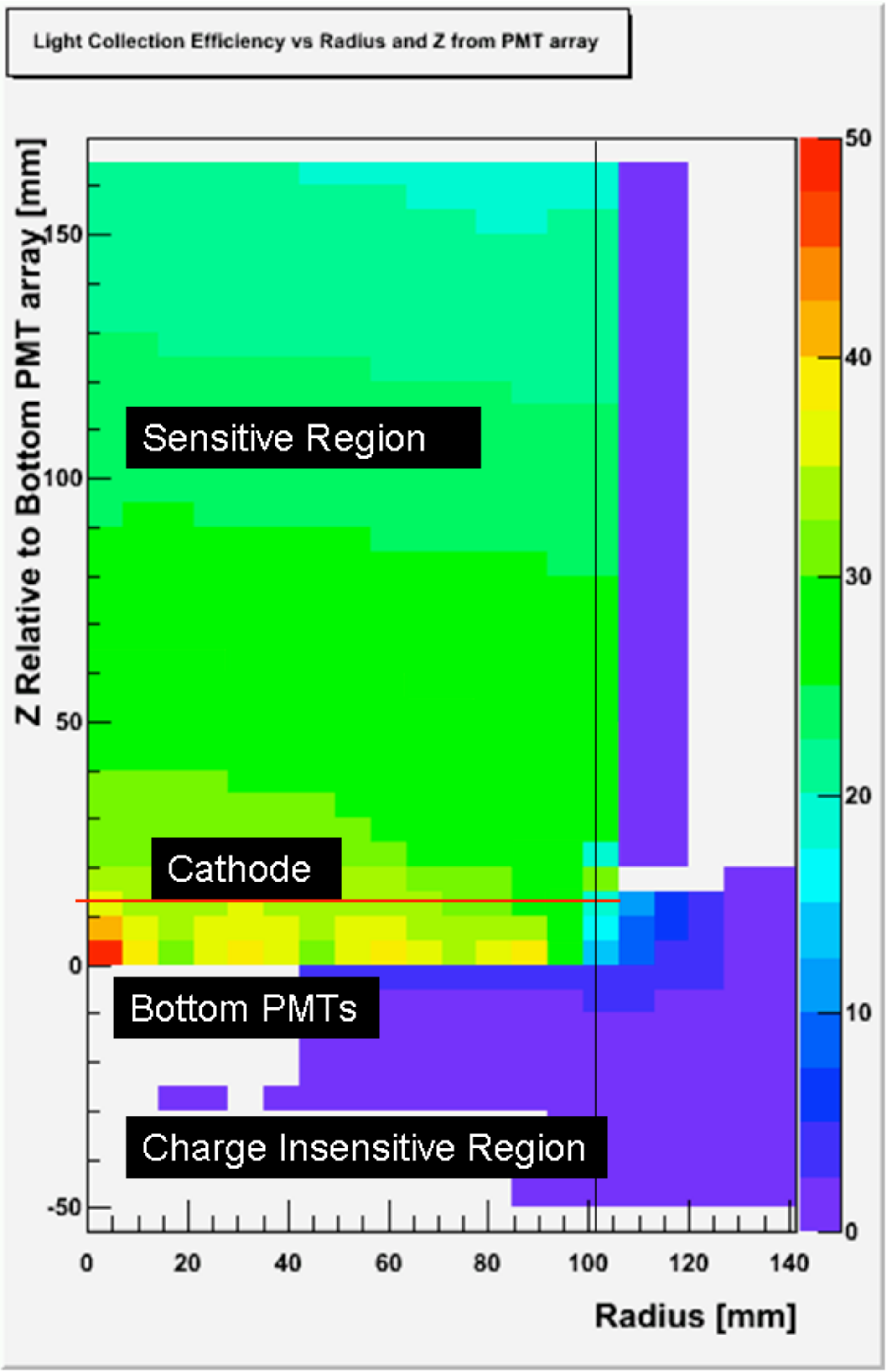}
	\caption{(Color online) Simulated S1 light collection efficiency throughout the XENON10 detector.}
	\label{fig:S1_LCE_sim}
\end{figure}

Fig.~\ref{fig:S1_LCE_sim} shows the simulated S1 LCE throughout the XENON10 detector, where we define LCE as
the probability of photons released in the detector to hit the photocathode of a PMT; it does not include the
QE of the PMTs. Some regions are light- but not charge-sensitive, most notably in the reverse field region
between cathode and bottom PMT array. Some light sensitivity also exists in the xenon around and below
the bottom PMT array, where stray light enters through openings between the Teflon cylinder and the PMTs.  The
primary light is predominantly detected by the bottom PMT array, due to total internal reflection at the
liquid-gas interface, where the index of refraction changes from 1.61~\cite{Bakov:1996} to 1. The ratio of top to bottom S1
signal ranges from $\sim 0.14$ to $\sim 0.3$ from the bottom to the top of the drift region. Some light is
absorbed by the meshes, by the PTFE walls, or by the liquid xenon. We calculated 92\% transparency of the meshes  and
assumed 92\% reflectivity for PTFE~\cite{Yamashita:04}. The absorption length of liquid xenon is taken as 100~cm and the scattering length
as 30~cm~\cite{ref:scatlength}.

\begin{figure}[htb]
	\includegraphics[width=0.4\textwidth]{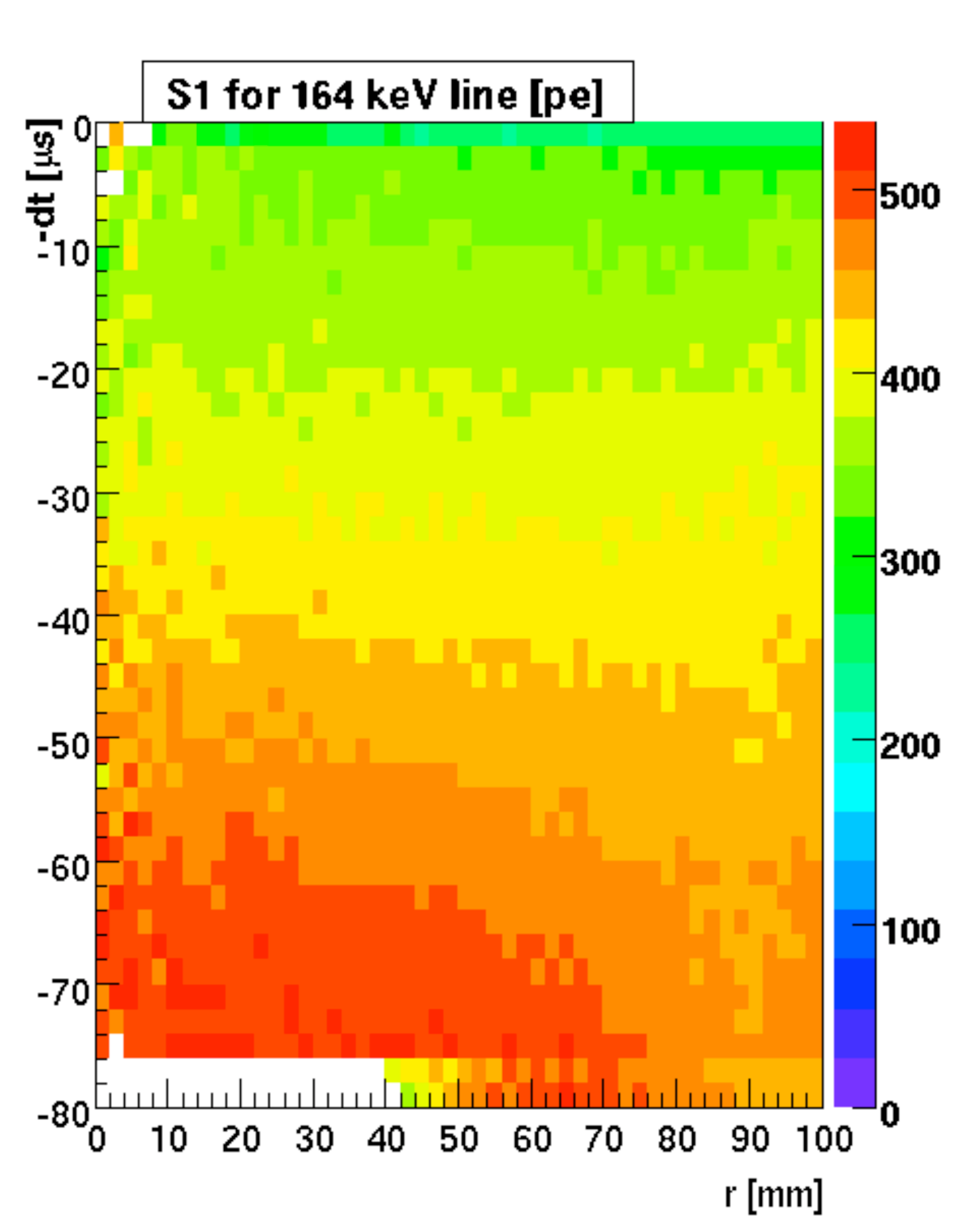}
	\caption{(Color online) S1 light yield throughout the sensitive liquid xenon target from activated xenon calibration with
		164~keV gamma rays.}
	\label{fig:s1dtr}
\end{figure}

\begin{figure}[htb]
	\includegraphics[width=0.45\textwidth]{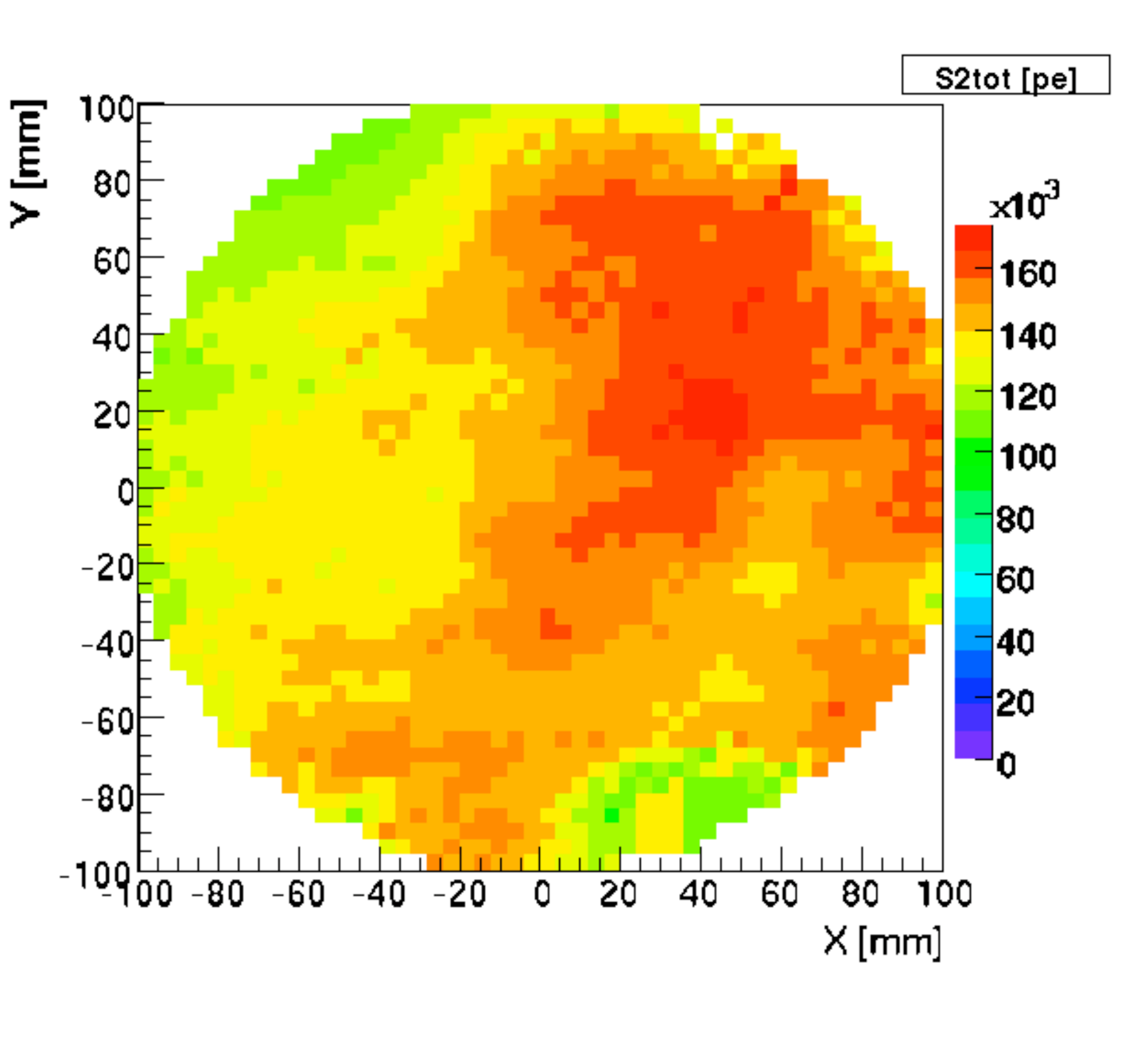}
	\caption{(Color online) S2 light yield response at different $XY$ positions from activated xenon calibration with 164~keV
		gamma rays.}
	\label{fig:s2xy}
\end{figure}

In addition to the simulations, we also used neutron activated xenon isotopes, primarily $\rm^{131m}Xe$
(164~keV gammas) and $\rm^{129m}Xe$ (236~keV gammas) to calibrate the detector's S1 response throughout the entire active volume. The activated xenon isotopes
were introduced into the XENON10 detector following several days of neutron activation of natural xenon
\cite{Ni:07act}. Unlike external calibration sources, activated Xe isotopes provided unique energy and position
calibration throughout the entire LXe sensitive volume. The excellent 3D position sensitivity of XENON10
allows event-by-event signal correction based on position. Fig~\ref{fig:s1dtr} shows the S1 light yield
for 164~keV gamma rays throughout the sensitive volume. The data can be compared with the simulated S1 response
(Fig.~\ref{fig:S1_LCE_sim}) to determine parameters such as PTFE reflectivity and liquid xenon absorption
length, that is relevant for the light collection. The response of the S2 signal at different
$XY$ positions was also studied using the 164~keV gamma calibration data (Fig.~\ref{fig:s2xy}). The signal
size differs by about 50\% from the edge to the center, due to the sagging
of the meshes which changes the gap where proportional scintillation light is generated. 
 
 We used the S1 and S2 light response maps from activated Xe calibration as a function of event position to
 correct the signals in the final data analysis. Better energy scale, electron/nuclear recoil discrimination
 and better energy resolutions were achieved after the position-dependent corrections of the two signals. 
 
\subsubsection{Combined energy scale}

\begin{figure}
\centering
\includegraphics[width=0.45\textwidth]{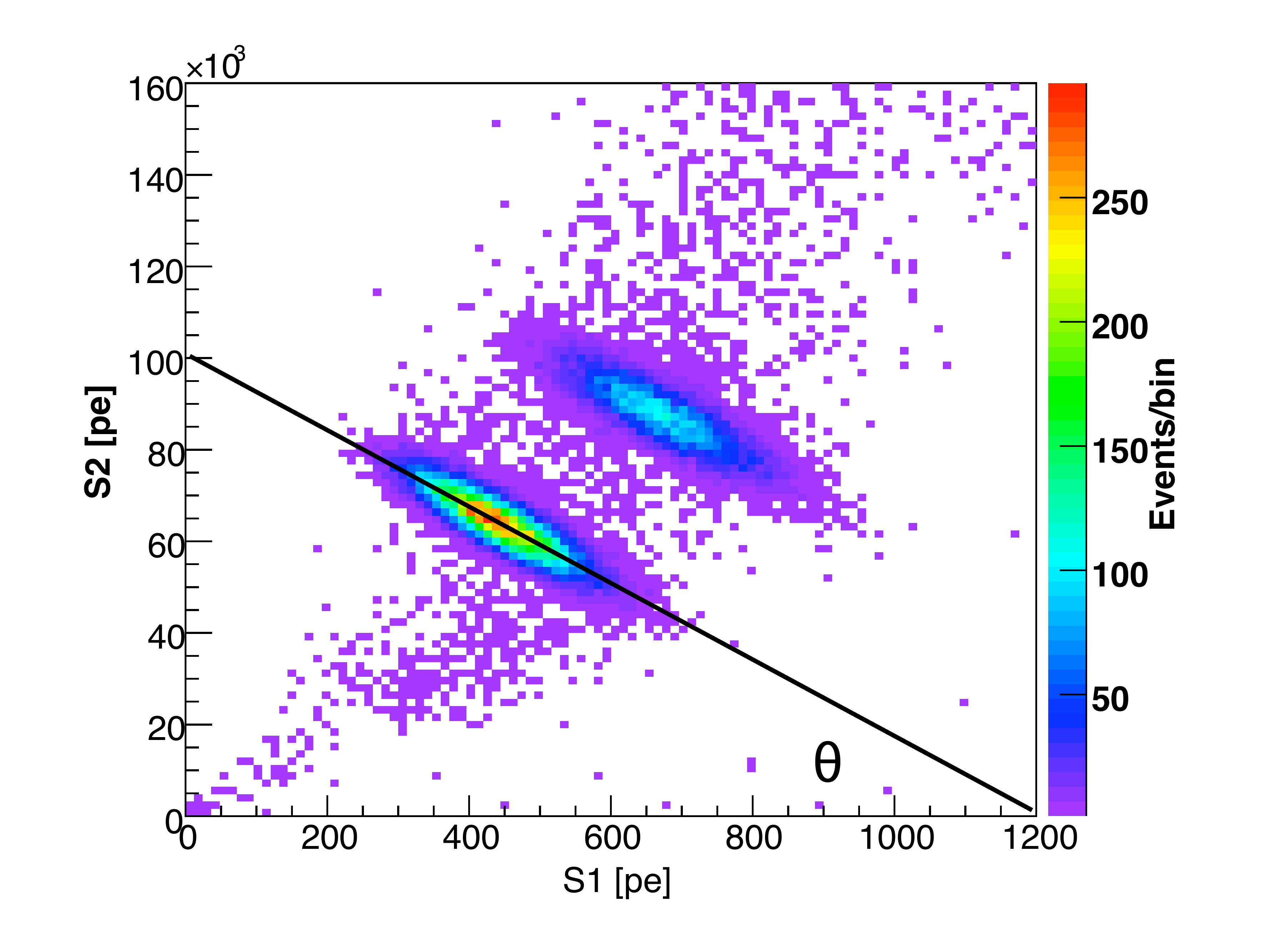}
\caption{(Color online) Anticorrelation between $S1$ and $S2$ for 164 keV and 236 keV $\gamma$ rays from activated xenon isotopes ($\rm^{131m}Xe$ and $\rm^{129m}Xe$).}
\label{fig:anticorr}
\end{figure}

\begin{figure}
\centering
\includegraphics[width=0.45\textwidth]{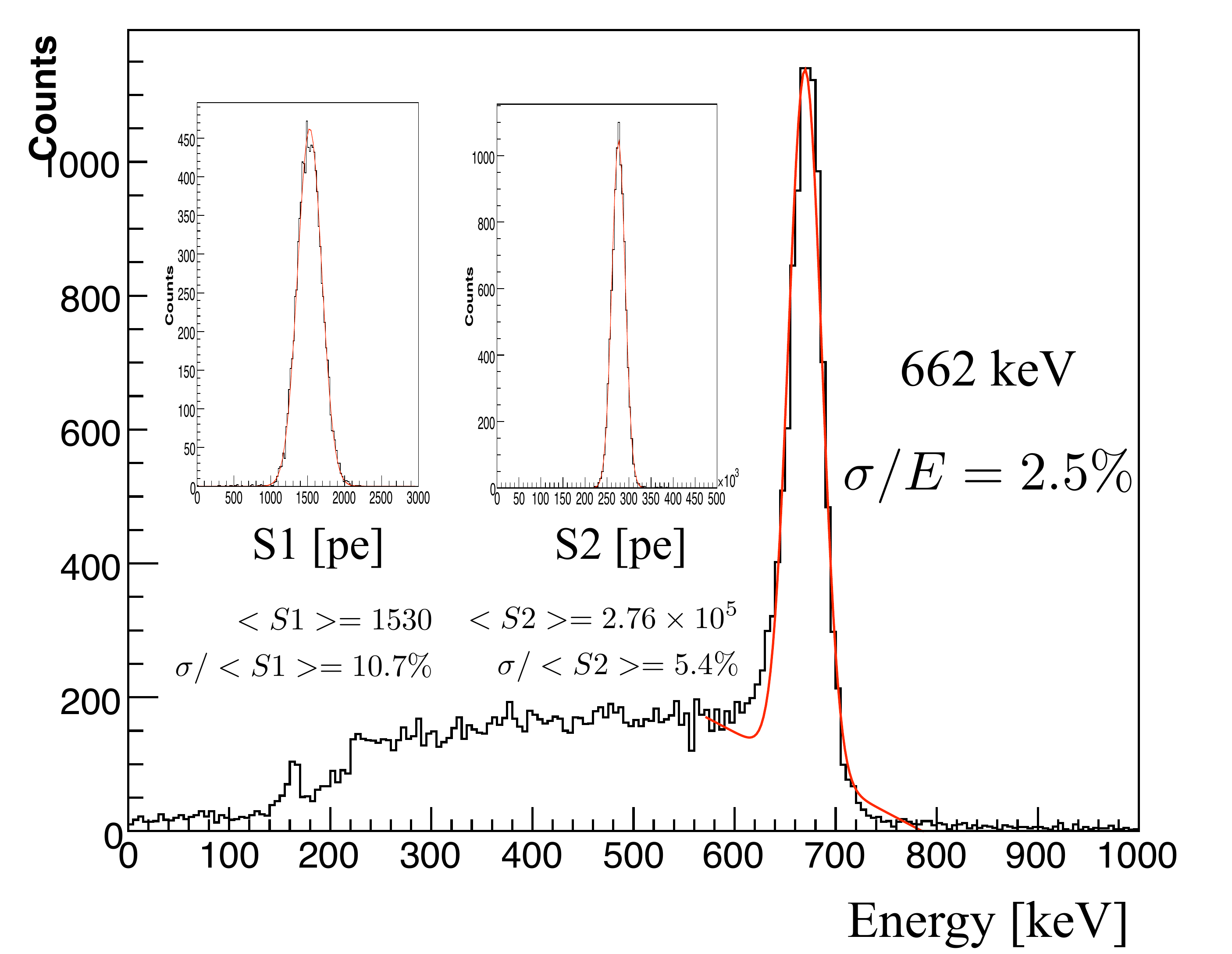}
\caption{Combined energy spectrum for single-scatter events from 662 keV gamma rays interacting  within the 5.4-kg fiducial mass of XENON10. Insets: S1 and S2 distributions for 662 keV photo-absorped events (events within 2-$\sigma$ around the 662 keV peak in the combined energy are selected). The energy resolution at 662 keV is 10.7\%, 5.4\% and 2.5\% for S1, S2 and the combined energy, respectively.}
\label{fig:662keVall}
\end{figure}

\begin{figure}
\centering
\includegraphics[width=0.45\textwidth]{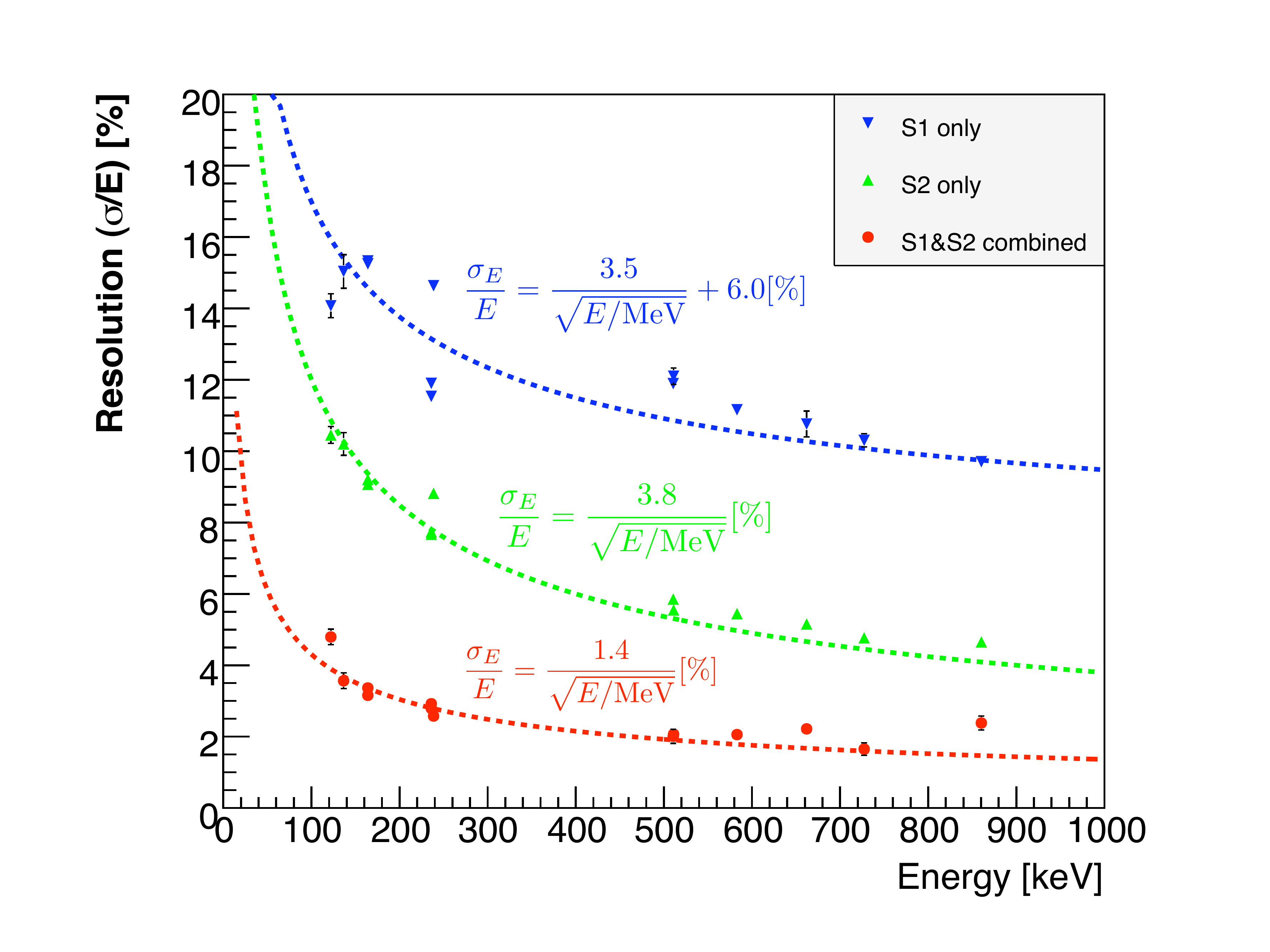}
\caption{(Color online) Energy dependence of resolutions in liquid xenon for $S1$ (scintillation) only, $S2$ (ionization) only and combined (scintillation plus ionization) energy spectra, obtained for gamma rays below 1 MeV in XENON10.}
\label{fig:cmres}
\end{figure}

\begin{figure}
\centering
\includegraphics[width=0.45\textwidth]{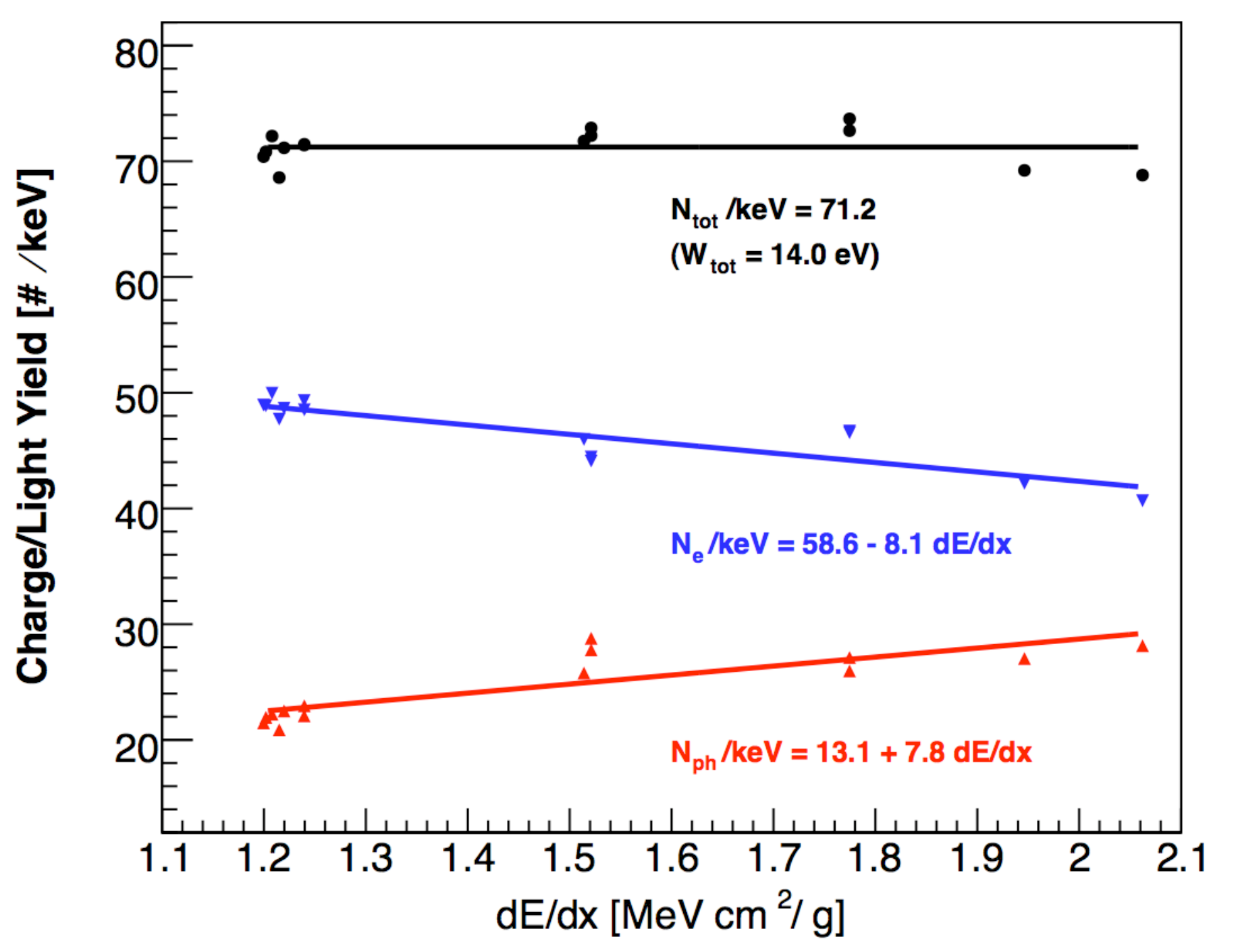}
\caption{(Color online) Number of scintillation photons ($N_{ph}$) and ionization electrons ($N_e$) per unit energy from gamma rays interaction in liquid xenon via photo-absorption for different stopping power ($dE/dx$) at the operating field (0.73 kV/cm) of XENON10. Values of $dE/dx$ corresponding to different energy of electrons (from photo-absorption of gamma rays) can be obtained from the ESTAR database \cite{ESTAR}. The total number of quanta (photons plus electrons) per unit energy does not depend on the energy of gamma rays.}
\label{fig:lyqy}
\end{figure}

A recombination-independent combined energy scale for electron recoils, as described in details
in~\cite{Conti:2003av,Aprile:2007qd}, can be produced from the appropriate sum of light (S1) and charge (S2).
This scale should be free of the nonlinearites present in the S1-based scale.  The energy deposited by each
event is determined by combining $S1$ and $S2$ signals as following,

\begin{equation}
	E = (\frac{S1}{\alpha} + \frac{S2}{\beta}) \cdot W_{tot}
	\label{eq:etot}
\end{equation}

where $S1$ and $S2$ are in units of number of photoelectrons ($N_{pe}$). $\alpha$ and $\beta$ are
experimentally determined parameters in units of $N_{pe}$/photon and $N_{pe}$/electron, respectively.
$W_{tot}$ is the average energy required to produce either a scintillation photon or an ionization electron in
LXe. In XENON10, $\beta$ is determined from the $S2$ corresponding to the single-electron
emission peak (see Fig.~\ref{F-singleelectron}). The anti-correlation between $S1$ and $S2$ is due to
electron-ion recombination fluctuation in LXe. Each recombined electron-ion pair will create
one UV photon. Thus, $\alpha$ in equation \ref{eq:etot} can be determined by $\beta$ and the slope $\theta$ in
Fig.~\ref{fig:anticorr}, according to the relation $\alpha = \beta / \tan\theta$. 
From the 164~keV calibration peak,
we obtain,

\begin{equation}
W_{tot} = 14.0~ \rm eV
\label{eqwtot}
\end{equation}

in good agreement with a study in a small LXe detector~\cite{Shutt:NIM07}. The energy resolution of the XENON10
detector was investigated with gamma ray sources ($\rm^{57}Co$, $\rm^{22}Na$, $\rm^{137}Cs$, $\rm^{228}Th$)
covering the energy range between 122~keV to 2.6~MeV. An example, from $\rm^{137}Cs$ 662~keV gamma rays, is shown in Fig~\ref{fig:662keVall}. For comparison, we also plot the energy resolution, in Fig.~\ref{fig:cmres}, obtained by using only $S1$, only $S2$ and the sum of these two signals. At 1~MeV, the resolution from the combined energy measurement is about a factor of seven
better than that from $S1$ alone, and a factor of three better than that from $S2$ alone. The energy determined from the combination of S1 and S2 signals show a much more linear response than that based on S1 or S2 alone (see Fig.~\ref{fig:lyqy}).


\subsection{Neutron Calibration}

To understand the XENON10 response to nuclear recoils, a neutron calibration was performed using a
3.7~MBq $^{241}$AmBe source, emitting $\sim$220 neutrons/second. The calibration was done by
exposing the XENON10 detector to the source for approximately 12 hours, with a live time fraction of 0.92. The $^{241}$AmBe
source (attached to a steel rod) was inserted through a 7~mm diameter hole in the XENON10 shield. The source
was positioned next to the cryostat, between two 5~cm thick Pb bricks used to block high energy (a few MeV)
gamma rays produced by the source. 
%

The emitted neutrons have energies ranging from 0.1~MeV to 11~MeV, with a mean at 4.3~MeV. The calibration
data were recorded at a constant rate of 6.5~Hz during the exposure. The trigger setup was the same as used for
the WIMP search, with the addition of a high energy veto to reject events with energies above $\sim$120~keV.

\begin{figure}[htpb]
	\begin{center}
		\includegraphics[scale=0.475]{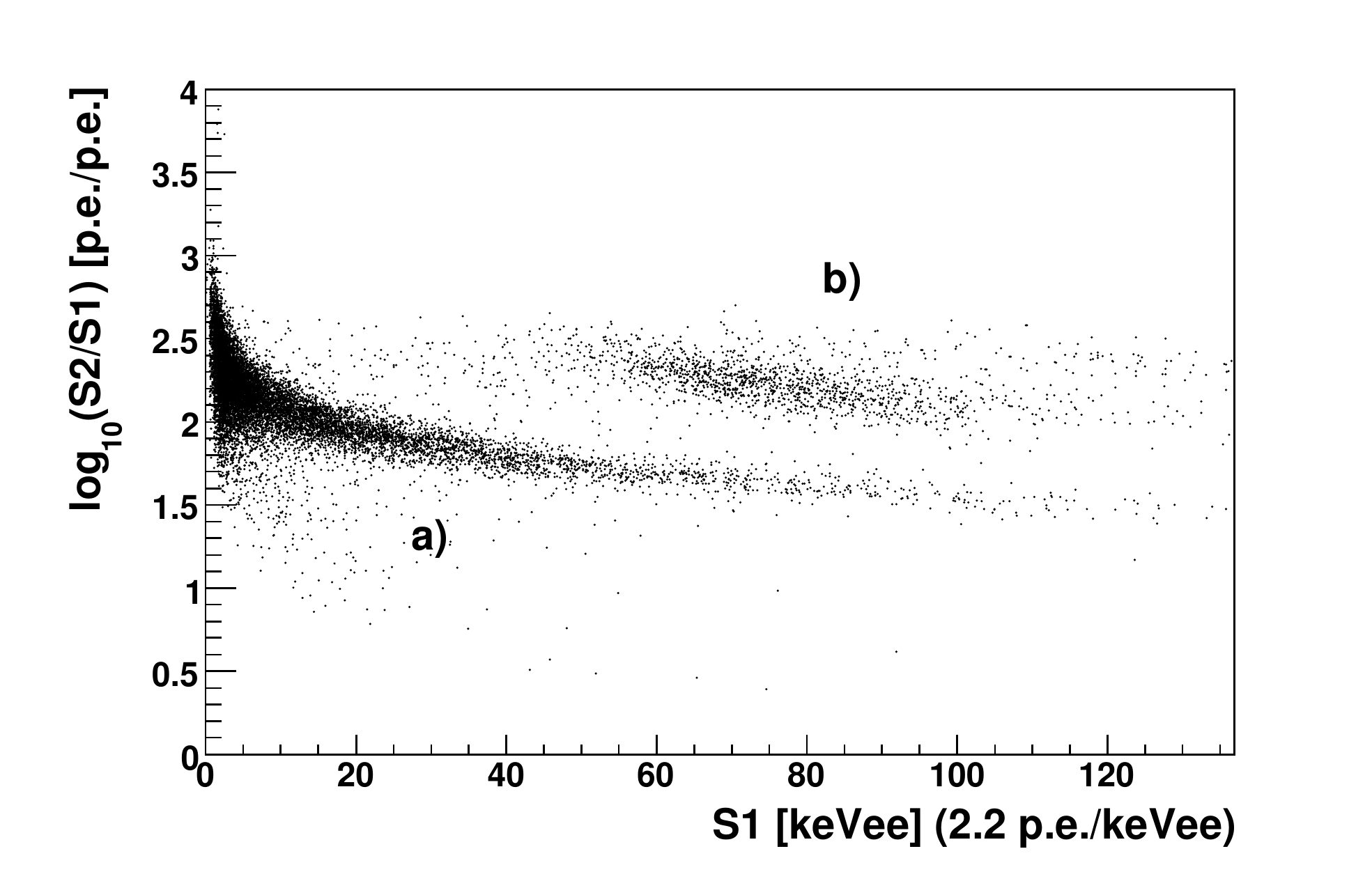}
	\end{center}	
	\caption{$\log_{10}(S2/S1)$ vs. $S1$ distribution from neutron interactions in the XENON10 detector. The
		data show two different regions: region a) which corresponds to  recoils from elastic scatters and defines the nuclear recoil band, and region b) which
		corresponds to recoils from inelastic scatters. }
	\label{fig:ambe_s2s1}
\end{figure}

Figure~\ref{fig:ambe_s2s1} shows the $\log_{10}(S2/S1)$ vs. $S1$ distribution from the neutron calibration
run, after applying the quality cuts discussed in Section~\ref{sec:qc}. Two regions are clearly distinguished in
Figure~\ref{fig:ambe_s2s1}: region a) which  defines the nuclear recoil band corresponding to single elastic
scatters, and region b) which corresponds to inelastic neutron scatters with $^{129}$Xe, which produce 40~keV
gamma rays. Neutron inelastic scatters with $^{131}$Xe will produce 80~keV gamma rays, however the data do not
show a peak around this value because of the events rejection by the high energy veto. The
nuclear recoil band is used to determine XENON10 discrimination power, described in
Section~\ref{sec:nrerdis}.

\subsection{Nuclear/Electron Recoil Discrimination}
\label{sec:nrerdis}

The success of XENON10 as a dark matter detector hinges in large part upon its ability to discriminate
electronic recoils from nuclear recoils, which in turn requires adequate definition of the detector response
to such events based on calibration data.  In addition to the formation of excitons, recoiling particles will produce a
population of ionized electrons, many of which promptly recombine with their parent ions.  Under an applied electric
field, the relative number of recombining electrons decreases.  However,
because nuclear recoils have a characteristically higher ionization density than electronic recoils,
fewer electrons escape recombination from recoiling nuclei than electrons, for a given energy and drift field.

\begin{figure}[h!]
\begin{center}
		\includegraphics[width=0.45\textwidth]{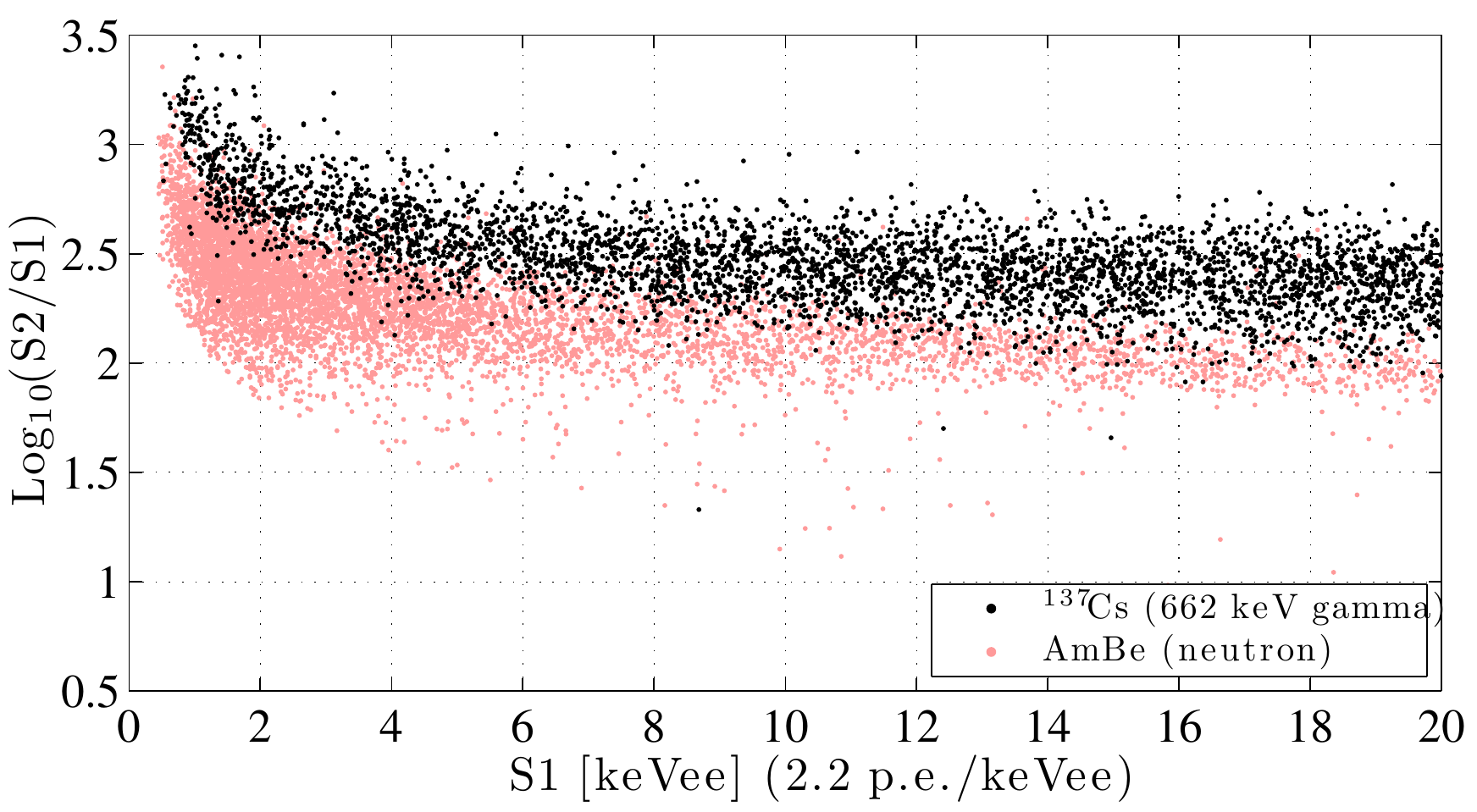} 
\caption{(Color online) The electronic and nuclear recoil bands shown in $\rm Log_{10}(S2/S1)$
vs. S1 space.}
\label{ERNRBands}
\end{center}
\end{figure}

Electrons which recombine contribute to the prompt scintillation signal (S1), while those which escape
recombination are drifted to the anode in the gas and produce the proportional signal (S2).  The
relative strength of recombination for a given event can be measured by the ratio S2/S1, and hence this
parameter can be used to discriminate between recoiling species.  Figure \ref{ERNRBands} shows the behavior of
$\rm Log_{10}(S2/S1)$ as a function of energy , in unit of keV electron-equivalent (keVee), for populations of both recoil types, called the electronic and
nuclear recoil bands, or ER and NR bands, respectively.  The main purpose of such ER and NR calibrations is to identify
a region in $\rm Log_{10}(S2/S1)$ vs. S1 space, called the WIMP acceptance window, which should be nearly free
of ER events while covering a significant portion of the NR band.  The lower bound of this window along the
horizontal axis is determined by the detector's S1 threshold, and the corresponding upper bound is chosen to
maximize the potential integrated WIMP rate while minimizing the effects of anomalous background events which occur mostly
at higher energies (see section~\ref{sec:gxcut}).  The choice of bounds along the vertical axis are discussed
here.

\begin{figure}[h!]
\begin{center}
		\includegraphics[width=0.45\textwidth]{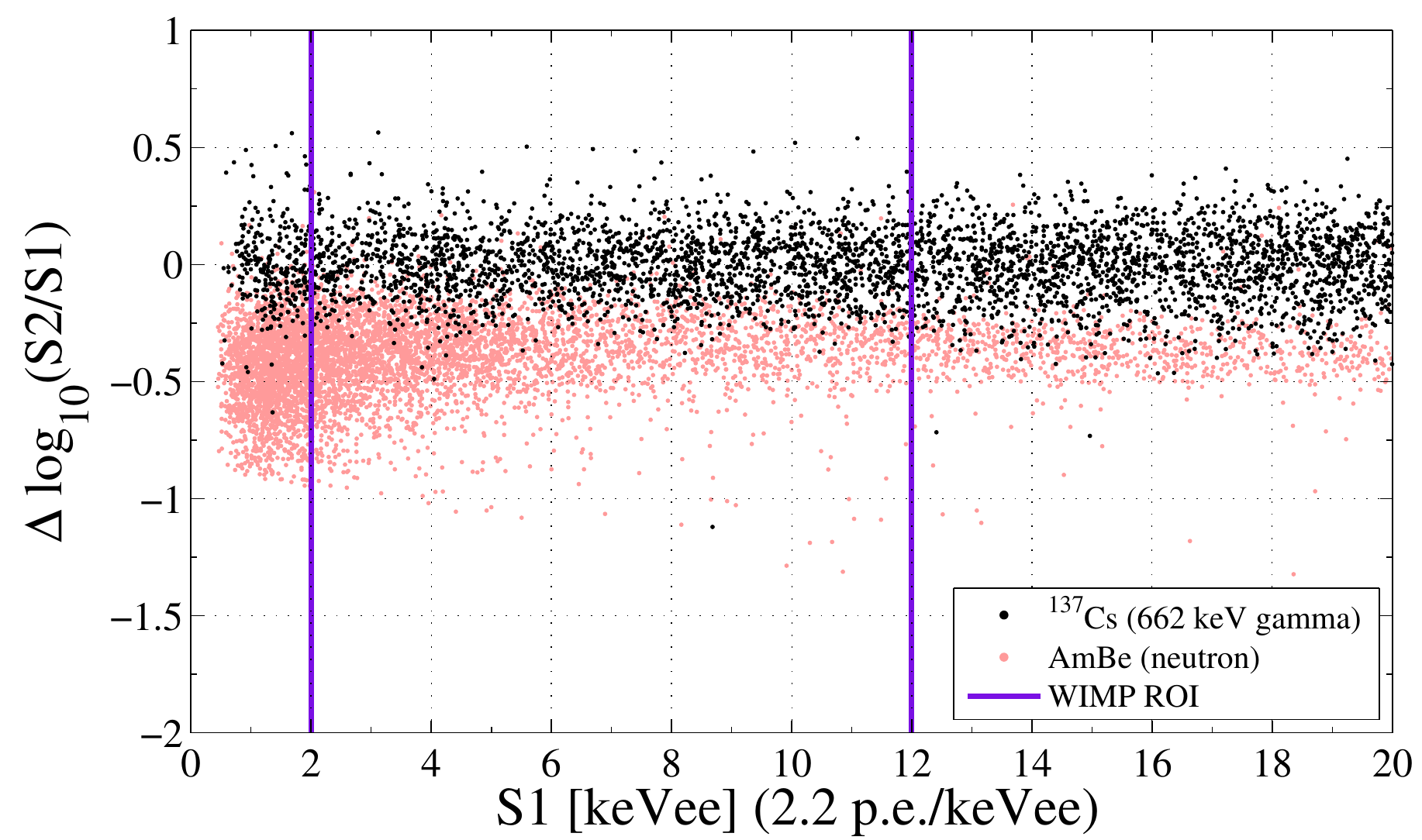} 
\caption{(Color online)The bands in figure \ref{ERNRBands} have been transformed to show the
distance in $\rm Log_{10}(S2/S1)$ space from the ER band center, giving the new discrimination
parameter, $\rm \Delta Log_{10}(S2/S1)$.  The vertical lines indicate the WIMP region of interest (ROI).}
\label{Delta_bands}
\end{center}
\end{figure}

Data were taken with the $^{137}$Cs source throughout all of November 2006, and intermittently from December 1
through February 14, 2007, accumulating a total of $\sim 2100$ events (after quality and fiducial cuts) in the WIMP
acceptance energy window, 4.4 p.e. $<$ S1 $<$ 26.4 p.e.  Fluctuations in $\rm Log_{10}(S2/S1)$ over most
of this range are dominated by recombination fluctuations, until the lowest energies where uncorrelated
statistical fluctuations take over.  The width of the electronic recoil band is very important for
gamma rejection because the two bands overlap.  Due mainly to the non-uniform S1 response at different
locations in the fiducial region, applying spatially-dependent corrections to S1 based on the $^{131m}$Xe
calibration (see section~\ref{sec:actxe}) improves the overall S1 resolution and thus helps to reduce the
variance of the bands.  Data with the AmBe neutron source were taken on December 1, 2006 for approximately
12 hours, accumulating a total of about 260,000 events.  The energy dependence of both bands makes it difficult to precisely measure the discrimination
power in the absence of extraordinarily large calibration datasets.  In an effort to remove this
energy-dependence, a one-dimensional transformation that ``flattens'' the ER band is applied to all data. The
ER band is broken up into 1 keVee-wide, vertical slices in S1.  For each, a Gauss fit is applied to the $\rm
Log_{10}(S2/S1)$ spectrum.  The mean of each fit now represents the center of the ER band in that particular
bin.  A high-order polynomial is fit to the Gauss means, which provides an analytic form for the ER band
centroid as a function of S1, and is subtracted from every data point in both bands.  This procedure flattens
the ER band (and to a large extent, the NR band as well), and introduces a new parameter, $\rm \Delta
Log_{10}(S2/S1)$, which represents the distance from the ER centroid in $\rm Log_{10}(S2/S1)$ space.  Figure
\ref{Delta_bands} shows the bands in $\rm \Delta Log_{10}(S2/S1)$ space.

\begin{figure}[h!]
\begin{center}
		\includegraphics[width=0.45\textwidth]{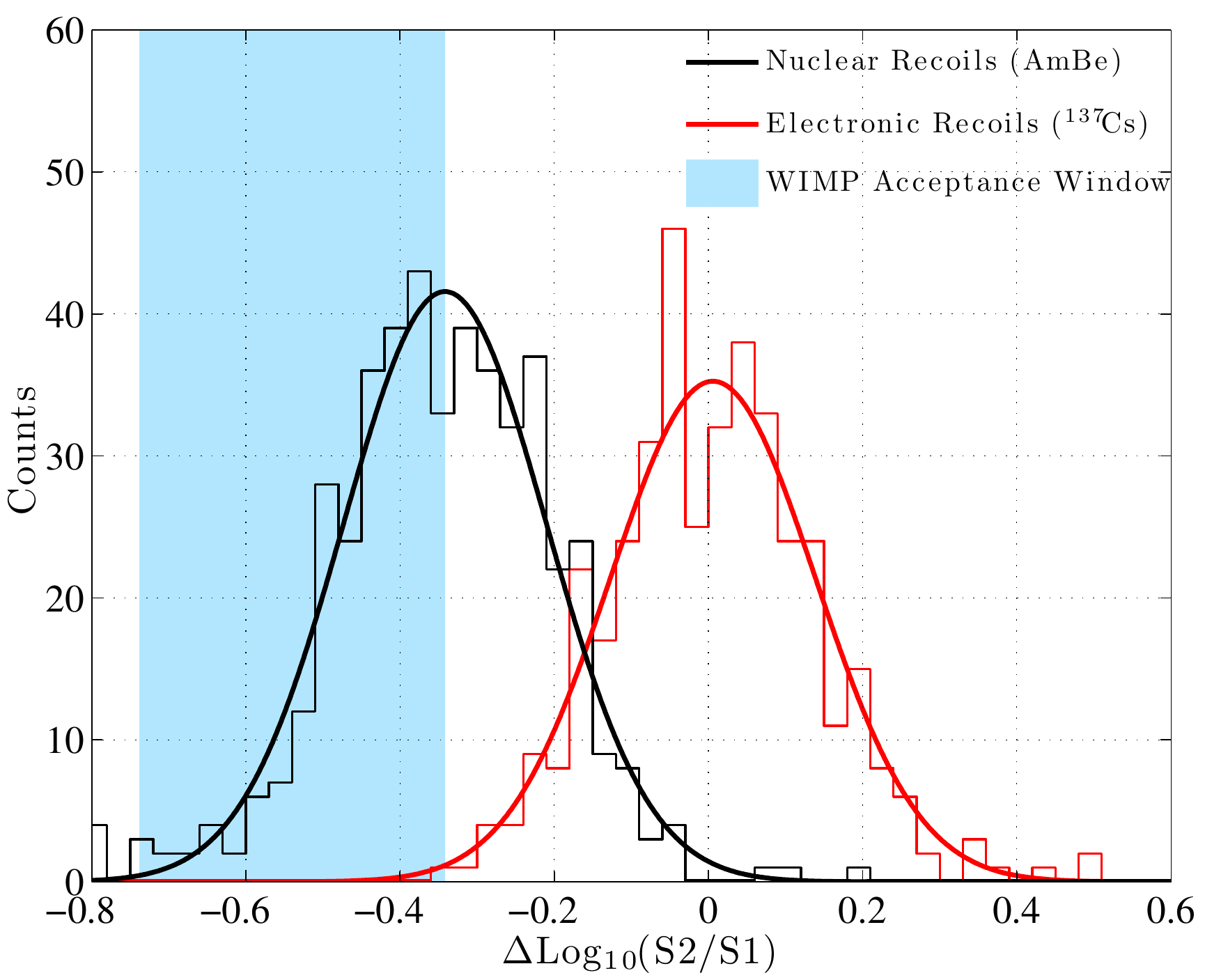} 
\caption{(Color online) Distributions of $\rm \Delta Log_{10}(S2/S1)$ for nuclear and electronic 
recoils in the range 13.4--17.2 keVr.  The WIMP acceptance window in this particular energy range
is defined by the blue, shaded rectangle which is between $\mu$ and $\mu-3\sigma$ of the NR band.}
\label{DLog_fit}
\end{center}
\end{figure}

Although the energy dependence of the ER band centroid has been removed, the NR band centroid and 
width still change with energy.  Again, the flattened bands are broken up into vertical S1 slices, 
only this time more coarse binning is used---seven bins in the WIMP energy region of interest (ROI)---in order to 
maximize the statistics in each slice, and a Gauss fit is applied to the $\rm \Delta Log_{10}(S2/S1)$
spectrum of both bands.  One such slice is shown in Figure \ref{DLog_fit}, for the range 13.4--17.2 keVr (1~keVr = 1.1~pe, according to~\cite{Angle:2007uj}) of nuclear recoil energy.  The WIMP acceptance 
window is defined to lie in the range $(\mu-3\sigma)<\rm{ \Delta Log_{10}(S2/S1)} < \mu$, where 
$\mu$ and $\sigma$ are the mean and sigma from the NR band Gauss fits, respectively.  The Gauss fits
were performed only to define the window bounds; the NR acceptance, $A_{nr}$, was calculated by 
counting the number of AmBe events that fall within this window, for each energy bin.

\begin{figure}[h!]
\begin{center}
		\includegraphics[width=0.45\textwidth]{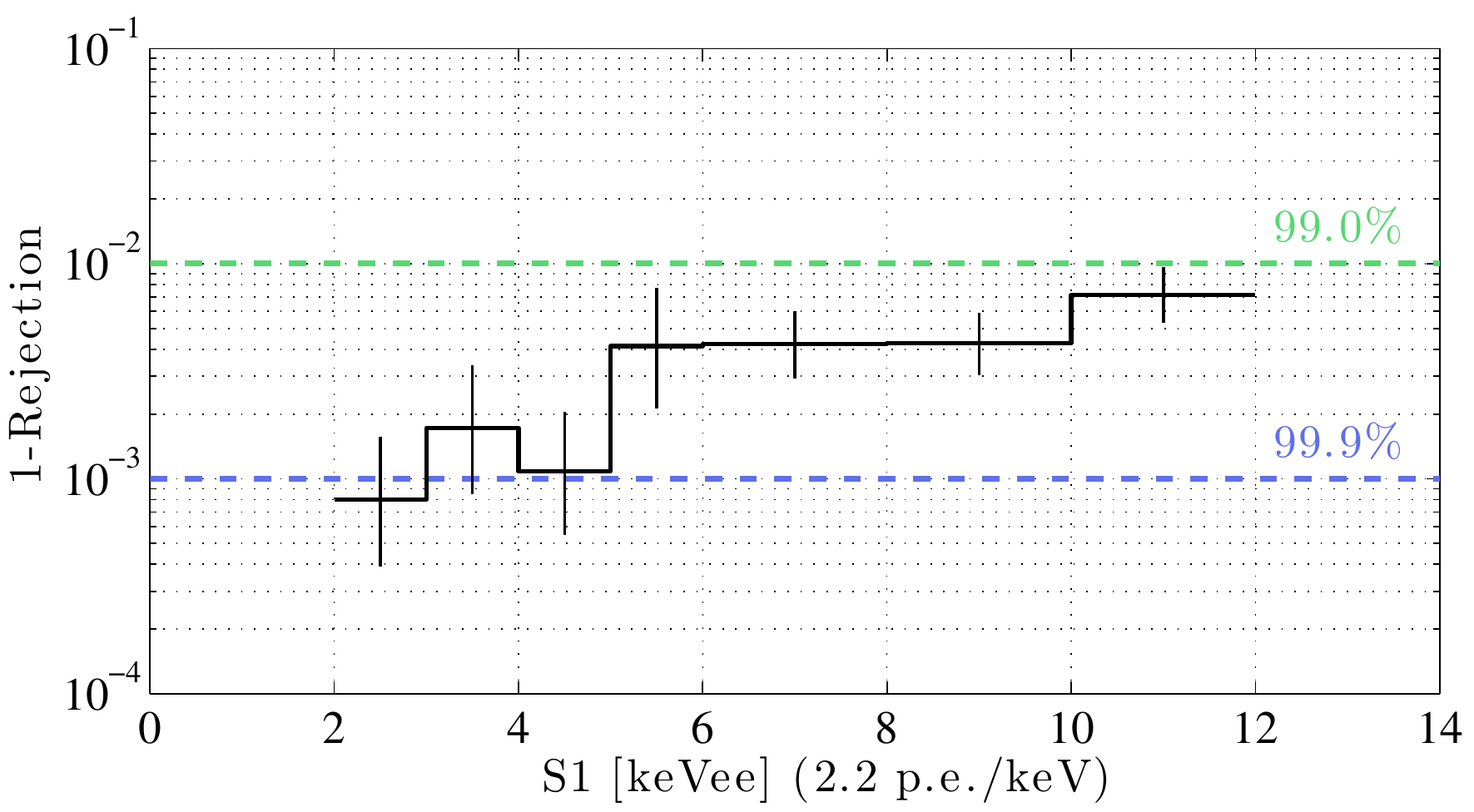} 
\caption{The ER rejection as a function of S1 for $\rm{\Delta Log_{10}(S2/S1)} < \mu$.
The rejection improves at lower energies, to better than 99.9\% in the range 2--3 keVee.}
\label{Rejection}
\end{center}
\end{figure}

The shape of the $\rm \Delta Log_{10}(S2/S1)$ fluctuations in the ER band are ``empirically'' Gaussian; 
that is, with the statistics available, they appear consistent with a Gaussian distribution.
As previously stated, the $\rm \Delta Log_{10}(S2/S1)$ spectrum is dominated by recombination 
fluctuations, which are poorly understood, and thus more cannot be said in the absence of a larger
calibration dataset.  We calculate the predicted ER rejection in the case that $\rm \Delta Log_{10}(S2/S1)$ 
fluctuations \emph{are} Gaussian.  That is, we use the Gauss fits to the $\rm \Delta Log_{10}(S2/S1)$ 
spectrum in each of the seven energy bins to determine the energy-dependent discrimination power.  
The results are shown in Table \ref{tab:table1} and Figure \ref{Rejection}.  Additionally, the 
expected number of background events in the WIMP acceptance window, $N_{leak}$, is shown,
calculated  on the basis of predicted rejection and background rate in the 58.6 live-days exposure (see Section~\ref{sec:bkgana}).

\begin{table}
\begin{center}
\caption{\label{tab:table1}{\footnotesize The nuclear recoil acceptance $A_{nr}$, the efficiency $\epsilon_{cut}$ of remaining nuclear recoil events after the $GammaX$ cuts (see section~\ref{sec:bkgana}), and the electron 
recoil rejection efficiency $R_{er}$  for each of the seven energy bins. The predicted number of 
leakage events, $N_{leak}$, is based on $R_{er}$ and the number of background events, $N_{evt}$, in
each energy bin, for the 58.6 live-days WIMP-search data. Errors are the statistical uncertainty 
from the Gaussian fits on the electron recoil $\rm \Delta Log_{10}(S2/S1)$ distribution.}}
\begin{tabular}{lllllrl}
$E_{nr}$ (keV) &   $A_{nr}$  & $\epsilon_{cut}$ &  1 - $R_{er}$ & $N_{evt}$ & $N_{leak}$ \\
   &   & &   ($10^{-3}$) & & \\
\hline
4.5 - 6.7 & 0.446 & 0.94  & 0.8$^{+0.7}_{-0.4}$ & 213 & 0.2$^{+0.2}_{-0.1}$\\
6.7 - 9.0 & 0.458 & 0.90  & 1.7$^{+1.6}_{-0.9}$  & 195 & 0.3$^{+0.3}_{-0.2}$ \\
9.0 - 11.2 & 0.457 & 0.89  & 1.1$^{+0.9}_{-0.5}$  & 183 & 0.2$^{+0.2}_{-0.1}$ \\
11.2 - 13.4 & 0.442 & 0.85  & 4.1$^{+3.6}_{-2.0}$ & 190 & 0.8$^{+0.7}_{-0.4}$\\
13.4 - 17.9 & 0.493 & 0.83  & 4.2$^{+1.8}_{-1.3}$ & 332 & 1.4$^{+0.6}_{-0.4}$\\
17.9 - 22.4 & 0.466 & 0.80  & 4.3$^{+1.7}_{-1.2}$  & 328 & 1.4$^{+0.5}_{-0.4}$ \\
22.4 - 26.9 & 0.446 & 0.77  & 7.2$^{+2.4}_{-1.9}$ &  374 & 2.7$^{+0.9}_{-0.7}$\\
\hline
Total &    & &  & 1815 & 7.0$^{+1.4}_{-1.0}$ \\
\end{tabular}
\end{center}
\end{table}

\begin{figure}[h!]
\begin{center}
		\includegraphics[width=0.45\textwidth]{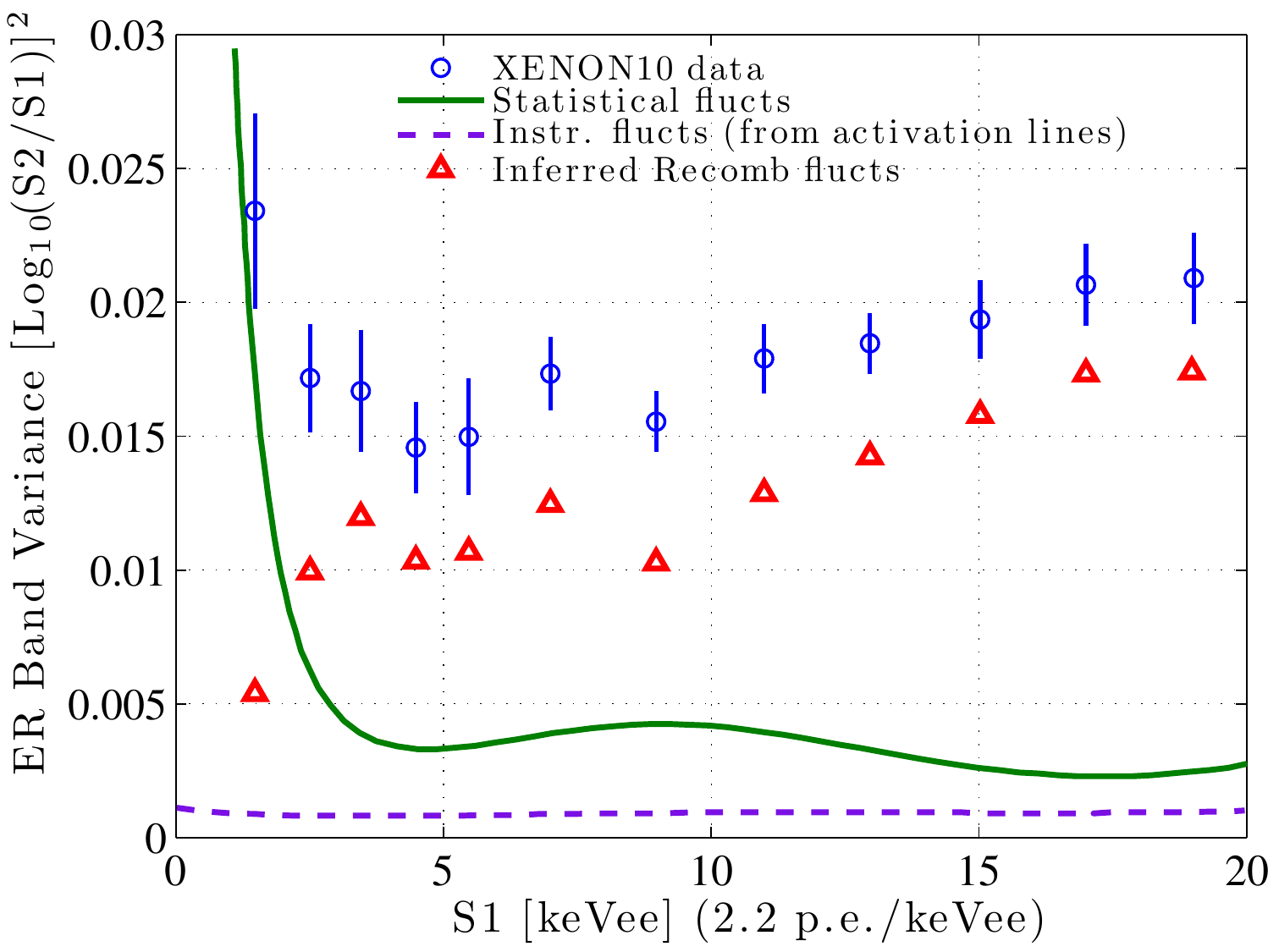} 
\caption{(Color online)  Decomposition of the ER band variance.  The recombination fluctuations are
inferred by comparing the expected statistical and instrumental fluctuations to the full observed
band variance.}
\label{variances}
\end{center}
\end{figure}

The observed trend of the ER rejection power with energy is unexpected.  If recombination 
fluctuations were flat at all energies, or if the band widths were dominated by binomial 
fluctuations from light collection, photoelectron emission, etc., one would expect the band 
widths to grow at low energies, and hence the ER rejection power would deteriorate.  The opposite 
is observed, and is due to two factors.  First, the ER and NR bands themselves diverge slightly at
lower energies, as first measured in~\cite{PRL06}.  Second, the width of the ER band does not grow at lower energies but instead 
remains relatively constant.  Figure \ref{variances} shows a decomposition of the ER band variance,
using assumed statistical and instrumental fluctuations.  It is quite evident that uncorrelated 
statistical and instrumental fluctuations cannot alone account for the observed degree of variance. 
Unfortunately, a model does not yet exist that successfully predicts recombination fluctuations in 
noble liquids, and hence more measurements are needed to reach a better understanding of the subject.

\section{The XENON10 Dark Matter Search Data}
\label{sec:bkgana}

\begin{figure}[htpb]
\centering
\includegraphics[height=6cm]{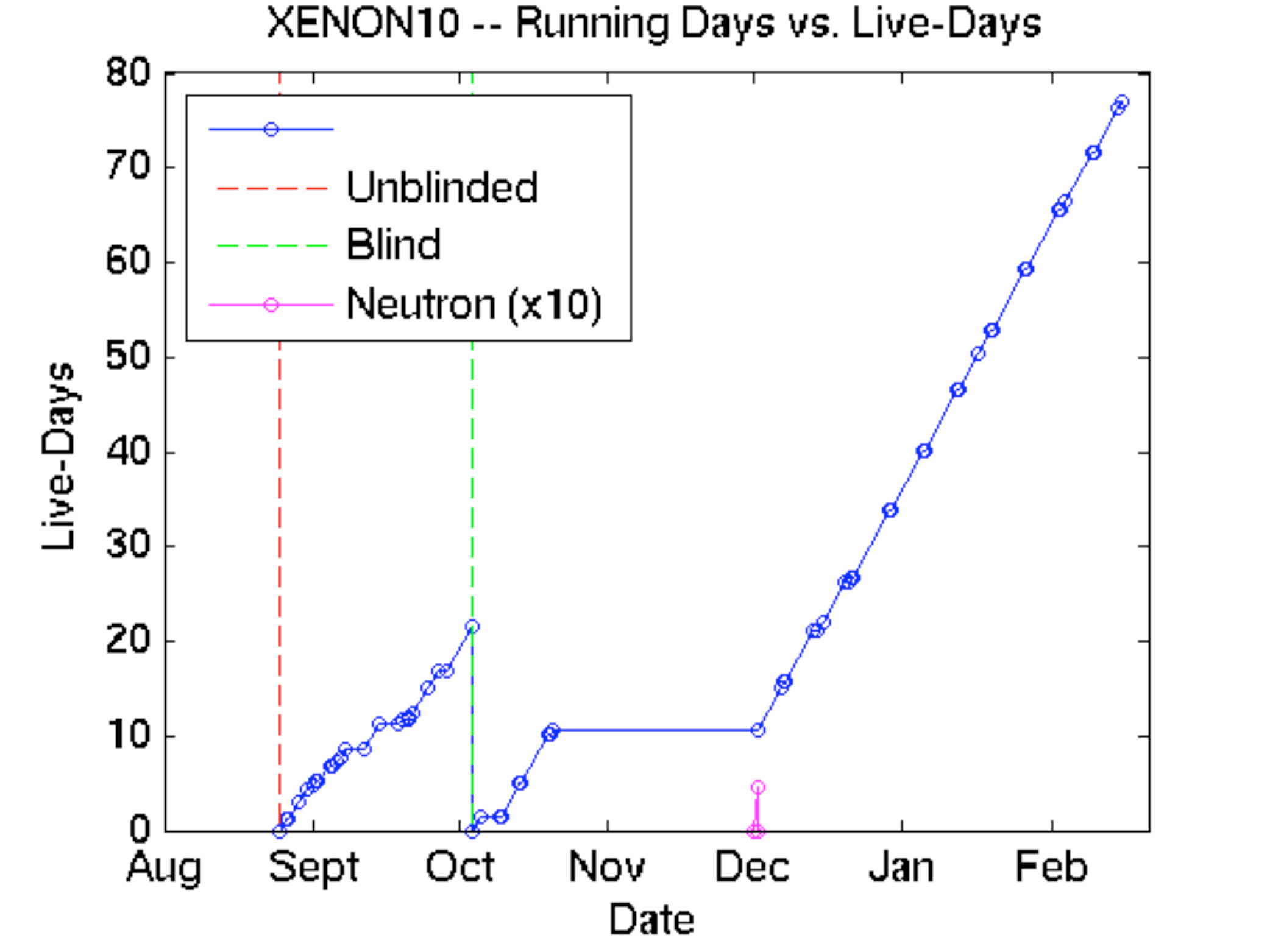} 
\caption{
	\label{xe10_livetime}(Color online) Running days vs live-days from XENON10 data-taking.}
\end{figure}

The XENON10 Dark Matter Search data covered the period between October 6th 2006
and February 14th 2007.  Figure~\ref{xe10_livetime} shows the accumulation of
measurement livetime, interrupted by calibration runs. The global trigger rate was
about 2.6~Hz with more than 92\% livetime. A total of 74.7 live-days of data were
collected. Of this, 16.3 were selected to be analyzed in order to define the event
selections and cuts. The remaining 58.4 live-days were embargoed in order to carry
out a blind analysis.  The XENON10 results from this blind analysis have been
published in \cite{Angle:2007uj,Angle:2008we}, and are referred to as blinded WIMP
search data in the following.


\subsection{Background reduction with fiducial volume selection}

\begin{figure}[htbp]
	\includegraphics[width =0.35\textwidth]{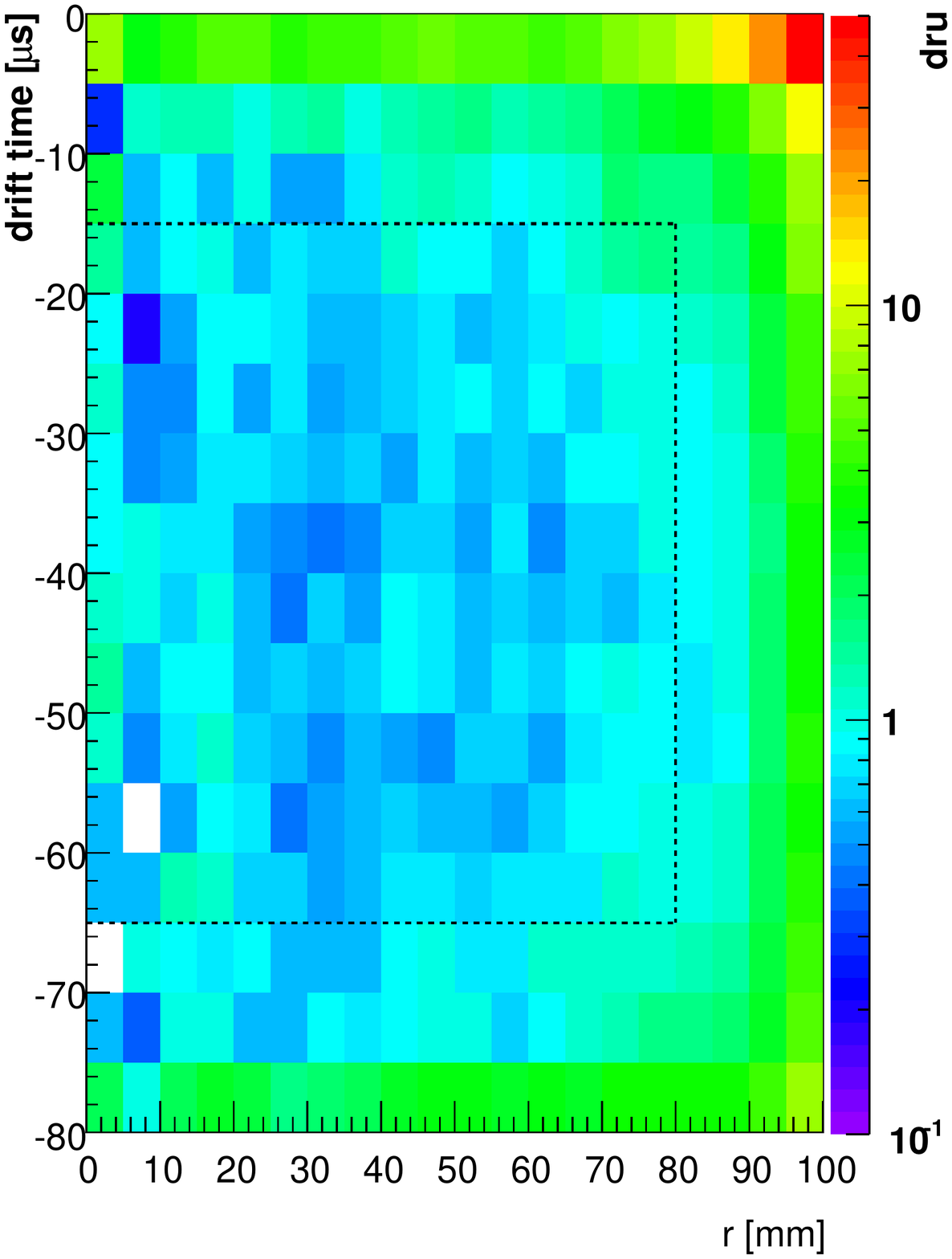}
	\caption{\label{rate_rz} (Color online) 2-30 keVee (based on S1 and 2.2 pe/keVee) event rate distribution in the fiducial volume as a function of the
		drift time and radius for the WIMP search run. The fiducial volume cut used in the final analysis is
		pre-defined based on the gamma calibration and unblinded WIMP search data and is indicated with the
		dashed lines.}
\end{figure}

\begin{figure}[htbp]
	\includegraphics[width =0.50\textwidth]{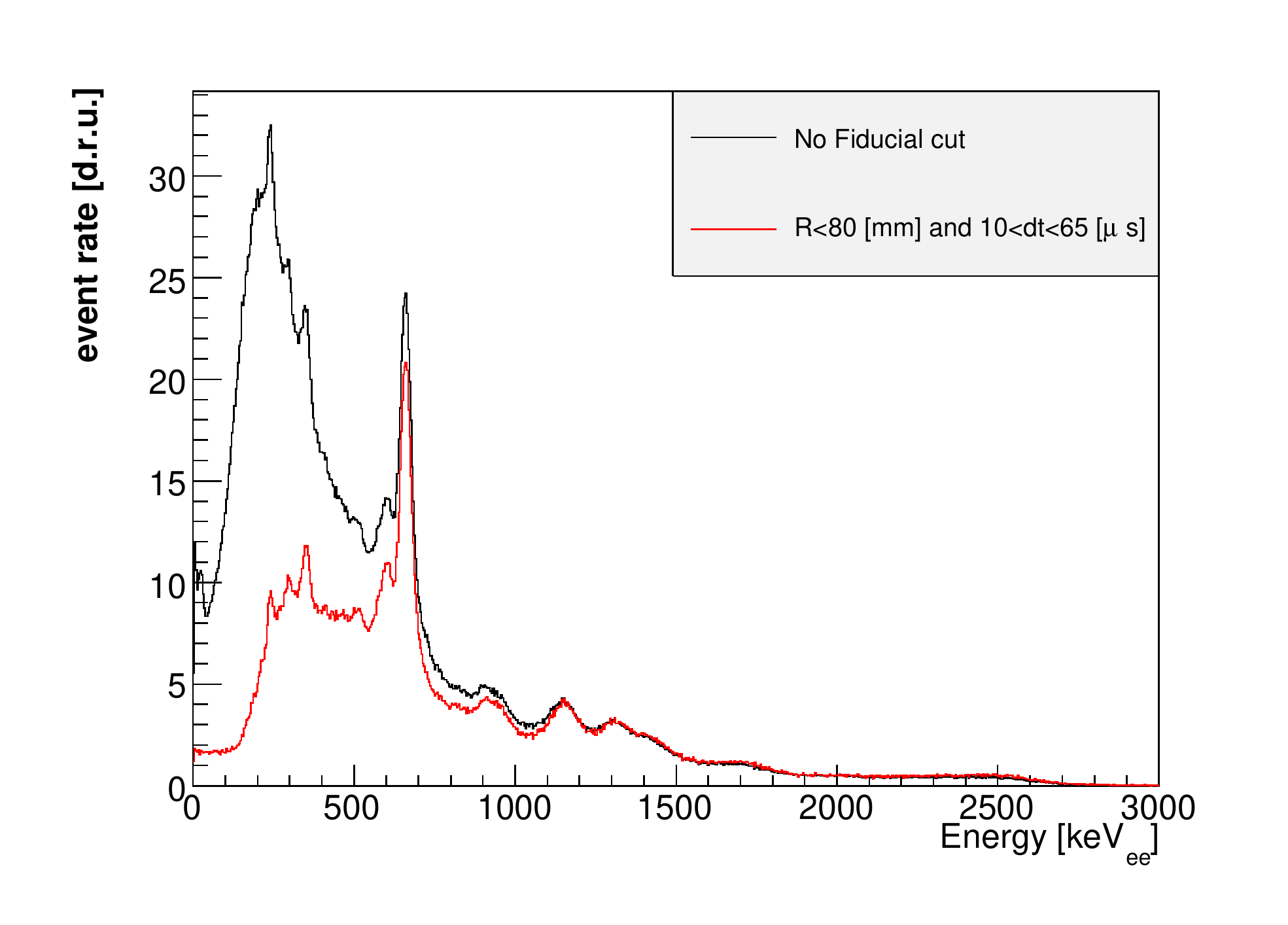}
	\caption{\label{fig:FVEffect}(Color online) Overall XENON10 background energy spectrum before (black) and after (red) the
		fiducial volume cut. Here we use the combined energy scale.}
\end{figure}

Background events in XENON10 are dominated by electron recoils from radioactivity
of detector materials. Due to the self-shielding of LXe, the background event rate
falls dramatically from the edges to the center of the target. Based on Monte
Carlo simulations and unblinded background data, a central fiducial volume was
chosen to minimize background while maintaining the largest possible target
mass. The fiducial volume was pre-defined for the blinded analysis with a radius
of 80~mm and a drift time between 15 and 65 $\mu$s, corresponding to a fiducial
mass of 5.4~kg.  Figure~\ref{rate_rz} shows the fiducial volume overlaid on the
background rate distribution of single recoil events between 2-30 keVee (2.2
pe/keVee).  Within the chosen fiducial volume, the overall background rate is
reduced by almost a factor of 10, as shown in Figure \ref{fig:FVEffect}.

\begin{figure}[htbp]
	\includegraphics[width =0.50\textwidth]{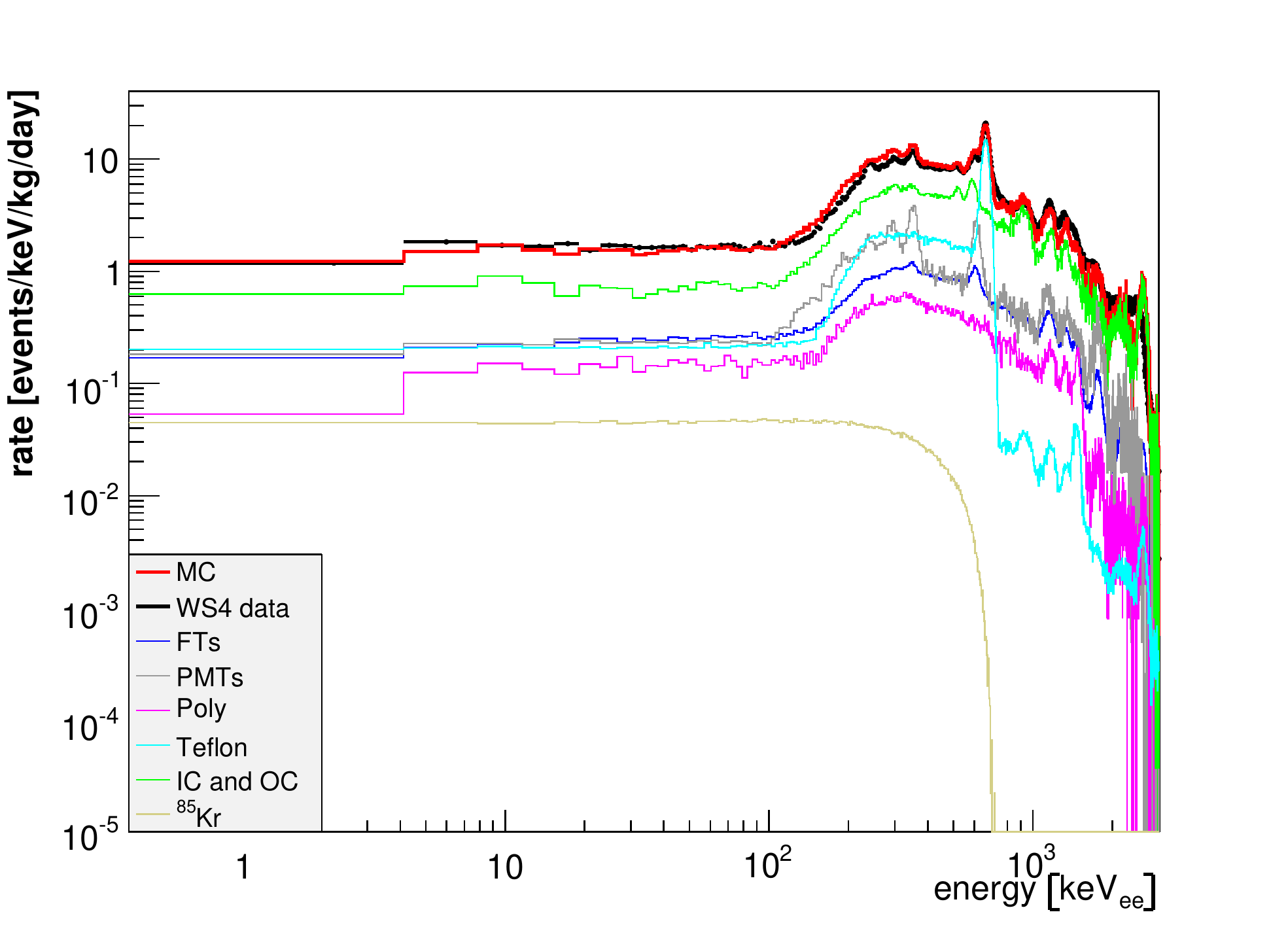}
	\caption{\label{fig:mc_data}(Color online) The measured XENON10 background event rate (black) compared to Monte Carlo
		simulations (red) in the 5.4-kg fiducial mass. The energy is based on combined energy scale. The contribution from different components from the main detector and shield are
		shown in different colors. FTs represent feedthroughs. Poly represents polyethylene in the shield. IC
		and OC represent internal cryostat and outer cryostat separately.}
\end{figure}

A comparison between the measured background rate and the projected background
from detailed Monte Carlo simulations, shown in Figure~\ref{fig:mc_data}, reveals
a dominant contribution from the stainless steel vessels, followed by
approximately equal contributions from PMTs, feedthroughs, and Teflon.  The
contribution from intrinsic contamination of $^{85}$Kr, Rn, $^{136}$Xe in the
liquid xenon, or nuclear recoils from ($\alpha$,n) reaction or fission of
$^{238}$U in the surrounding environment is subdominant or negligible in the
entire WIMP search run.  Details of the simulations and understanding of
background contributions are discussed in Appendix~\ref{sec:bkg}.

\subsection{Definition of anomalous events and cuts}
\label{sec:gxcut}

During the analysis of both gamma calibration data and unblinded background data
from XENON10, a class of anomalous single scatter events with smaller $S2/S1$
ratio than that of normal single scatter electron recoil events were
observed. These events produce a non-Gaussian tail extending into the acceptance
region for nuclear recoils and to even lower $S2/S1$ ratios. Most of these
anomalous events are likely due to the coincidence of two gamma-ray interactions,
where a single interaction occurs in the sensitive volume and at least one other
interaction occurs in a region of the detector that is light-sensitive but not
charge-sensitive. Such events, dubbed \textit{Gamma-X}, may comprise two Compton
scatters of a high-energy gamma-ray, or, e.g., coincident gamma-rays from a
cascade, such as $^{60}$Co from the stainless steel cryostat. There are a few
regions of the XENON10 detector that are light-sensitive but not charge-sensitive,
such as the region between cathode and bottom PMTs or the region around and below
the bottom PMT array, as was shown in Fig.~\ref{fig:S1_LCE_sim}.

\begin{figure}[htbp!]
	\begin{center}
		\includegraphics[height=6cm]{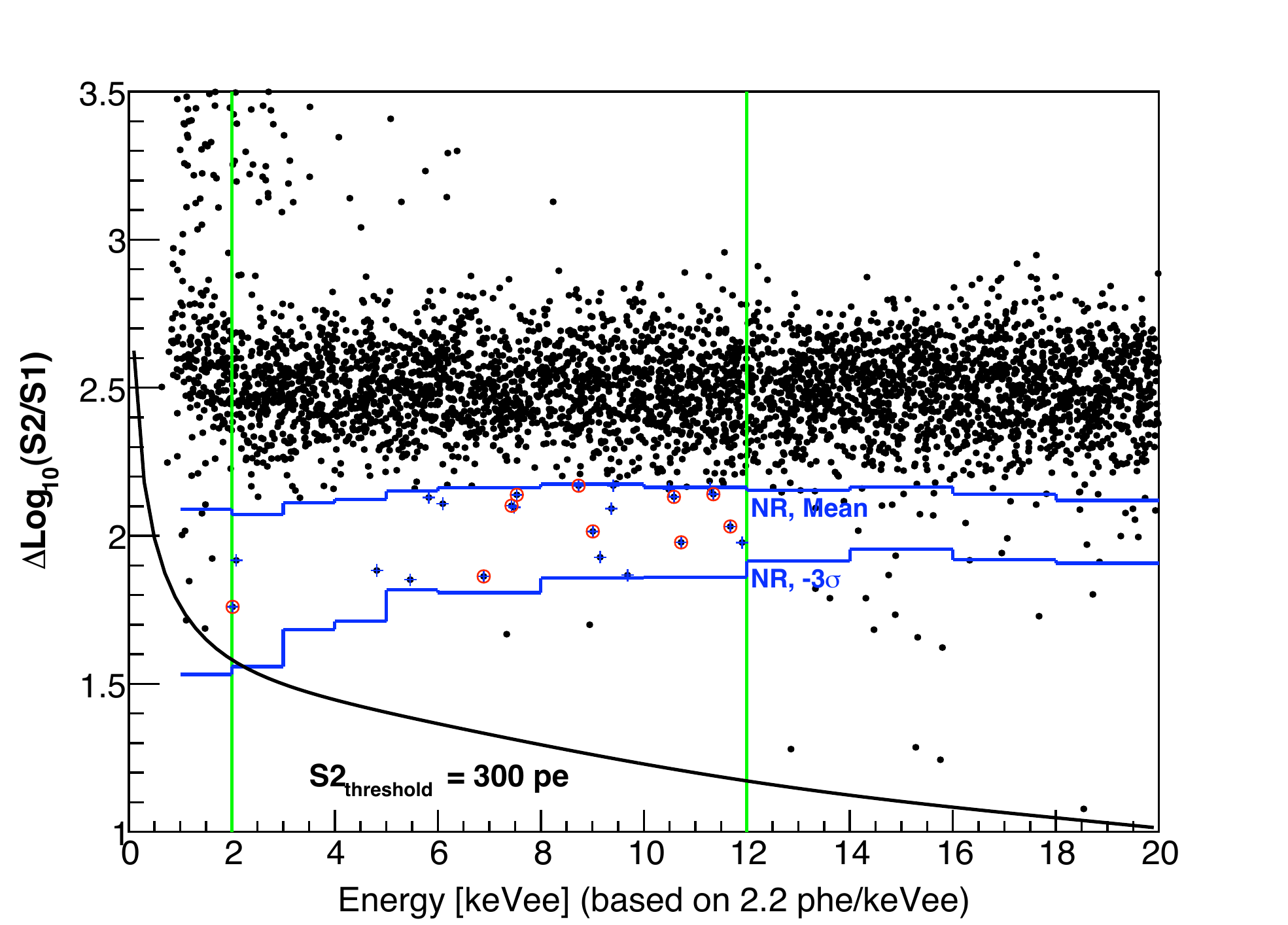}
		\caption{(Color online) Event distribution in the $\Delta$Log$_{10}$(S2/S1) vs. Energy space for the 58.4-day WIMP search data.
			The crosses indicate the events that ``leak" into the nuclear recoil acceptance region, which is
			within the mean and $3\sigma$ of the nuclear recoil band. Red circles indicate the remaining
			events after all software cuts, discussed in the text.}
		\label{fig:WS4}
	\end{center}
\end{figure}

\begin{figure}[htbp]
	\begin{center}
		\includegraphics[height=6cm]{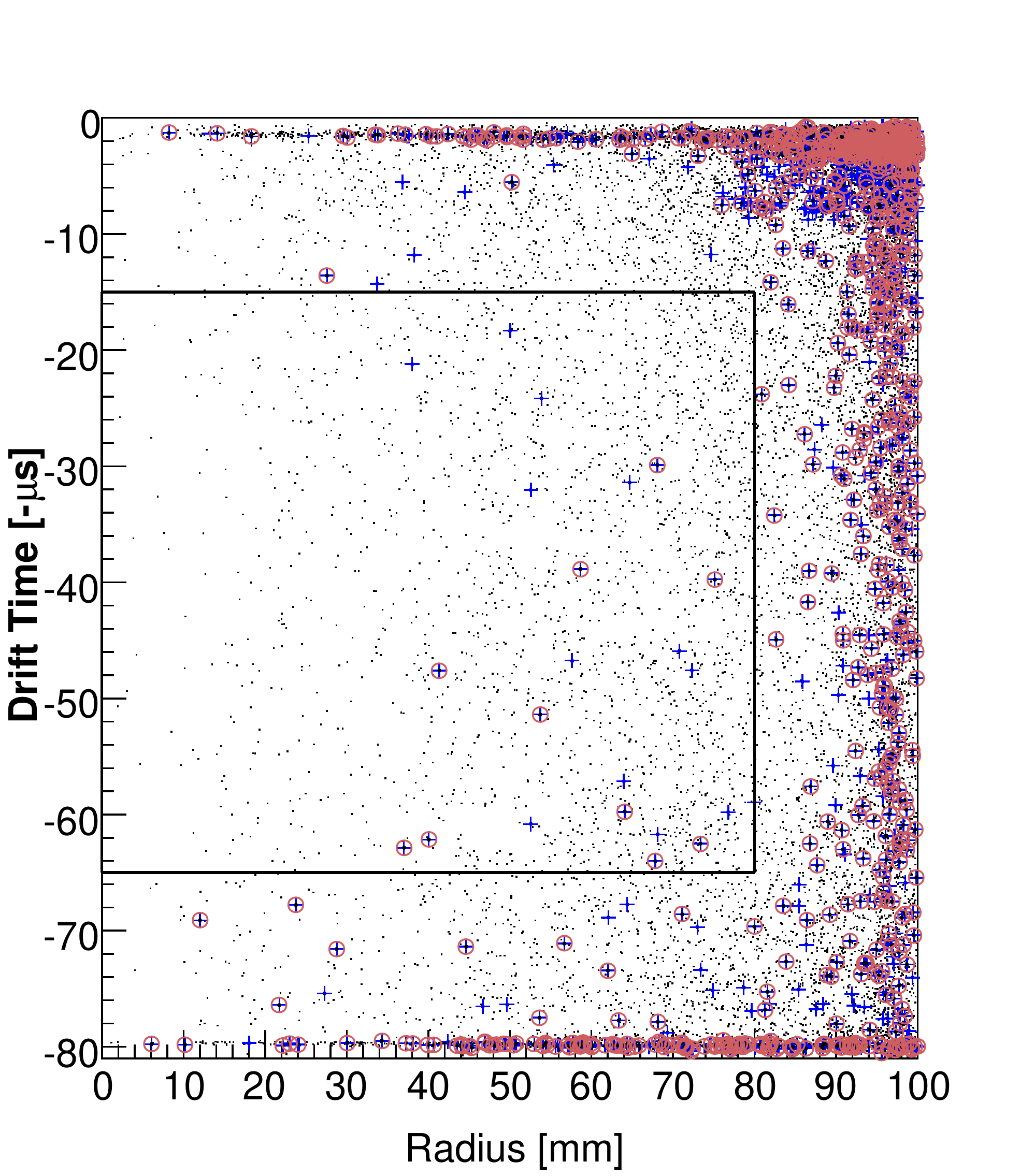}
		\caption{(Color online) Event distribution in the 58.4-day WIMP search data. The crosses indicate the events that
			``leak" into the nuclear recoil acceptance region, as shown in Fig~\ref{fig:WS4} in the energy
			window (2-12 keVee) of interest. Red circles indicate the remaining events after all software
			cuts, discussed in the text.}
		\label{fig:WS4FD}
	\end{center}
\end{figure}

For illustration, Fig~\ref{fig:WS4} shows the $\Delta$Log$_{10}$(S2/S1) vs.\ energy event
distribution of the blinded WIMP search data. Prior to a final set of cuts,
designed to remove these anomalous events, a total of 22 events is observed in the
WIMP acceptance region. Only some of these events are expected from the statistical leakage of the
electron recoil band. The spatial distribution of all events in the energy region of interest is shown
in Fig~\ref{fig:WS4FD}. 

\begin{figure}[htpb]
	\centering
	\includegraphics[height=6cm]{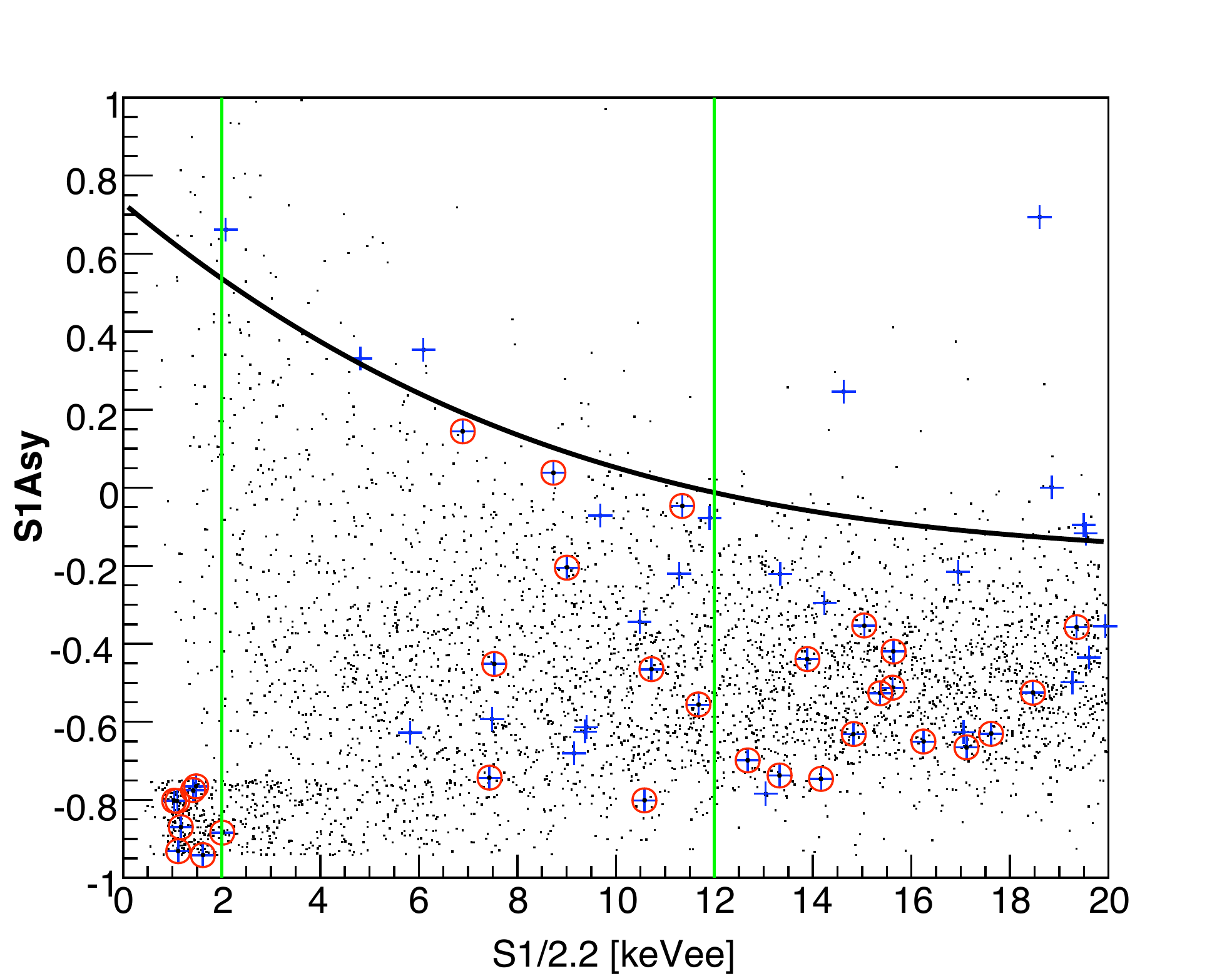} 
	\caption{\label{mk:asys1}(Color online) A cut based on the asymmetry of the S1 signal, $p_{asy}$, as discussed in the
		text.  The crosses indicate the events that ``leak" into the nuclear recoil acceptance region, as
		shown in Fig~\ref{fig:WS4}. The two vertical lines are the energy window of WIMP search. Red circles
		indicate the remaining events after all software cuts. The solid black curve is pre-defined to keep
		98\% electron recoil events. It removes 3 out of 22 ``leakage events" in the energy window of interest
		(2-12 keVee).}
\end{figure}

Two cuts, based on the S1 hit pattern, were identified to remove these anomalous
events. One cut is based on the asymmetry of S1 between the top and bottom array
of PMTs, i.e. $p_{asy} = (S1_{t}-S1_{b})/(S1_{t}+S1_{b})$.  $p_{asy}$ is larger
for anomalous events with some scatters in the region close to the top PMTs, but
not in the sensitive LXe target. This cut also removes most of the events with
misidentified S1 in software, e.g. a small S2 signal, or a micro-discharge pulse
in the gas. Figure~\ref{mk:asys1} shows the distribution of this parameter versus
energy.

\begin{figure}[htpb]
	\centering
	\includegraphics[height=5.5cm]{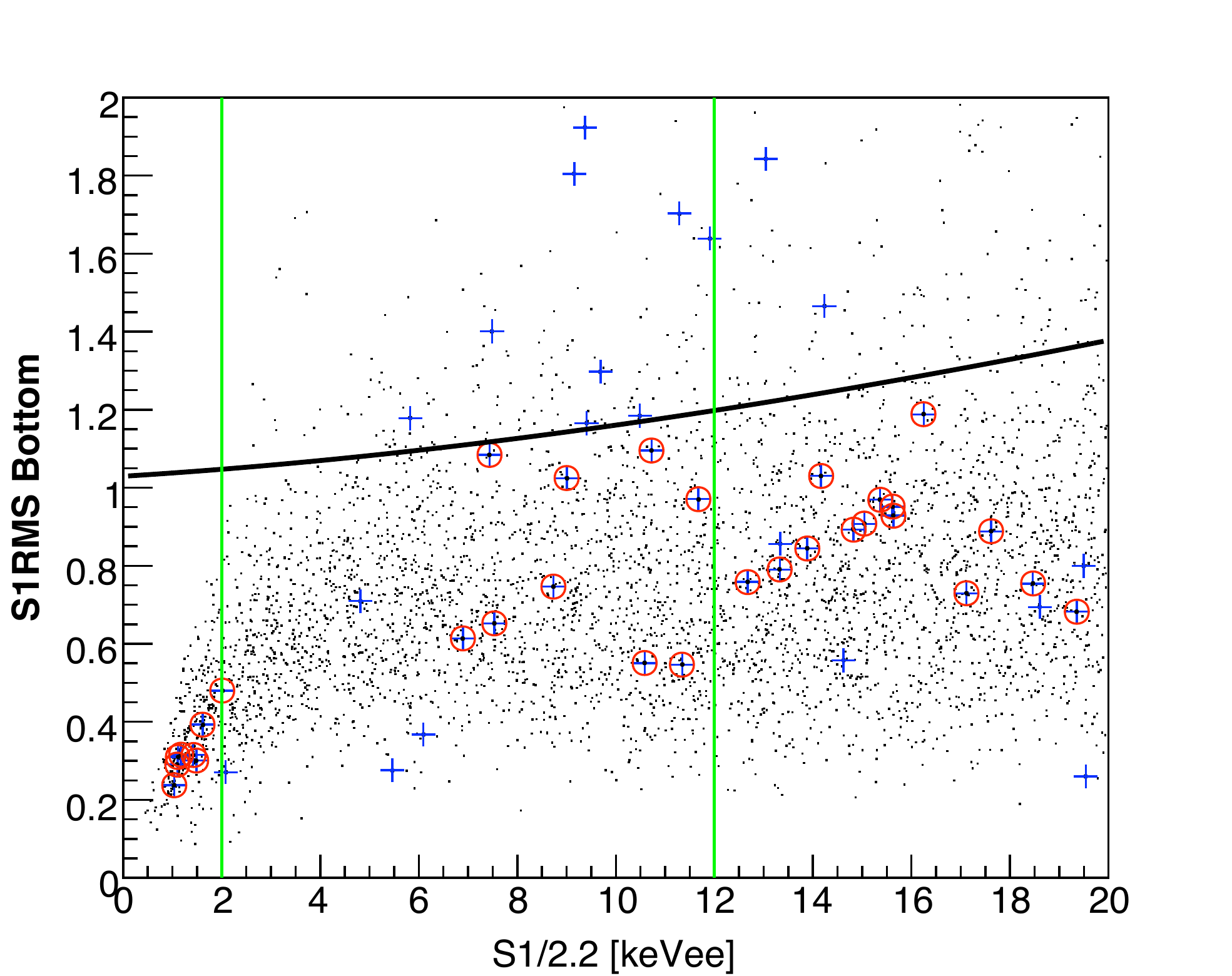} 
	\caption{\label{mk:rms}(Color online) Cuts based on $p_{RMSb}$ for the bottom PMTs, as defined in the text. The crosses
		indicate the events that ``leak" into the nuclear recoil acceptance region, as shown in
		Fig~\ref{fig:WS4}.  The two vertical lines are the energy window of WIMP search. Red circles indicate
		the remaining events after all software cuts.  The solid black curve is pre-defined, which varies from
		99\% at 2 keVee and 90\% at 20 keVee cut efficiency and fitted by the third-order polynomial function.
		It removes 9 out of 22 ``leakage events" in the energy window of interest (2-12 keVee).}
\end{figure}

A second set of cuts is based on the S1 signal distribution on the top and bottom PMT arrays. We define a parameter
$p_{RMS} = \sqrt{\frac{1}{n}\sum(S1_i -\overline{S1})^2}$ from the Root-Mean-Square (RMS) value
of the 5 (for top) or 10 (for bottom) PMTs receiving the largest hits. 
Anomalous events tend to give non-uniform S1-signal distributions, e.g., if one of
the interactions occurs near the bottom PMT array. Figure~\ref{mk:rms} shows the
effect of this cut for the bottom PMTs.

The above software cuts also remove a small fraction of genuine nuclear recoil events. By comparing the AmBe
nuclear recoil data before and after the software cuts, the estimated cut efficiency $\epsilon_{cut}$ was calculated and shown in
Table~\ref{tab:table1}. 


\appendix
\section{The XENON10 Backgrounds}
\label{sec:bkg}

The particles potentially contributing to the background of the XENON10 detector
are alphas, betas, gammas, and neutrons. Neutrons scattering with Xe nuclei yield
nuclear recoils, possibly mimicking a WIMP signal. Unless they are detected as
multiple interactions in the sensitive volume, neutrons constitute an unrejectable
background in the nuclear recoil band.  Gamma and beta rays mostly populate the
electron recoil band and constitute the most prominent background of the
experiment. Alphas can contribute to background in the WIMP search region as they
give rise to neutrons via $\alpha$-n reactions in the detector materials.

\subsection{Gamma and Neutron Backgrounds from Data and MC Simulations}
\label{sec:bkgmc}

Gamma and neutron backgrounds from the environment were studied to optimize the
shield, and have been discussed in section~\ref{sec:shield}. After shielding, the
dominant backgrounds in XENON10 result from radioactive
contaminants in the detector and shield materials. These internal backgrounds were
studied with material screening, using gamma-ray spectroscopy
in a low count rate facility, with Monte Carlo simulations, and with the analysis of
delayed coincidences in decay chains of radio-nuclides inside LXe.
Simulations were based on the GEANT4
simulation toolkit \cite{geant4}, where the complete geometry of the detector and
surrounding material was included.

%

As detailed in Table~\ref{tab:radio01}, the shielding material and many detector
construction components were screened with HPGe gamma spectrometers at the LNGS
and SOLO screening facility \cite{SOLO}. The radioactive contamination of the
screened material are listed in Table \ref{tab:radio01} (for the shield) and in
the first row of table \ref{tab:radio02} (for the PMTs + bases). However, since
XENON10 was a prototype experiment, the radioactivity of some detector components
was not known.  We used the minimum $\chi^2$ method to extract the individual
activities, by comparing the measured energy spectrum with the sum of the
energy spectra for each radioactive decay chain within each material:

\begin{equation}
\chi^2=\sum_{i=n}^N\frac{d_i-\sum_{j=1}^m F_{ij}(u_jU_j,th_jTh_j,k_jK_j,co_jCo_j,cs_jCs_j)}{d_i}
\label{eq:chi2}
\end{equation}

\noindent
where $n$ and $N$ are the first and last bin of the energy spectra, defining the
fit range. Here we use a bin width of 3.75 keV, $n=50$, and $N=450$. $d_i$ is the
rate in the $i^{th}$ bin of the measured energy spectrum, and $F_{ij}$ is the
corresponding expected rate from simulated backgrounds in material $j$; $m=41$ is
the total number of simulated materials; $u_j$, $th_j$, $k_j$, $co_j$ and $cs_j$
are the scaling parameters related to nominal activities of material $j$ in
$^{238}$U ($U_j$), $^{232}$Th ($Th_j$), $^{40}$K ($K_j$), $^{60}$Co ($Co_j$), and
$^{137}$Cs ($Cs_j$), respectively. The minimization is done by the MINUIT routines
in ROOT~\cite{rootwebsite}.

\begin{figure}[htbp] \includegraphics[width =0.50\textwidth]{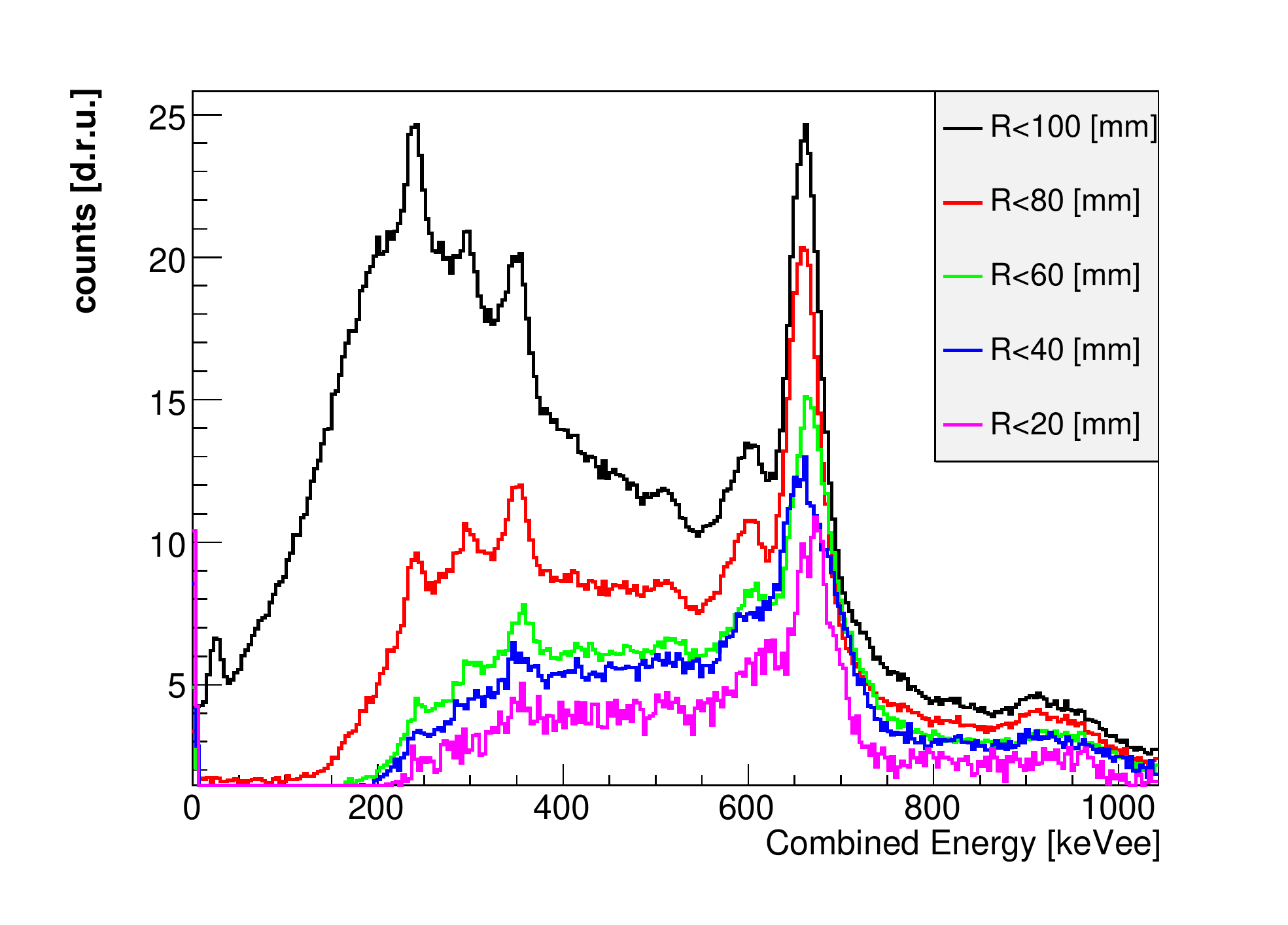}
	\caption{\label{fig:cs}(Color online) Energy spectrum of background events in 
	XENON10 with different radial cuts, with a focus on the $^{137}$Cs peak.}
\end{figure}

The results of this fit procedure are summarized in table~\ref{tab:radio02}.
The main component to the background energy spectrum of XENON10 originates from
a  $^{137}$Cs peak at 662 keV. Fig.~\ref{fig:cs} shows the
662 keV peak decreasing at smaller radii, a clear indication that it is not
diluted in the LXe. A study of the spacial distribution of events
in this peak shows that the $^{137}$Cs source is all around the detector edges. Thus a surface contamination of the anthropogenic $^{137}$Cs has to be assumed, either on the inner cryostat can or on the Teflon.  The best
fit results localize the  $^{137}$Cs on the Teflon, with a total activity of 5.73$\pm$0.57 Bq.

\begin{table*}
\caption{\label{tab:radio02} Radioactive contaminations of material used in
the construction of the XENON10 detector, as obtained from screening (PMTs,
bases and polyethylene shield) and by comparing the low-background data to MC
simulations.} 
\begin{tabular}{lccccccc}
Sample & Mass or & $^{238}$U & $^{232}$Th & $^{40}$K & $^{60}$Co & $^{137}$Cs & $^{85}$Kr \\
		& number (M)	& mBq/M	&	mBq/M	&	mBq/M	 & 	mBq/M	& mBq/M	& mBq/M	\\
\hline 
\hline 
PMTs + bases & 89 pieces & $0.32$ & $0.23$ & $8.6$ & $1.7$ & $1.0$ & \\
Inner cryostat & 37.4 kg & $130$ & $470$ & $310$ & $65$ & $8.1$ & \\
Outer cryostat & 144.7 kg & $80$ & $120$ & $39$ & $24$ & $8.0$ & \\
Polyethylene & 1540 kg & $1.2$ & $0.3$ & $11.6$ & $0.01$ & $1.0$ & \\
Teflon (TPC) & 6.26 kg & $1.1$ & $4.8$ & $48$ & $0.01$ & $910$ & \\
3 Feedthroughs & 1.758 kg & $130$ & $28$ & $270$ & $30$ & $13$ & \\
Xenon & 14.5 kg & & & & & & 1.0 \\
\hline 
\hline 
\end{tabular}
\end{table*}

The main source of neutrons in XENON10 are ($\alpha$,n) and spontaneous fission
reactions from $^{238}$U and $^{232}$Th in the detector and shield material.
Another source of neutrons are spallation and photo-nuclear reactions of cosmic
ray muons in the rock and shield. As explained in Section~\ref{sec:shield}, the
neutrons from outside the shield are stopped or thermalized by the
polyethylene. 

Using the  $^{238}$U and $^{232}$Th activities as given in Table
\ref{tab:radio02}, the energy spectra and number of expected neutrons from
($\alpha$,n) and spontaneous fission reactions  in each material has been
calculated with the modified SOURCES4A code \cite{Carson}.  These are given in
Table \ref{tab:neutrons} (second column).  



\begin{figure}[htbp]
	\includegraphics[width =0.50\textwidth]{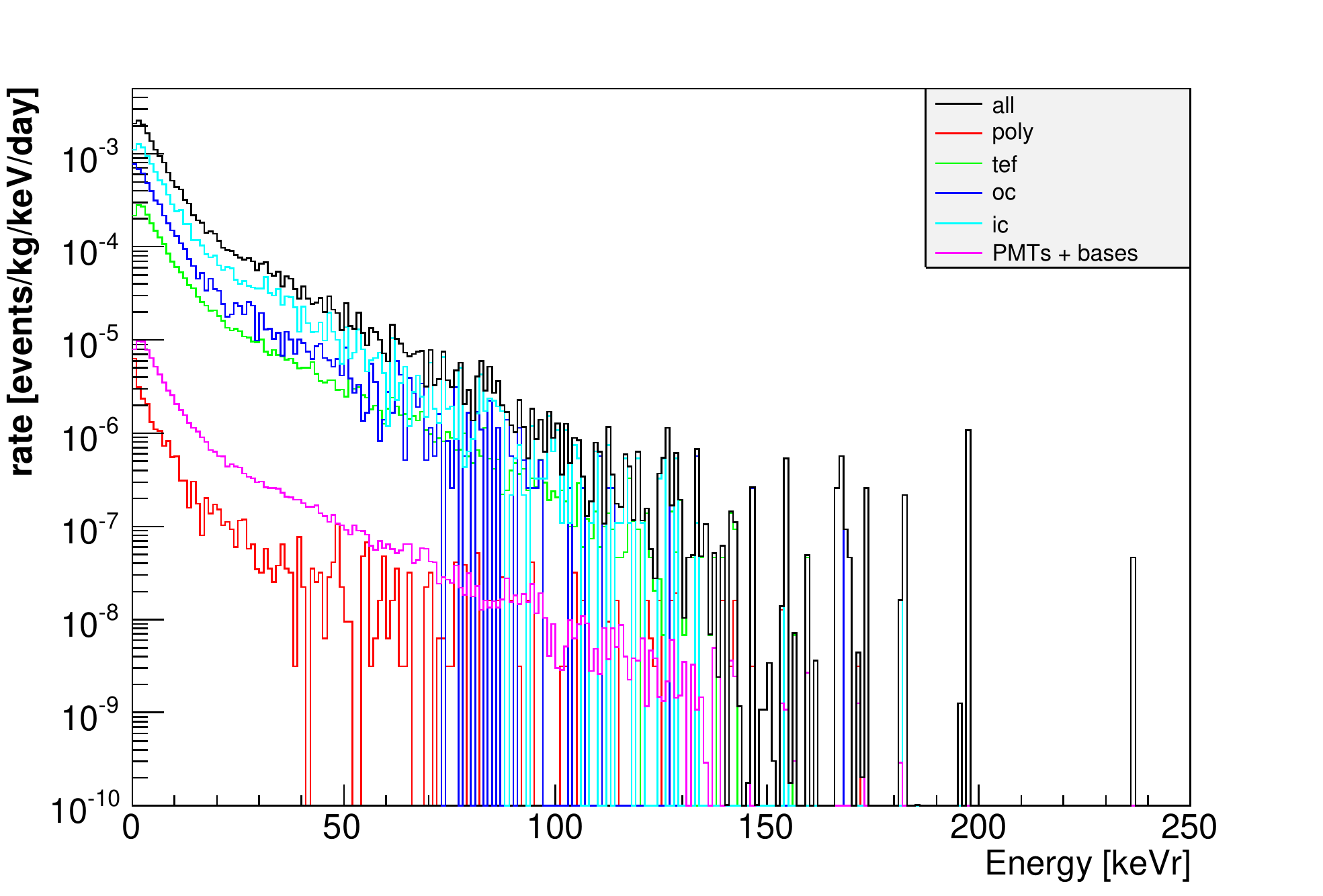}
	\caption{\label{fig:pmt_n_spec}(Color online) Monte Carlo simulation of total nuclear recoil spectrum expected from the
	neutrons produced in all the material considered (black), along
	with the individual contributions from each material.}
\end{figure}

The neutrons are then propagated
into the sensitive region using the GEANT4 code and detector geometry. In figure \ref{fig:pmt_n_spec}
the predicted single scattering nuclear recoil energy spectrum from the
neutrons from all the material considered in the simulation is shown, together
with the individual contributions from all the material. The
number of predicted single nuclear recoils in the WIMP search region is shown
in Table \ref{tab:neutrons} (third column), along with the total number of
single nuclear recoils (fourth column).

The total rate of single nuclear recoils in the energy region from 2 to 12 
keV$_{\mbox{ee}}$ is expected to be 1.67$\times10^{-3}$ event/kg/day, i.e.\ for
an exposure of 5.4\,kg$\times$58.6\,days, the
total number of nuclear recoils expected from the contamination of the XENON10
detector material is less than 0.53 events \cite{Angle:2007uj}.

\begin{table*}
\caption{\label{tab:neutrons}Neutron fluxes (second column) calculated with the modified
SOURCES4A code, using the activities in $^{238}$U and $^{232}$Th given in
table \ref{tab:radio02}. Column 3 lists the number of single nuclear recoils
expected in the WIMP search region of interest, inside the fiducial volume, for
each material. Column 4 lists the total number of
single nuclear recoils expected in the whole volume.}
\begin{tabular}{lccc}
Material & Flux 		&	nuclear recoil  &	total nuclear recoil  \\
		 & [neutron/sec]&	evt/kg/day		&	evt/kg/day		\\
\hline 
\hline 
Teflon	& 9.60$\times10^{-6}$	&	5.34$\times10^{-4}$ &	3.69$\times10^{-4}$ \\
Poly	& 5.22$\times10^{-6}$	&	2.38$\times10^{-6}$ &	7.32$\times10^{-7}$ \\
OC		& 5.64$\times10^{-5}$	&	2.82$\times10^{-4}$ &	1.88$\times10^{-4}$ \\
IC		& 4.38$\times10^{-5}$	&	5.57$\times10^{-4}$ &	4,02$\times10^{-4}$ \\
PMT		& 2,41$\times10^{-8}$	&	1.13$\times10^{-4}$ &	5.16$\times10^{-4}$ \\
Base	& 3.81$\times10^{-6}$	&	1.80$\times10^{-4}$ &	6.39$\times10^{-5}$ \\
\hline 
\hline 
\end{tabular}
\end{table*}

\subsection{Intrinsic backgrounds in LXe}
\label{sec:bkgkr85}

Delayed coincidence analyses to determine the Radon, Thoron, U/Th and $^{85}$Kr
concentration in the LXe have been carried out by looking at the specific decay
signatures. For Radon, the consecutive beta-alpha decays occurring in the decay
of $^{214}$Bi (shown in Eq. \ref{eq:Rn}) are required to occur within the same
waveform, in the same time order as in the decay scheme and with the proper
energies ($E_{\beta}<3000~$keV and $E_{\alpha}>6500$~keV). With these cuts, the
residual accidental coincidences are negligible and the events surviving the
cuts are taken as real $^{214}$Bi decays. The resulting Rn concentration in LXe,
considering the detection efficiency and the efficiency of above cuts, is
 (59$\pm$2)$\mu$Bq/kg.  For $^{220}$Rn arising from the
$^{232}$Th chain, the time window for the events
is much shorter, namely 3 $\mu$s versus the maximum time acquisition window, i.e. 80
$\mu$, and the energy of the two particles has to be $E_{\beta}<2400$~keV and
$E_{\gamma}>$7000~keV.  The resulting concentration of $^{220}$Rn in LXe is
 (4.7$\pm$0.2)$\mu$Bq/kg. For both Radon and Thoron the events are
distributed mainly in the outermost LXe region, suggesting that both
radioisotopes emanate mainly from the surface of the teflon vessel or inner stainless steel chamber.



\begin{figure}[htbp]
	\includegraphics[width =0.45\textwidth]{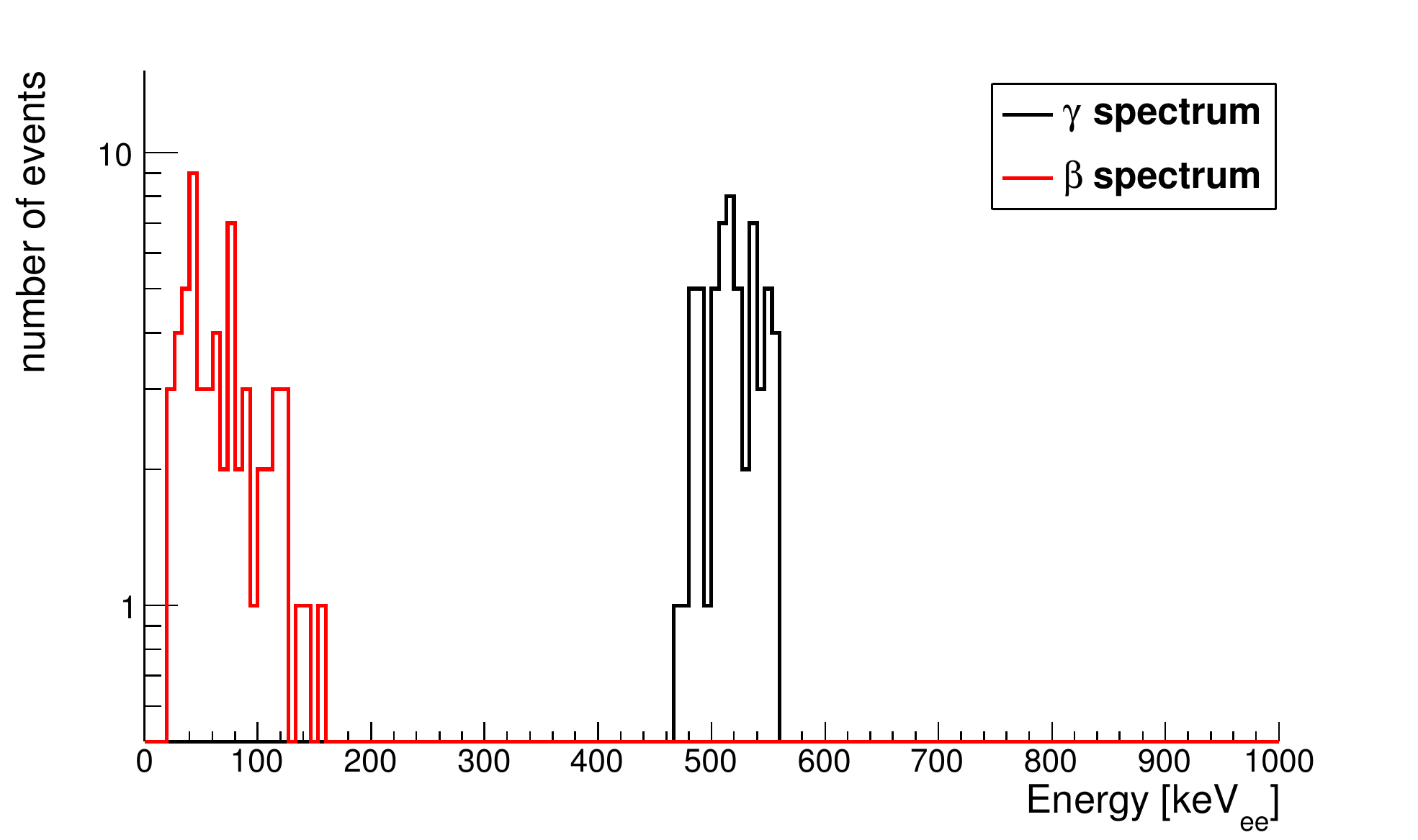}
	\caption{\label{fig:85Kr}(Color online) Gamma and beta energy spectra of the $^{85}$Kr
	$\beta-\gamma$ decay (0.434\% BR) from the XENON10 data, after delayed
	coincidence cuts.}
\end{figure}

The delayed coincidence technique has also been used to estimate the Krypton
concentration in LXe, by tagging the beta-gamma decay mode (0.434 \%
branching ratio) of the $^{85}$Kr diluted in the liquid. $^{85}$Kr, which has a concentration about 10$^{-11}$ in Kr, is a beta emitter that produces background in the LXe volume. XENON10 experiment requires the Kr/Xe contamination to be less than 10~ppb, which gives about 0.2~dru electron recoil background. To have this low Kr/Xe contamination level, the Xe gas used for XENON10 was purified  by the Spectra Gases company~\cite{spectra} in several passages through  a large cryogenic distillation column.  Figure \ref{fig:85Kr} shows the energy spectra of the gamma and beta particles selected by the cuts. A
total of 59 delayed-coincidence events were found in the entire WIMP-search
data. Taking into account the detection efficiency of these events, we determine
an $^{85}$Kr activity of (1.0$\pm$0.06)\,mBq/kg, which translates into
(5.0$\pm$0.3)\,ppb of natural Kr/Xe contamination for typical atmospheric
$^{85}$Kr abundances. This value is compatible with the value quoted by the Xe gas supplier~\cite{spectra} and is sufficiently low for XENON10.

\begin{eqnarray*}
^{214}Bi\;(Q_{\beta}:3.272 MeV) 
& \stackrel{\beta}{\longrightarrow} & ^{214}Po\;(E_{\alpha}:7.833 MeV) \\
&\stackrel{\alpha}{\longrightarrow} & ^{210}Pb\; (\tau=164.3 \mu s) \label{eq:Rn} \\
^{212}Bi\;(Q_{\beta}:2.254
MeV) & \stackrel{\beta}\longrightarrow & ^{212}Po\;(E_{\alpha}:8.954 MeV)\\
& \stackrel{\alpha}\longrightarrow& ^{208}Pb\; (\tau=0.299\,\mu s) \label{eq:212Bi} \\
^{85}Kr\;(Q_{\beta}:173 keV) & \stackrel{\beta}\longrightarrow & ^{85}Rb^{*}(E_{\gamma}:514 keV)\\
& \stackrel{\gamma}\longrightarrow & ^{85}Rb\; (\tau=1.015 \, \mu s)\label{eq:85Kr}
\end{eqnarray*}

\begin{acknowledgments}
This work was supported by the National Science Foundation under grants No.~PHY-03-02646 and PHY-04-00596, and by the Department of Energy under Contract No.~DE-FG02-91ER40688, the CAREER Grant No.~PHY-0542066, the Volkswagen Foundation (Germany) and the FCT Grant No.~POCI/FIS/60534/2004 (Portugal). We thank the Director of the Gran Sasso National Laboratory, Prof. E. Coccia, and his  staff for support throughout this effort. Special thanks go to the Laboratory engineering team, led by P. Aprili, and to F.~Redaelli of COMASUD for their contribution to the XENON10 installation. We are also thankful to M. Laubenstein for the radioactivity screening of the many XENON10 materials.
\end{acknowledgments}


\end{document}